\documentclass[twocolumn]{aastex62}

\usepackage{multirow}
\usepackage{afterpage}
\usepackage{booktabs}

\graphicspath{{./}{figures/}}

\received{}
\revised{}
\accepted{\today}
\submitjournal{ApJ}

\shorttitle{{\it UVIT} Study of M67}
\shortauthors{Jadhav et al.}

\begin{document}

\title{{\it UVIT} Open Cluster Study. II. Detection of extremely low mass white dwarfs and post-mass transfer binaries in M67}

\author[0000-0002-8672-3300]{Vikrant V. Jadhav}
\affiliation{Joint Astronomy Program and Physics Department, Indian Institute of Science, Bangalore - 560012, India}
\affiliation{Indian Institute of Astrophysics, Koramangala II Block, Bangalore-560034, India}
\email{vikrant.jadhav@iiap.res.in}

\author[0000-0001-6006-1727]{N. Sindhu}
\affiliation{Indian Institute of Astrophysics, Koramangala II Block, Bangalore-560034, India}
\affiliation{Department of Physics, School of Advanced Science, VIT, Vellore -632014, India}

\author{Annapurni Subramaniam}
\affiliation{Indian Institute of Astrophysics, Koramangala II Block, Bangalore-560034, India}

\begin{abstract}

A detailed study of the ultraviolet (UV) bright stars in the old open star cluster, M67 is presented based on the far-UV observations using the {\it Ultra Violet Imaging Telescope} ({\it UVIT}) on {\it ASTROSAT}. The UV and UV-optical colour-magnitude diagrams (CMDs) along with overlaid isochrones are presented for the member stars, which include blue straggler stars (BSSs), triple systems, white dwarfs (WDs) and spectroscopic binaries (SB). The CMDs suggest the presence of excess UV flux in many members, which could be extrinsic or intrinsic to them. We construct multi-wavelength spectral energy distribution (SED) using photometric data from the {\it  UVIT, Gaia} DR2, 2MASS and WISE surveys along with optical photometry. We fitted model SEDs to 7 WDs and find 4 of them have mass $>$ 0.5 \(M_\odot\) and cooling age of less than 200 Myr, thus demanding BSS progenitors. SED fits to 23 stars detect extremely low mass (ELM) WD companions to WOCS2007, WOCS6006 and WOCS2002, and a low mass WD to WOCS3001, which suggest these to be post mass transfer (MT) systems. 12 sources with possible WD companion need further confirmation. 9 sources have X-ray and excess UV flux, possibly arising out of stellar activity. This study demonstrates that UV observations are key to detect and characterise the ELM WDs in non-degenerate systems, which are ideal test beds to explore the formation pathways of these peculiar WDs. The increasing detection of post-MT systems among BSSs and main-sequence stars suggest a strong MT pathway and stellar interactions in M67. 

\end{abstract}

\keywords{ultraviolet: stars --- (stars:) blue stragglers --- (stars:) Hertzsprung–Russell and C–M diagrams --- (stars:) white dwarfs --- (Galaxy:) open clusters and associations: individual:(M67)}

\section{Introduction} \label{sec:intro}
The evolution of stars in close binaries and multiple stellar systems within star clusters will be different due to the interactions with their neighbours. The high stellar density in globular clusters causes collisions leading to mergers, creation and disruption of binary systems. Open clusters (OCs), on the other hand, provide ample examples of binaries which have more chance of remaining relatively undisturbed due to the low stellar densities.

The evolution of a binary star occurs in multiple pathways in OCs as it depends on their orbital parameters. Very long period binaries are likely to evolve independent of each other, while closer binaries may merge or undergo mass transfer (MT) \citep{Perets2015}. Among contact binaries, W UMa-type binaries evolve into a contact configuration from initially detached systems by angular momentum loss via magnetic torques from a stellar wind in which the spin angular momentum and the orbital angular momentum are coupled through tides \citep{Vilhu1982, Guinan1988, Eggen1989}. Estimates based on the level of chromospheric and coronal activity exhibited by components of short-period main-sequence binaries suggest that systems with initial orbital periods of a few days may evolve into a contact configuration on a time-scale of a few Gyr. W UMa systems ultimately coalesce into single stars \citep{Webbink1985}, which provide a natural pathway for the formation of BSSs.  

The stars formed from the binary evolution in OCs are generally detected in ultra-violet (UV) and X-rays \citep{Eaton1980, Geske2005}. On the other hands, young WDs emit in the UV region due to their high-temperature \citep{Sindhu2018}. Hot-spots in contact and semi-detached binaries also show enhanced UV flux. \citet{Kouzuma2019} gave examples of stellar hot-spots in contact binaries showing hot-spots with $T_{eff}$ of 4500 to 11000 K where the hotter hot-spots can give significant UV flux. Single and binary stars that show magnetic activity contribute to the total UV flux emitted by intermediate-age star clusters. Chromospheric activity on the stellar surface can reach temperature of 7000 to 8000 K \citep{Linsky2017, Hall2008} which could also produce UV flux. Flares on the stars are also sources of transient UV radiation. Many of these systems also contribute to the X-ray flux. Coronal emissions at very high temperature, capable of producing X-rays, can emit in UV region. It is important to note that the hot-spots, flares, coronal activity and very hot WDs also produce X-rays as well as a significant flux in the UV \citep{Dempsey1993, Mitra2005}. Some active stars like the RS CVn type stars have spots resulting in excess emission in UV and X-rays \citep{Walter1981}. Among contact binaries, W-Uma type stars are the most common and are found in intermediate-age OCs like M67 and NGC 188 \citep{Geller2015, Chen2016}. These systems are known to have excess UV flux, along with detectable X-ray flux. Semi-contact binaries may also develop hot spots resulting in excess UV flux \citep{Polubek2003}. Therefore, it is important to identify the source of UV flux in known binary systems, as it could be due to the intrinsic property of the star, or due to the presence of a hot companion. This is particularly important in the case of single-lined spectroscopic binaries (SB1), where a sub-luminous companion is expected.

Two intermediate age OCs with well-identified member stars, along with well-studied binary properties, through proper motion and radial velocity studies, are NGC 188 and M67. These clusters are well known to have a large fraction of various type of binaries including contact binaries. The M67 star cluster is well studied through photometry in several wavelength bands covering from the X-rays to the Infra-Red (IR) regions \citep{Belloni1993, Belloni1998, Van2004, Landsman1997, Mathieu1986, Sarajedini2009}, and through spectroscopy \citep{Mathieu1990, Shetrone2000, Bertelli2018}. The cluster has a rich population of exotic stellar types, that do not follow the standard single stellar evolutionary theory. 

Old OCs are also ideal sites to study the properties of white dwarfs (WDs) \citep{Kalirai2010}.
In general, WDs detected in OCs are the end products of single star evolution. Hence, typically the mass of a WD that is recently formed in OCs is from a progenitor with the MS turn-off (MSTO) mass of the cluster. \citet{Williams2018} detected $\sim$ 50 WD candidates in M67 and estimated their mass and spectral type, where many WDs required a progenitor more massive than a single star at MSTO of M67. Therefore, they concluded that these high mass WDs are likely to be evolved from blue stragglers stars (BSSs).  Similarly, \citet{Sindhu2019} detected WDs with mass $<$0.3 M$_\odot$ as a companion to a BSS in M67. They suggest that the formation pathway MT in a binary produces a BSS with an initially hot companion, such as a WD. As single star evolution takes more time than the age of the universe to form such extremely low-mass WDs (ELMs), they must have undergone significant mass loss during their evolution in close binary systems \citep{Brown2010} and have never ignited helium in their cores. 
WDs of mass up to 0.1 \(M_\odot\) were found to be part of contact binaries \citep{Marsh1995, Benvenuto2005}, whose progenitors lost mass due to Roche lobe overflow. 
ELM WDs are generally found in binary systems where the companions are neutron stars/pulsars \citep{Driebe1998, Lorimer2008}, WDs \citep{Brown2016}, or A/F MS stars in EL CVn-type systems \citep{Maxted2014, Wang2018}. Recently two R CMa-type eclipsing binaries are suggested to have precursors of low mass He WDs \citep{Wang2019}.

M67 has been studied in UV by \citet{Landsman1998}, \citet{Siegel2014}, \citet{Sindhu2018} and  \citet{Sindhu2019}. \citet{Landsman1998} studied 11 BSSs, 7 BSS and 1 yellow giant (YG) using \textit{Ultraviolet Imaging Telescope} (\textit{UIT}) using 1210 s exposure in single FUV filter. They found the integrated UV spectrum was dominated by BSSs and some stars indicate hot subluminous companions. \citet{Siegel2014} used Ultraviolet
Optical Telescope (UVOT) aboard the \textit{Swift Gamma-Ray Burst Mission} to study the NUV CMD of M67 encompassing 10 BSSs and a WD.

We have recently started a program to understand the UV properties of binary and single stars in intermediate and old OCs. The first paper in this series was a study of the UV properties of M67 stars by \citet{Sindhu2018}, which identified a number of UV bright stars using GALEX. Some were found to be bright in the far-UV (FUV), whereas a larger number were found to be bright in the near-UV (NUV), which comprised of a wide variety of stars. They found two RGs to be bright in the FUV, which was explained due to the presence of chromospheric activity in them, as traced by the MgII line emission in the \textit{International Ultraviolet Explorer (IUE)} spectra. The authors also speculated that many UV bright stars located near the MSTO could be chromospherically active.

In order to understand the properties of the FUV bright stars detected by \citet{Sindhu2018}, we carried out an imaging study of M67 in FUV, using the \textit{Ultra Violet Imaging Telescope} (\textit{UVIT}) on \textit{ASTROSAT}. \textit{UVIT} has superior spatial resolution (1.5\arcsec) when compared to \textit{Galaxy Evolution Explorer} \textit{GALEX} \citep{Martin05} ($>$4\arcsec) and hence will be able to obtain uncontaminated photometry of individual members of M67 that have detectable flux in the FUV. We observed M67 using three FUV filters to comprehensively study the UV bright population.
\citet{Sindhu2019} detected an ELM WD companion to one of the BSS in M67 using these observations. We further this study by analysing other UV sources in M67 and to ascertain the connection between the stellar type and UV/X-ray emission.
We created colour-magnitude diagrams as well as spectral energy distribution (SEDs) to estimate the fundamental parameters such as luminosity, temperature and radius. We analyse 30 members of M67 individually to assess the source of the UV flux.

The paper is arranged as follows: Observations and archival data is in section~\ref{sec:obs}. Membership, Colour Magnitude Diagrams (CMD), SED fitting and mass estimation in section~\ref{sec:Analysis}. Results and discussion is given in section~\ref{sec:results} and section~\ref{sec:Discussion} respectively.

\section{Observations and Analysis}
\label{sec:obs}
\subsection{{\it UVIT} Data}
\label{sec:UVIT_Data} 
\begin{figure}
\centering
\includegraphics[width=0.45\textwidth]{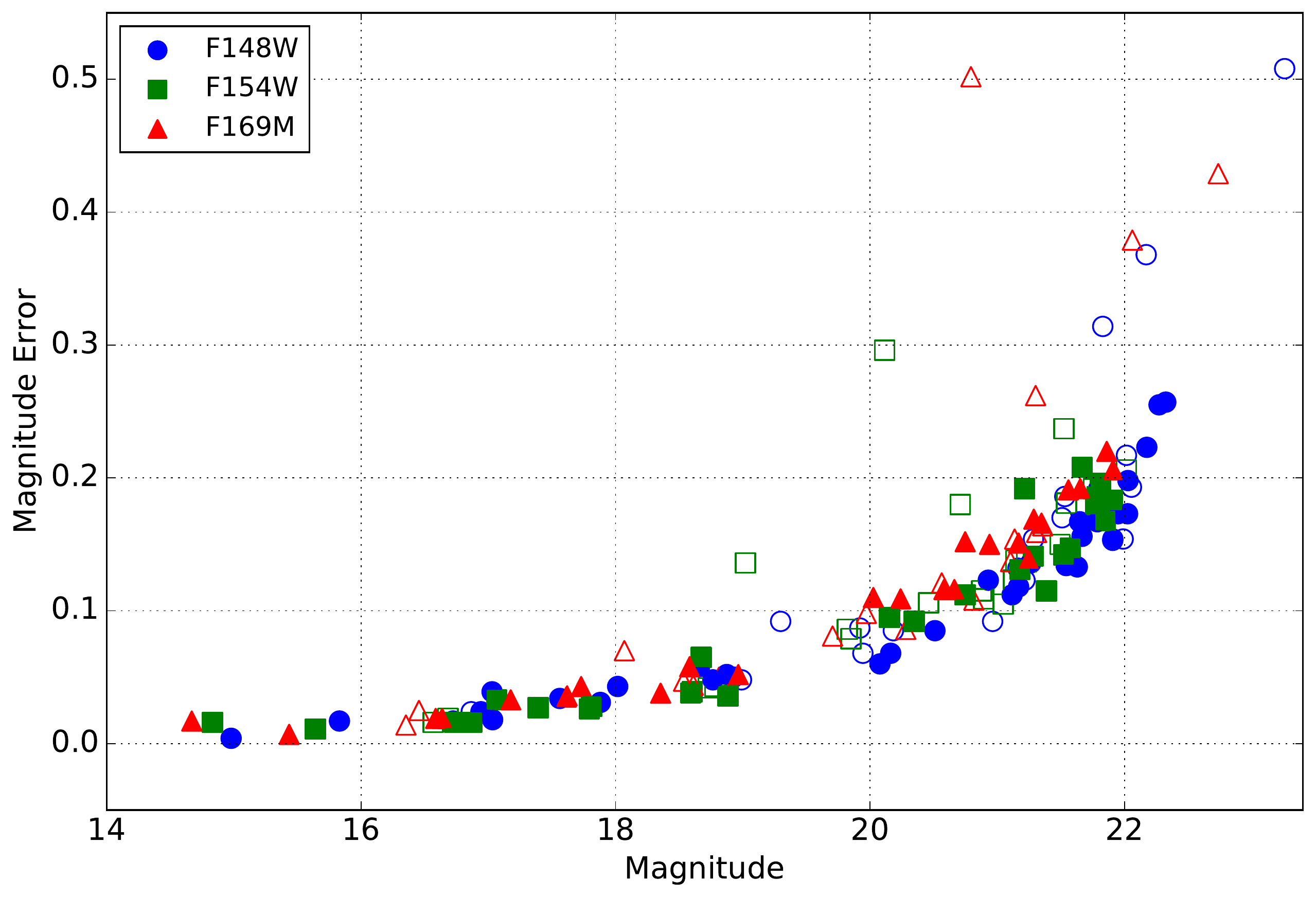}
    \caption{The photometric error in magnitudes for all three filters. Solid points are M67 members while hollow points are other detected stars.}
    \label{fig:Err_V_mag}
\end{figure}

\begin{figure}
\centering
\includegraphics[width=0.45\textwidth]{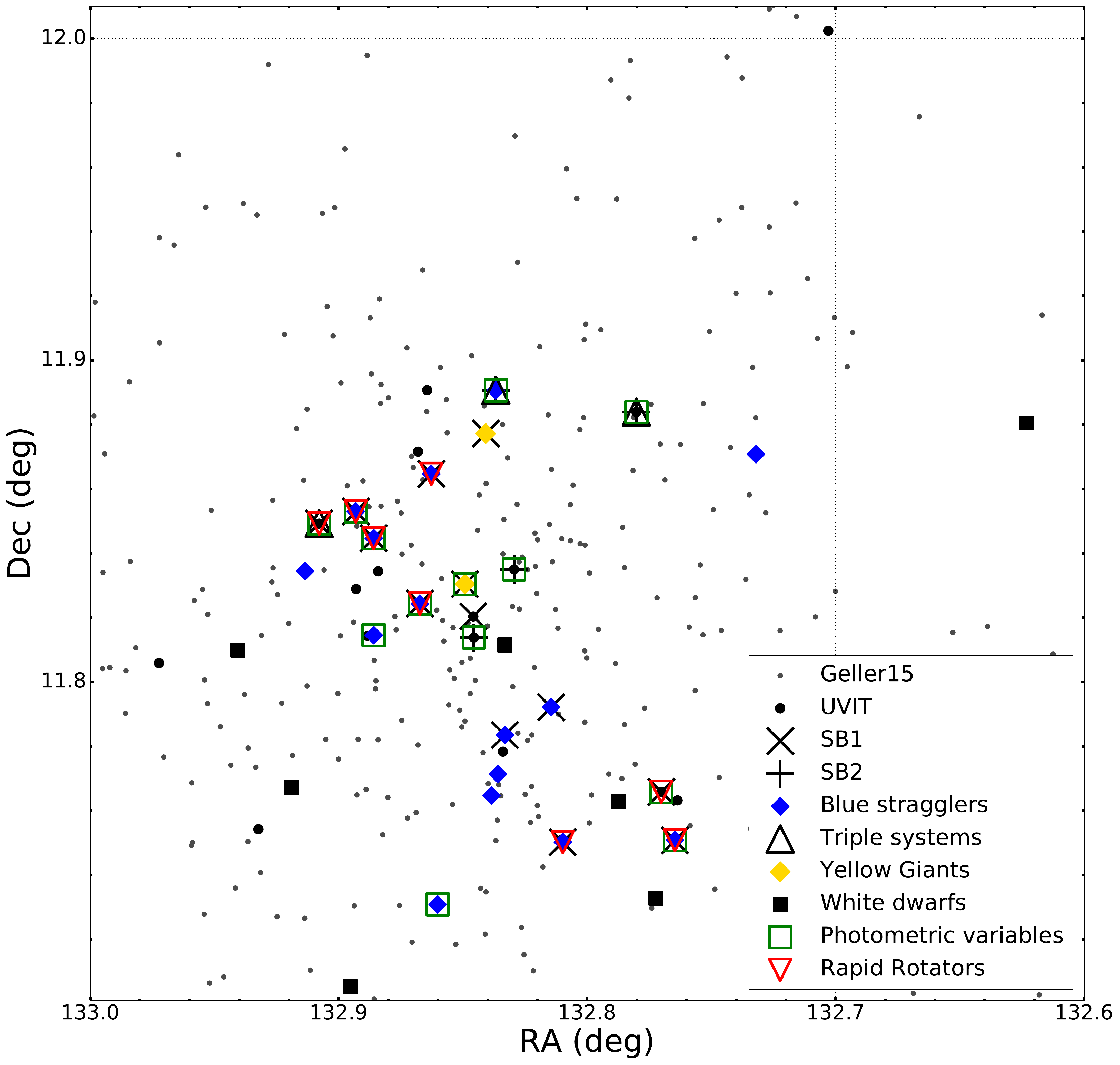}
    \caption{Spatial distribution of M67 members as observed by {\it UVIT} along with the members from \citet{Geller2015} catalogue. The M67 members according to \citet{Geller2015} are shown grey, while the members detected by \textit{UVIT} are stylised according to their known classification.}
    \label{fig:Members_UVIT_and_Geller}
\end{figure}

\begin{table*}
\centering
\caption{The measured magnitudes (not corrected for extinction) of WD and other stellar sources of M67 from three {\it UVIT} filters. Probability of Proper Motion membership (PPM) and Probability of Radial Velocity membership (PRV)  are obtained from $^a$ \citet{Yadav2008}, $^b$ \citet{Zhao1993}, $^c$ \citet{Girard1989}, $^d$ \citet{Geller2015}, $^e$ \citet{Williams2018}, $^f$ Photometric member \citep{Williams2018}.}
\label{tab:photometry}
\begin{tabular}{lcccc cccr} 
\toprule
\textbf{Name}	&	\textbf{RA} (deg)	&	\textbf{Dec} (deg)	&	\textbf{F148W} (AB mag)	&	\textbf{F154W} (AB mag)	&	\textbf{F169M} (AB mag)	&	\textbf{PPM}	&	\textbf{PRV$^d$}	\\ \toprule
WOCS1001	&	132.84560	&	11.81378	&	21.89$\pm$0.18	&	21.67$\pm$0.21	&	21.65$\pm$0.19	&	99$^a$	&	98	\\
WOCS1006	&	132.86270	&	11.86466	&	14.98$\pm$0.00	&	14.83$\pm$0.02	&	14.67$\pm$0.02	&	99$^c$	&	--	\\
WOCS1007	&	132.89310	&	11.85297	&	17.03$\pm$0.02	&	16.87$\pm$0.02	&	16.64$\pm$0.02	&	99$^c$	&	90	\\
WOCS11005	&	132.83390	&	11.77832	&	21.91$\pm$0.15	&	--	&	--	&	100$^b$	&	98	\\
WOCS11011	&	132.77015	&	11.76581	&	22.02$\pm$0.17	&	--	&	--	&	98$^c$	&	--	\\
WOCS2002	&	132.84918	&	11.83040	&	18.66$\pm$0.06	&	18.67$\pm$0.07	&	18.58$\pm$0.06	&	99$^c$	&	98	\\
WOCS2003	&	132.82937	&	11.83498	&	--	&	21.85$\pm$0.17	&	--	&	100$^a$	&	98	\\
WOCS2007	&	132.83850	&	11.76469	&	21.54$\pm$0.13	&	21.52$\pm$0.14	&	21.56$\pm$0.19	&	99$^c$	&	89	\\
WOCS2008	&	132.84077	&	11.87721	&	20.93$\pm$0.12	&	20.75$\pm$0.11	&	20.58$\pm$0.12	&	99$^c$	&	97	\\
WOCS2009	&	132.83674	&	11.89064	&	17.56$\pm$0.03	&	17.39$\pm$0.03	&	17.73$\pm$0.04	&	99$^c$	&	98	\\
WOCS2011	&	132.86013	&	11.73081	&	15.83$\pm$0.02	&	15.64$\pm$0.01	&	15.43$\pm$0.01	&	99$^c$	&	97	\\
WOCS2012	&	132.76361	&	11.76317	&	22.32$\pm$0.26	&	--	&	--	&	100$^a$	&	98	\\
WOCS2015	&	132.73204	&	11.87076	&	21.63$\pm$0.13	&	--	&	--	&	99$^c$	&	98	\\
WOCS3001	&	132.84580	&	11.82036	&	21.67$\pm$0.16	&	21.39$\pm$0.12	&	21.29$\pm$0.17	&	100$^a$	&	98	\\
WOCS3005	&	132.88590	&	11.81452	&	16.94$\pm$0.02	&	16.74$\pm$0.02	&	16.59$\pm$0.02	&	99$^c$	&	94	\\
WOCS3009	&	132.91350	&	11.83443	&	21.80$\pm$0.19	&	21.57$\pm$0.15	&	21.17$\pm$0.15	&	99$^c$	&	98	\\
WOCS3010	&	132.80980	&	11.75018	&	18.76$\pm$0.05	&	18.59$\pm$0.04	&	18.35$\pm$0.04	&	90$^d$	&	98	\\
WOCS3012	&	132.78017	&	11.88390	&	21.65$\pm$0.17	&	21.28$\pm$0.14	&	20.66$\pm$0.12	&	100$^a$	&	98	\\
WOCS3013	&	132.76464	&	11.75078	&	18.02$\pm$0.04	&	17.79$\pm$0.03	&	17.62$\pm$0.04	&	99$^c$	&	61	\\
WOCS4003	&	132.86730	&	11.82434	&	20.51$\pm$0.09	&	20.35$\pm$0.09	&	20.03$\pm$0.11	&	100$^a$	&	--	\\
WOCS4006	&	132.88590	&	11.84466	&	17.88$\pm$0.03	&	17.81$\pm$0.03	&	17.61$\pm$0.04	&	99$^c$	&	--	\\
WOCS4015	&	132.97231	&	11.80585	&	--	&	21.81$\pm$0.20	&	--	&	99$^c$	&	98	\\
WOCS5005	&	132.83309	&	11.78349	&	20.08$\pm$0.06	&	--	&	--	&	99$^c$	&	98	\\
WOCS5007	&	132.86807	&	11.87159	&	--	&	21.77$\pm$0.18	&	21.35$\pm$0.17	&	100$^a$	&	98	\\
WOCS5013	&	132.93230	&	11.75419	&	22.03$\pm$0.20	&	--	&	--	&	100$^a$	&	93	\\
WOCS6006	&	132.89300	&	11.82889	&	21.79$\pm$0.17	&	21.79$\pm$0.19	&	21.25$\pm$0.14	&	100$^a$	&	98	\\
WOCS7005	&	132.88410	&	11.83439	&	--	&	--	&	21.86$\pm$0.22	&	99$^a$	&	91	\\
WOCS7009	&	132.90787	&	11.84922	&	--	&	21.90$\pm$0.18	&	--	&	97$^c$	&	--	\\
WOCS7010	&	132.86439	&	11.89071	&	--	&	21.81$\pm$0.18	&	--	&	100$^a$	&	98	\\
WOCS8005	&	132.88843	&	11.81432	&	22.27$\pm$0.26	&	--	&	--	&	100$^a$	&	98	\\
WOCS8006	&	132.83588	&	11.77128	&	21.26$\pm$0.14	&	--	&	20.75$\pm$0.15	&	100$^a$	&	98	\\
WOCS8010	&	132.81020	&	11.75099	&	--	&	21.21$\pm$0.19	&	--	&	--	&	91	\\
WOCS9005	&	132.81447	&	11.79214	&	--	&	21.18$\pm$0.13	&	--	&	99$^c$	&	98	\\
WOCS9028	&	132.70291	&	12.00243	&	--	&	--	&	21.91$\pm$0.21	&	99$^c$	&	94	\\ \hline
Y1168	&	132.83310	&	11.81147	&	17.03$\pm$0.04	&	17.07$\pm$0.03	&	17.18$\pm$0.03	&	Y$^d$	&	--	\\
Y563	&	132.89530	&	11.70523	&	18.87$\pm$0.05	&	18.88$\pm$0.04	&	18.96$\pm$0.05	&	Y$^d$	&	--	\\
Y886	&	132.91900	&	11.76718	&	20.16$\pm$0.07	&	20.15$\pm$0.10	&	20.24$\pm$0.11	&	Y$^d$	&	--	\\
Y1157$^f$	&	132.94060	&	11.80987	&	21.12$\pm$0.11	&	--	&	--	&	--	&	--	\\
Y701	&	132.77230	&	11.73277	&	21.95$\pm$0.17	&	--	&	--	&	Y$^d$	&	--	\\
Y856	&	132.78730	&	11.76274	&	22.17$\pm$0.22	&	--	&	--	&	Y$^d$	&	--	\\
Y1487	&	132.62320	&	11.88052	&	21.17$\pm$0.12	&	--	&	20.94$\pm$0.15	&	Y$^d$	&	--	\\ \hline
\bottomrule
\end{tabular}
\end{table*}

We observed M67 with the \textit{UVIT} on {\it ASTROSAT}, the first Indian space observatory, launched on 2015 September 28. The near-simultaneous observations of M67 were carried out by the {\it UVIT} on 2017 April 23. The telescope has three channels: FUV (130-180 nm), NUV (200-300 nm) and VIS (350-550nm), where the VIS channel is mainly used to correct the drift of the spacecraft. We used three filters in the FUV region viz. F148W (1481$\pm$250 \AA), F154W (1541$\pm$190 \AA), F169M (1608$\pm$145 \AA). 
The details of effective area curves, \textit{UVIT} calibration and instrumentation can be found in \citet{Tandon2017a} and  \citet{Kumar2012}. The band-passes are shown in the top panel of Fig.~\ref{fig:SED_3009} (c).

NUV data could not be obtained due to some instrument-related issues. The data reduction and drift correction were performed using {\sc CCDLAB} \citep{2017PASP..129k5002P}. We created science ready images with the following exposure times: F148W = 2290 s; F154W = 2428 s; F169M = 2428 s. The {\it UVIT} images (shown in figure 1 of \citet{Sindhu2019}) were analysed using the {\sc daophot} package of {\sc IRAF} by performing point spread function (PSF) photometry on all three images. The PSF magnitudes were corrected for aperture and saturation corrections. The astrometry of the detected stars was done using co-ordinate cross-match within 1$\arcsec$ with {\it GALEX} and \citet{Montgomery1993} catalogues. 

We detected a total of 133, 114 and 92 stars in F148W, F154W and F169M filters respectively. The variation of error with magnitudes in all three filters is shown in Fig.~\ref{fig:Err_V_mag}. It shows that we have detected objects up to 22 mag with a maximum error of 0.25 mag for the faintest members. \textit{GALEX} is about twice as sensitive compared to \textit{UVIT} in the FUV. Also, long exposure observations of M67 by \textit{GALEX} are available \citep{Sindhu2018}.

\subsection{Archival Data}
\label{sec:Archival_Data}

\begin{figure*}
  \centering
  \begin{tabular}{c c}
    \includegraphics[width=.45\textwidth]{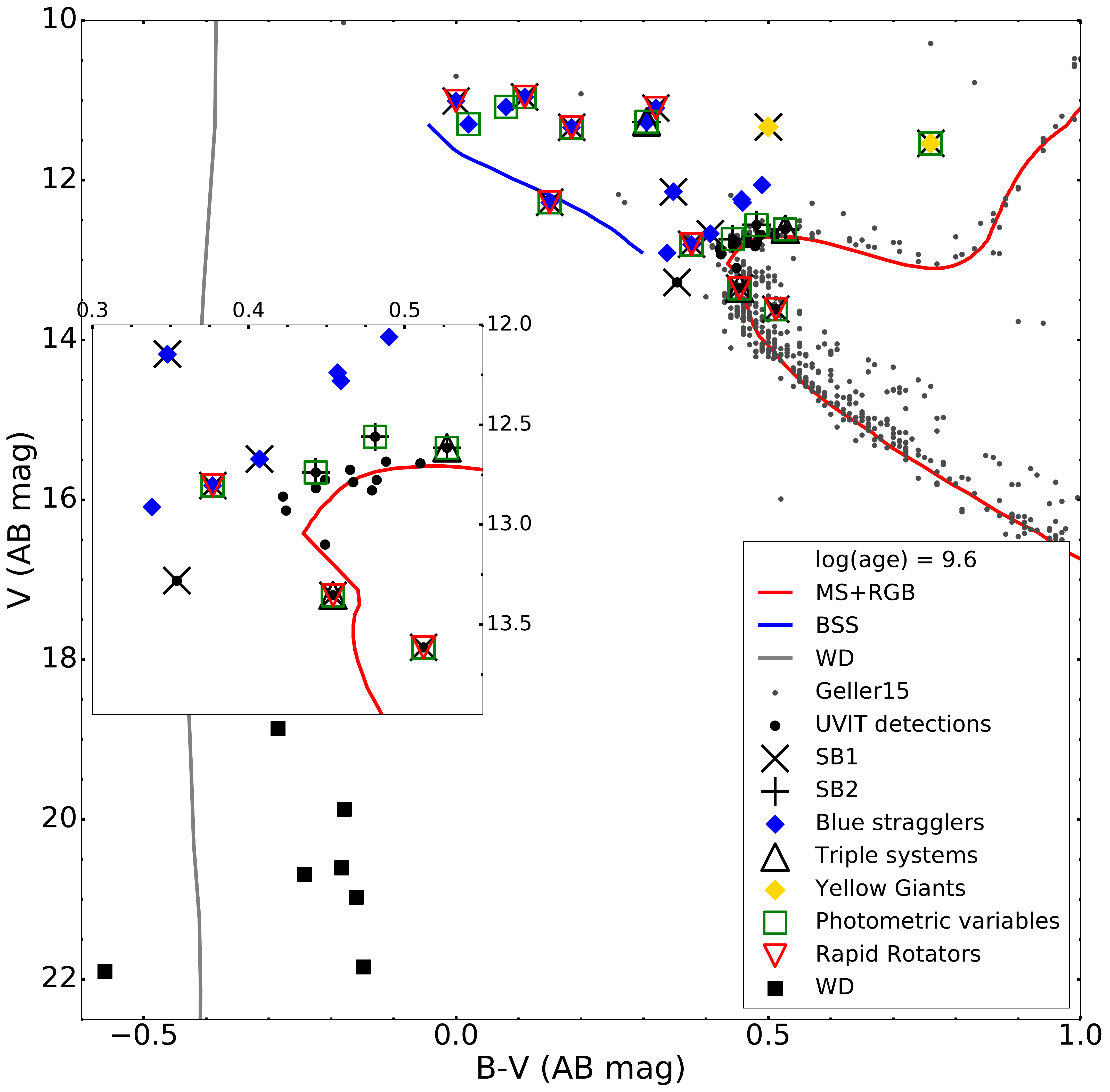} &
    \includegraphics[width=.45\textwidth]{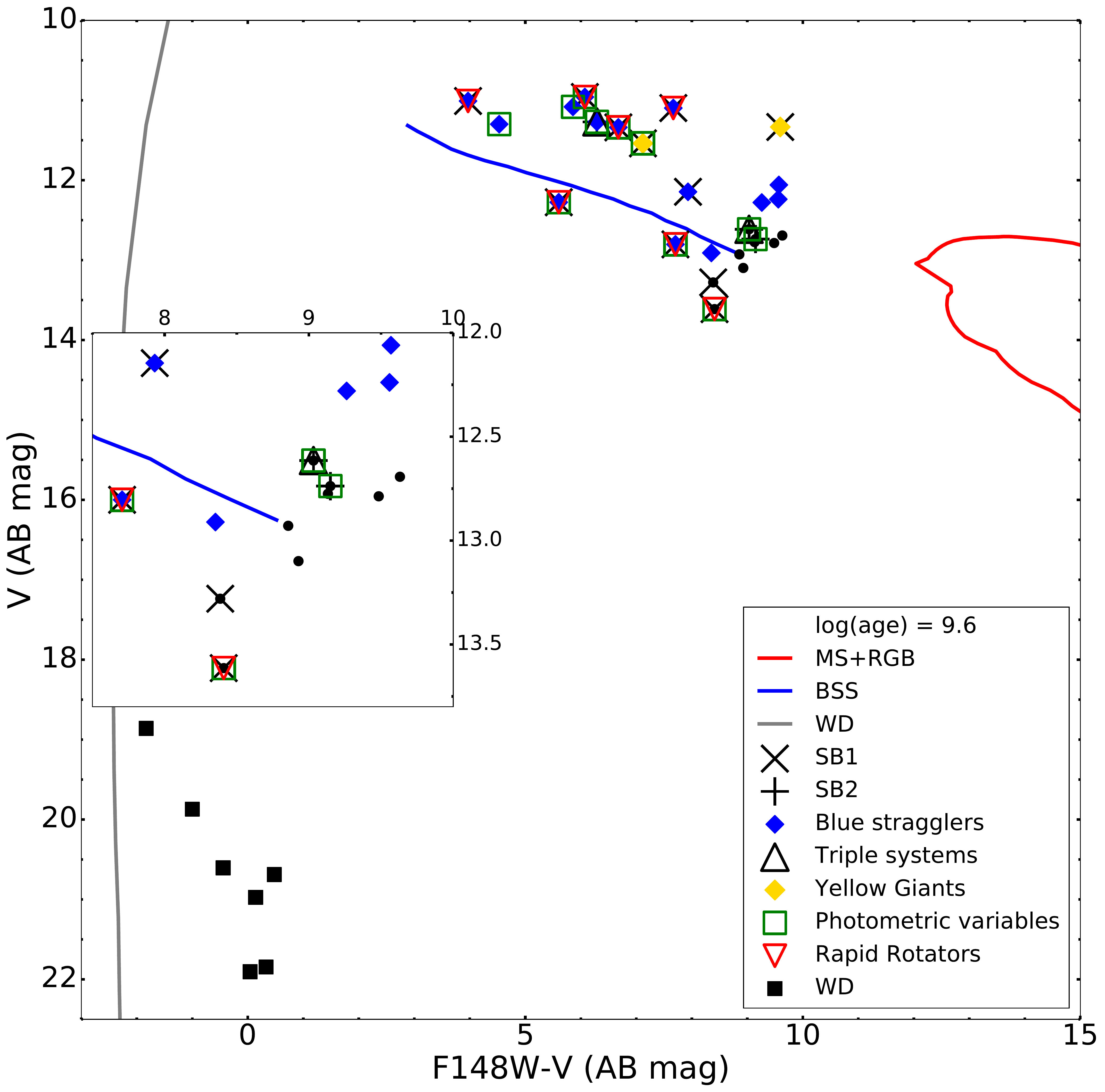} \\
    (a) & (b) \\
    \includegraphics[width=.45\textwidth]{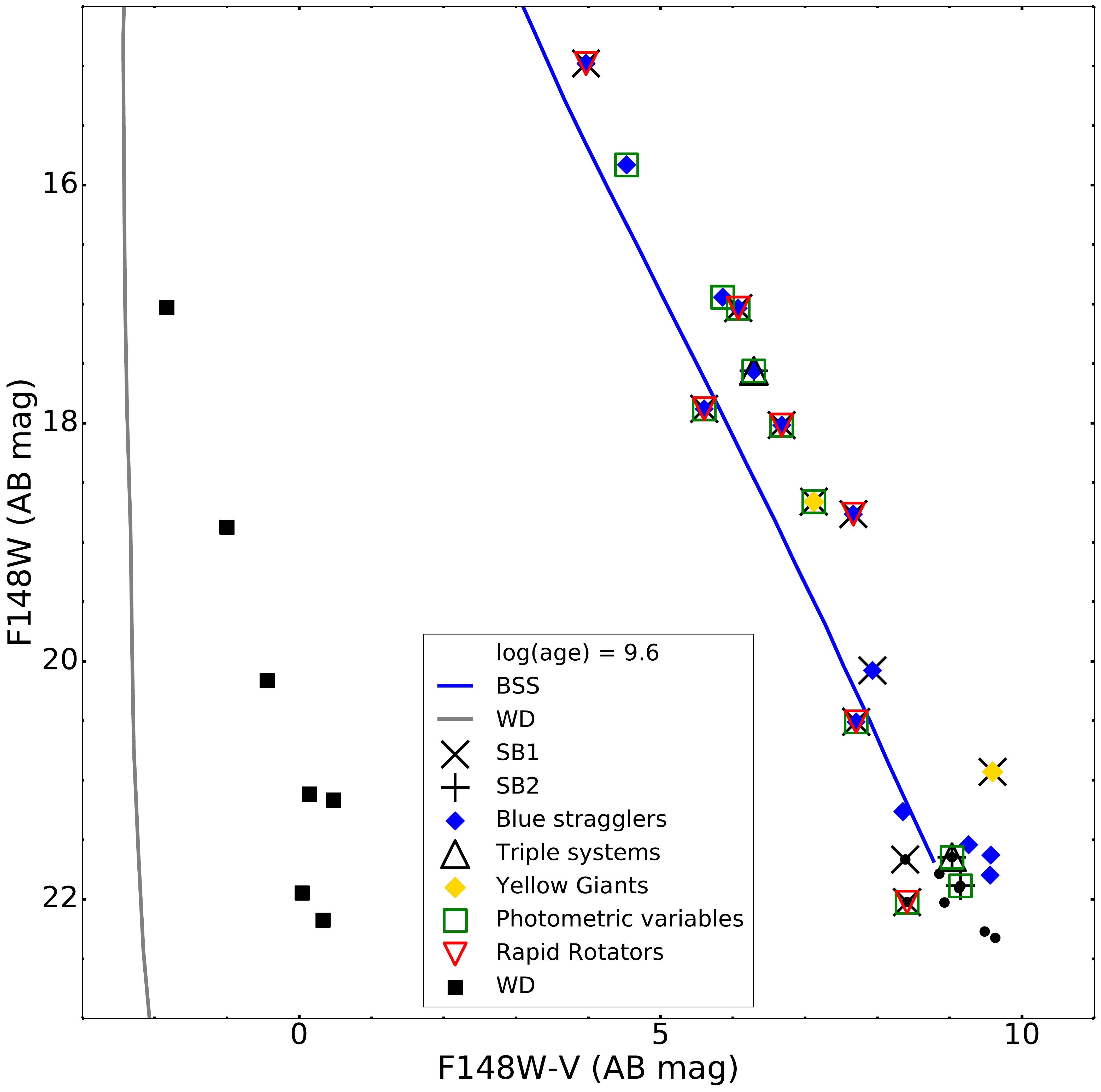} &
    \includegraphics[width=.45\textwidth]{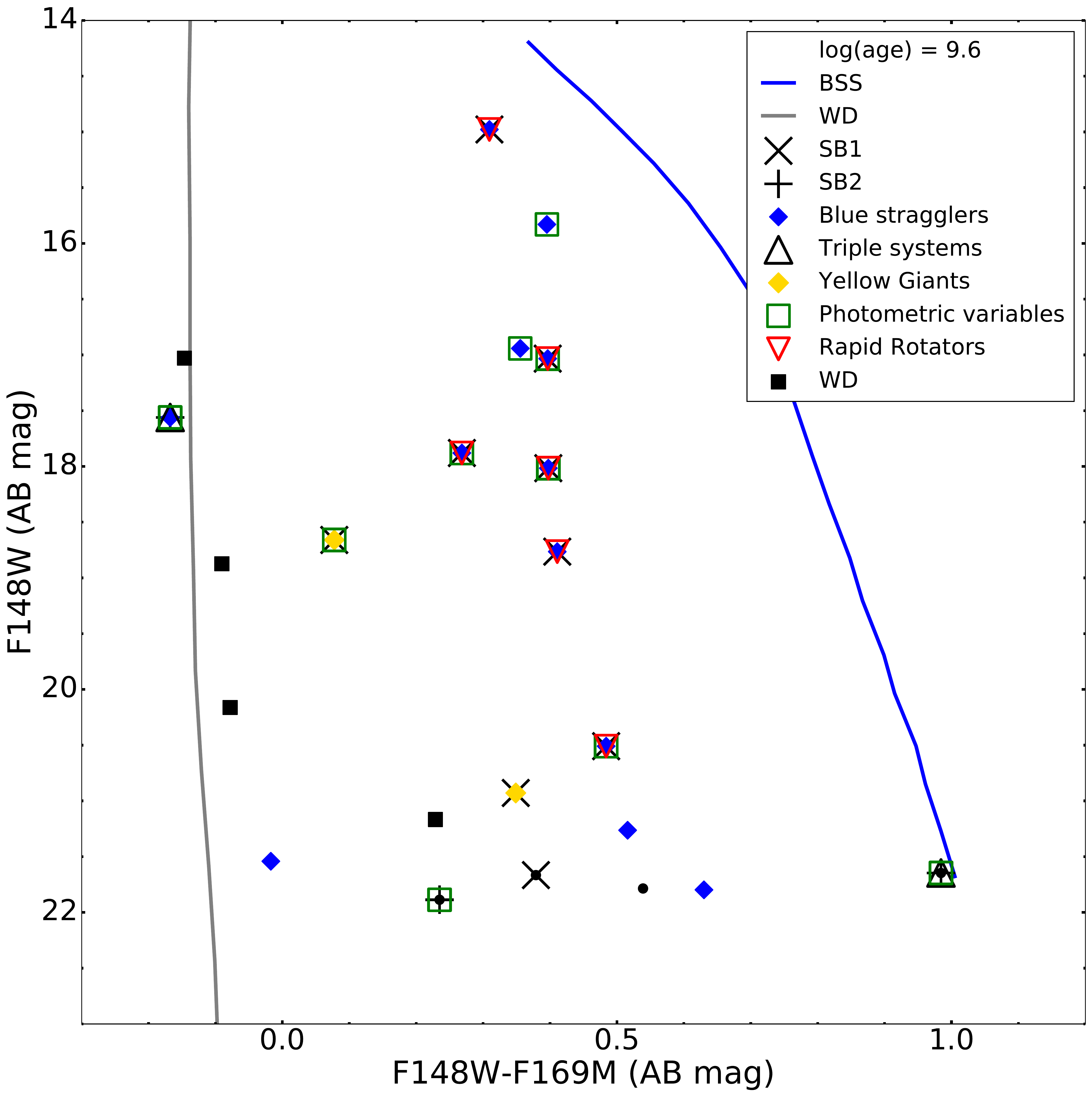} \\
    (c) & (d) \\
  \end{tabular}
  \caption{(a) Optical CMD of M67. The members listed by \citet{Geller2015} are shown in grey dots while the unclassified members detected by {\it UVIT} are shown as black dots. The isochrone (age = 3.98 Gyr) is generated with FSPS code using BaSTI model. The main sequence-subgiant-red giant phase is shown with red colour, expected location of BSS and WDs are shown in blue and grey colour respectively.
  The inset shows the expanded view of  the turn-off region. (b) UV-Optical CMD (V, F148W$-$V) of M67. The inset shows the expanded view of the fainter end of the BS sequence. (c) UV-Optical CMD (F148W, F148$-$V) of M67. The inset shows the expanded view of  the turn-off region. (d) UV CMD (F148W, F148W$-$F169M) of M67}
  \label{fig:CMD}
\end{figure*}

We combine the {\it UVIT} data with the data in the longer wavelengths to identify and characterise the detected stars. All cross-matches were done with a maximum separation of 3\arcsec.

Flux measurements from UV to IR bands were obtained as follows: FUV (1542$\pm$200 \AA), NUV (2274$\pm$530 \AA) from \textit{GALEX}; U (3630$\pm$296 \AA), B (4358$\pm$502 \AA), V (5366$\pm$470 \AA), R (6454$\pm$776 \AA) \& I (8100$\pm$912 \AA) from Kitt Peak National Observatory (KPNO, \citealt{Montgomery1993}); Gbp (5050$\pm$1172 \AA), G (6230$\pm$2092 \AA) \& Grp (7730$\pm$1378 \AA) from \textit{Global Astrometric Interferometer for Astrophysics} {\it Gaia} DR2 \citep{Gaia2018a}; J (12350$\pm$812 \AA), H (16620$\pm$1254 \AA) \& Ks (21590$\pm$1309 \AA) from Two Micron All-Sky Survey (2MASS, \citealt{Skrutskie2006}, \citealt{Ochsenbein2000}); W1 (33526$\pm$3313 \AA), W2 (46028$\pm$5211 \AA) \& W3 (115608$\pm$27528 \AA) from Wide-Field Infrared Survey Explorer (WISE, \citealt{Ochsenbein2000}, \citealt{Wright2010}). 

For WDs, we included photometry from two M67 catalogues: B (4525$\pm$590 \AA), V (5340$\pm$520 \AA), I (9509$\pm$2000 \AA) from LaSilla \citep{Yadav2008}; G (4581$\pm$830 \AA), U (3550$\pm$350 \AA), R (6248$\pm$800 \AA) from MMT \citep{Williams2018}. IR photometry is not available for the WDs.

\section{Analysis}
\label{sec:Analysis}

\subsection{Membership}
\label{sec:Membership}

Among the detected stars, we identified 34 members by cross-matching with \citet{Geller2015} catalogue with proper motion membership probability \citep{Yadav2008} or radial velocity membership probability over 90\%. Among these members, 16 stars are catalogued by \citealt{Geller2015} as BSSs. However, it is to be noted that not all sources labelled BSS in their catalogue are confirmed BSSs, some stars are BS candidates. We also identify 2 YGs and 2 triple systems (WOCS2009, WOCS3012) (WOCS: WIYN Open Cluster Study) as classified by \citet{Geller2015}. These stars are further categorised as 13 SB1 and 4 double-lined Spectroscopic Binaries (SB2). We used \citet{Yadav2008} and \citet{Williams2018} catalogues to identify 6 WDs with proper motion membership and 1 WD with photometric membership.

The spatial distribution of all member stars identified by \citet{Geller2015} along with 41 stars detected by \textit{UVIT} is shown in Fig.~\ref{fig:Members_UVIT_and_Geller}. The photometry in the three \textit{UVIT} FUV filters, their probabilities of proper motion, radial velocity membership are tabulated in Table~\ref{tab:photometry}.

\subsection{Colour-Magnitude Diagrams}
\label{sec:CMD}
CMDs are very useful to detect stars in various evolutionary phases. As we have three filters in the FUV, we can use UV CMDs as well as UV-optical CMDs to identify UV bright stars. 
\citet{Sindhu2018} demonstrated that the UV and UV-optical CMDs, along with the optical CMDs are good tools to identify the UV bright stars. 
We have overlaid the isochrones generated from Flexible Stellar Population synthesis (FSPS) code \citep{Conroy2009, Conroy2010} on the CMDs. FSPS code can generate modified isochrones models of BaSTI and Padova to include multiple phases of the stellar evolutionary track such as Horizontal Branch (HB), Asymptotic Giant Branch (AGB), BSS, WD etc. We have used FSPS code to generate both optical and UV isochrones of BaSTI model (\citealp{Pietrinferni2004}, \citealp{Cordier2007}) by providing the input parameters of the cluster viz. distance modulus $V-M_{v}= 9.57 \pm 0.03$ mag \citep{Stello2016}, solar metallicity, reddening of $E(B-V)= 0.05 \pm 0.01$  mag \citep{Montgomery1993} and age of $\sim$4 Gyr.
The isochrones are corrected for reddening and extinction.
The FSPS-generated locus of BSS, assuming them to be MS stars with masses in excess of the turn-off mass, which uniformly populates 0.5 magnitudes above the MSTO to 2.5 magnitudes brighter than the MSTO is also shown. 

We present an optical CMD, two UV-optical CMDs and a UV CMD in Fig.~\ref{fig:CMD}. The optical CMD shows all 41 members detected by {\it UVIT} according to their respective known categories along with the members identified in \citet{Geller2015} shown as grey dots in Fig.~\ref{fig:CMD} (a). The optical photometry for non-WD and WD sources are adopted from \citet{Montgomery1993} and \citet{Yadav2008} respectively. As we have marked the {\it UVIT} detected sources, it can be seen that we have detected only stars near the MSTO and hotter stars including BSSs and WDs. We do not have {\it UVIT} detections for most of the main sequence, as these stars are relatively cooler and are much fainter in the FUV. The blown-up view near the MSTO is shown in the inset.

In Fig.~\ref{fig:CMD} (b), we have shown the V, (F148W$-$V) CMD of the detected members (hereafter referred to as UV-optical CMD). This figure has the same y-axis as the figure (a), but the x-axis uses the F148W flux. We also note that the  colour spread of BSSs increases from 0.5 in (B$-V$) in Fig.~\ref{fig:CMD} (a) to 6 magnitudes in (F148W$-$V) as seen Fig.~\ref{fig:CMD} (b). We can see that the BSSs follow the model BSS line, whereas the stars located near the MSTO in the optical CMD get bluer and are located close to the red end of the BSS model line. We can also notice that the MSTO of the isochrone is at $(F148W-V)$ $\sim$ 12 mag, much redder than the detections, which have a colour range of 4-10 mag. The inset shows the blown-up view of the red end of the BSS model line. It is clear that stars near the MSTO in the optical CMD have an excess of at least 2 magnitudes in the (F148W$-$V) colour with respect to their expected location in this CMD. 
This indicates a possibility of a significant amount of excess flux in FUV, for some of the \textit{UVIT} detected stars.

In Fig.~\ref{fig:CMD} (c), we have plotted the F148W, (F148W$-$V) CMD. Excluding the WDs, we find that all detected members are located close to the BSS model line. The limiting magnitude of our observations is $\sim$ 22 mag and the tip of the MS is found to be at $\sim$ 25 mag in F148W. This demonstrates that our observations are not suited to detect the MS stars in the F148W filters as they are at least 3 magnitudes fainter than the limiting magnitude. Therefore, the {\it UVIT} observations cannot detect any normal MS star due to the detection limit. Noticeably, a few stars on the MSTO in the optical are detected in \textit{UVIT} filters, suggesting that these stars have excess flux in the F148W filter. Similar brightening of stars in the FUV was found by \citet{Sindhu2018} in the UV-optical CMDs constructed using the {\it GALEX}  data.

In order to compare the flux of the detected members in the F169M filter with respect to the F148W filter, we created the F148W, (F148W-F169M) CMD. In Fig.~\ref{fig:CMD} (d) we have shown the UV CMD for stars detected in both the {\it UVIT} filters. The y-axes for the plots (c) and (d) are the same, but the colour axis in (d) is a UV colour. The {\it UVIT} detected stars belong to various classes and they are identified in the CMDs, which include BSSs, YG, photometric variables, rapid rotators, MSTO stars, WDs etc. We detect 7 WDs and these are located close to the WD model line. The BSS model line shows a slope in the UV colour, suggesting a range in temperature. We note that the stars which were located along the model BSS line in (c) no longer fall on the BSS model line. The stars have a range in (F148W$-$F169M) colour, suggesting a difference of up to 1.0 magnitude between the F148W and F169M magnitudes. Many stars have colour around $\sim$ 0.4 mag, suggesting that these are likely to have a similar $T_{eff}$. 

We note that in figure (d), one triple system and a YG are located close to the WD region appearing as hot as the WDs.
In the CMDs presented here, the excess UV flux for the detected stars could be due to either intrinsic or extrinsic factors. In order to investigate this further, we estimate the properties of these stars using SED in the next section.

Note that, not all 41 members are detected in all the 3 {\it UVIT} filters, thus CMDs in Fig.~\ref{fig:CMD} (b), (c) and (d) will not have all 41 members. The number of stars present in each CMD will depend on whether they were detected in the respective filters.

\subsection{Spectral Energy Distribution}
\label{sec:SED}

\begin{table}
\centering
\caption{The $\chi^2_{red}$ comparison between single fits and double fits. Single fits are done by fitting single Kurucz model SED to all available points ($\chi^2_{S1}$) or excluding UV points ($\chi^2_{S2}$). Double fits are the combination of one Kurucz SED ($T_{A}$) and one WD SED ($T_B$ ) at log $g=$ 7 ($\chi^2_{Dob}$)}
\label{tab:chi}
\begin{tabular}{lccccr}
\toprule
        \multirow{2}[3]{*}{\textbf{WOCS}}  & \multicolumn{2}{c}{\textbf{Single Fit}} & \multicolumn{3}{c}{\textbf{Double Fit}}\\ 
        \cmidrule(lr){2-3}         \cmidrule(lr){4-6} 
&  $T_{S1}$ (K) & $\chi^{2}_{S1} (\chi^{2}_{S2})$ & $T_{A}$ (K) &  $T_B$ (K) & $\chi^2_{Dob}$ \\ 
        \toprule
1001	&	6750	&	2.4(1.4)	&	6250	&	11500	&	0.31	\\
11005	&	6500	&	4.3(1.1)	&	6250	&	11500	&	0.41	\\
11011	&	6000	&	3.8(3.3)	&	6000	&	11500	&	0.65	\\
2002	&	5250	&	78(4.3)	&	5250	&	14750	&	1.6	\\
2003	&	6500	&	2.5(4.8)	&	6250	&	9250	&	0.34	\\
2007	&	6500	&	11(12)	&	6000	&	11500	&	6.3	\\
2008	&	6500	&	7.6(1.1)	&	6000	&	11500	&	0.25	\\
2012	&	6000	&	4.4(1.1)	&	6000	&	11500	&	0.6	\\
2015	&	6500	&	5.4(14)	&	6250	&	9750	&	2.2	\\
3001	&	7000	&	2(2.9)	&	6750	&	12500	&	0.65	\\
3009	&	6750	&	2.6(1.3)	&	6250	&	10000	&	0.28	\\
4003	&	7250	&	8.2(3)	&	6500	&	10250	&	1.2	\\
4015	&	6500	&	1.9(2.7)	&	6250	&	11500	&	0.59	\\
5007	&	6750	&	2.9(1.5)	&	6250	&	9750	&	0.21	\\
5013	&	6500	&	2.4(12)	&	6250	&	11500	&	3.2	\\
6006	&	6750	&	1.8(3.2)	&	6250	&	10250	&	0.57	\\
7005	&	6500	&	1.5(6.4)	&	6000	&	11500	&	0.71	\\
7010	&	6500	&	3(3)	&	6250	&	11500	&	0.062	\\
8005	&	5750	&	17(23)	&	6000	&	10750	&	3.9	\\
8006	&	7000	&	9.9(13)	&	6750	&	11500	&	1.1	\\
9005	&	6750	&	1.3(5.6)	&	6500	&	11500	&	0.43	\\

\bottomrule
\end{tabular}
\end{table}

\begin{figure*}
  \centering
  \begin{tabular}{c c}
    \includegraphics[width=.45\textwidth]{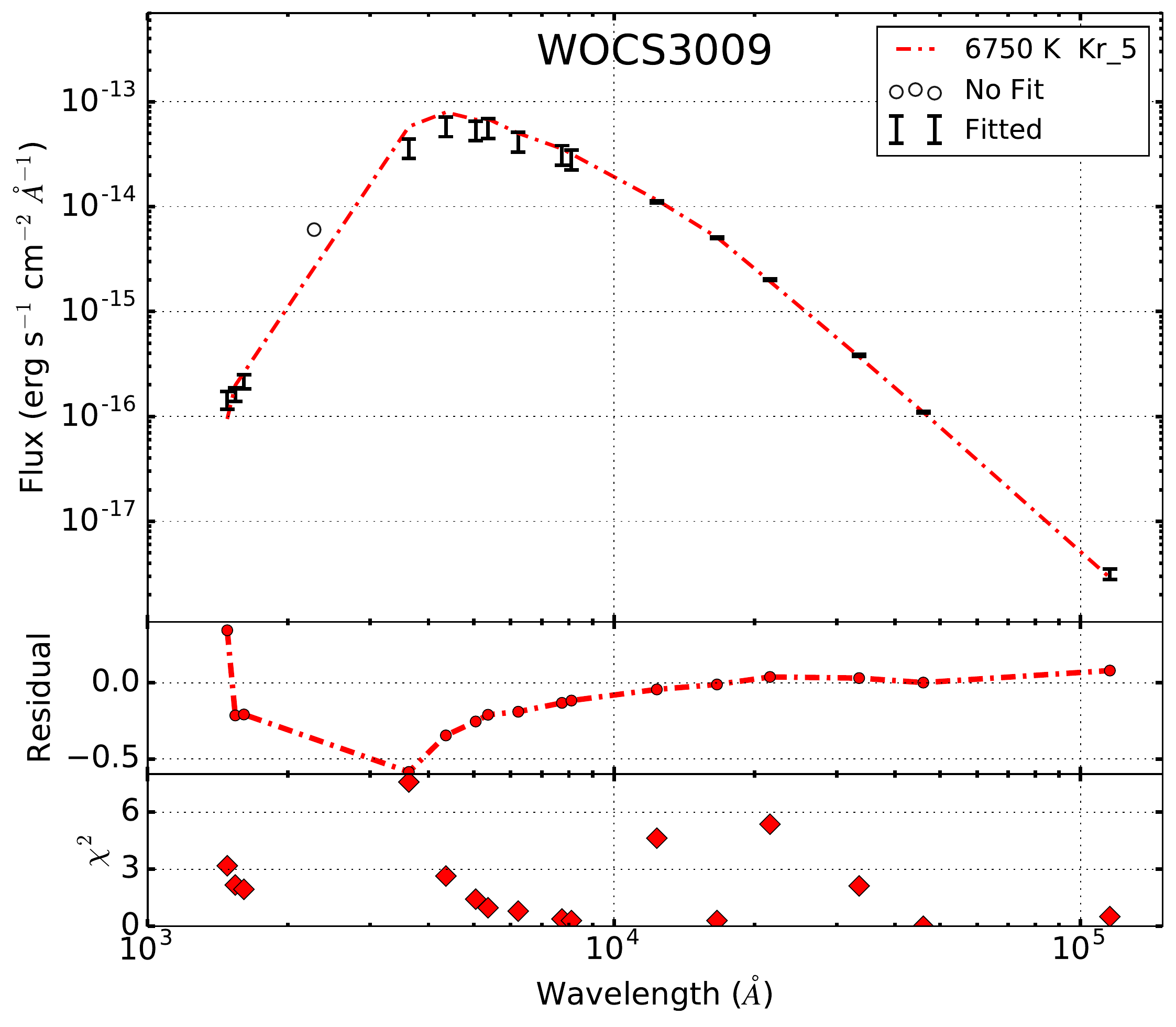} &
    \includegraphics[width=.45\textwidth]{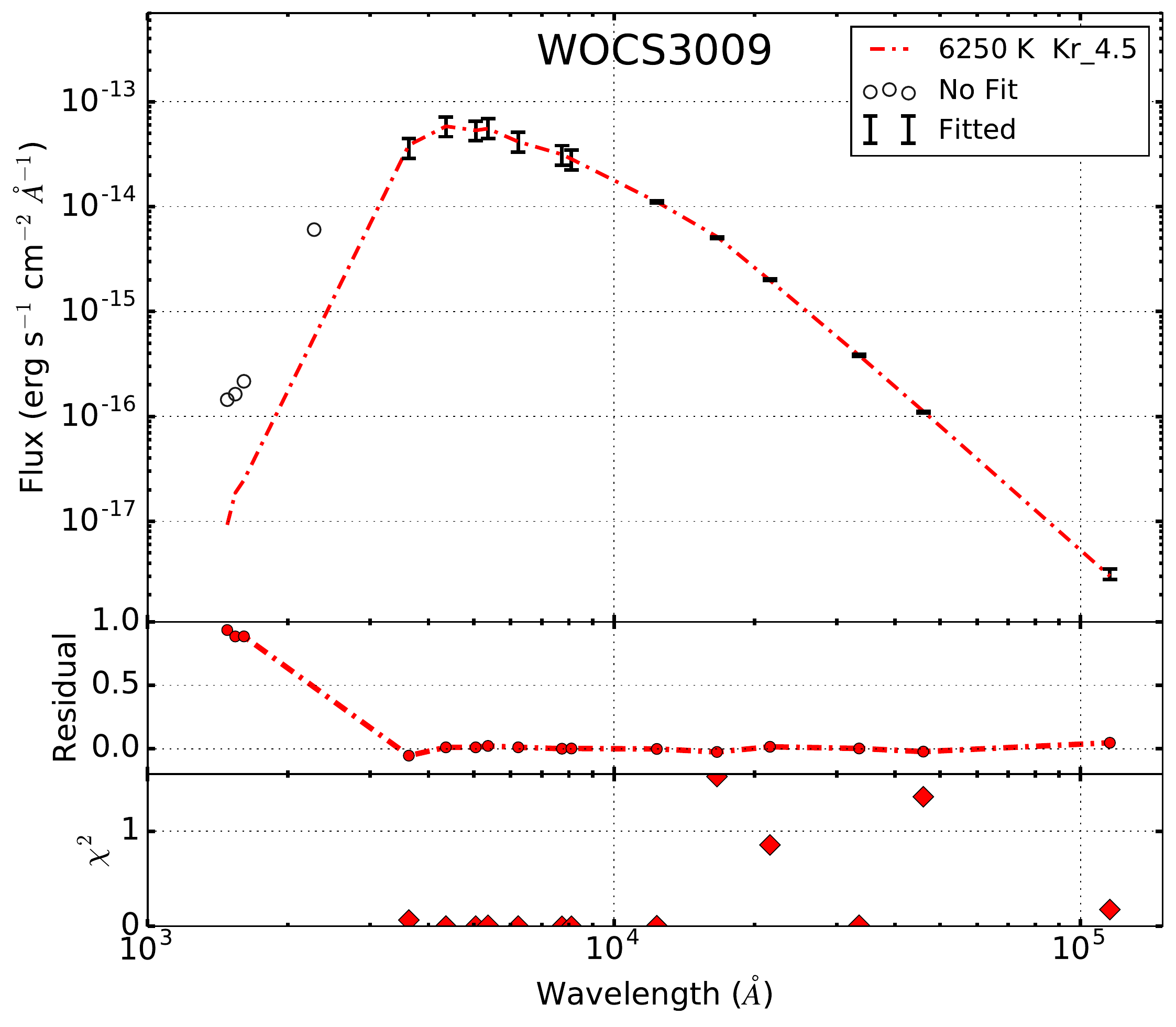}\\
    (a) & (b) \\
    \multicolumn{2}{c}{\includegraphics[width=.95\textwidth]{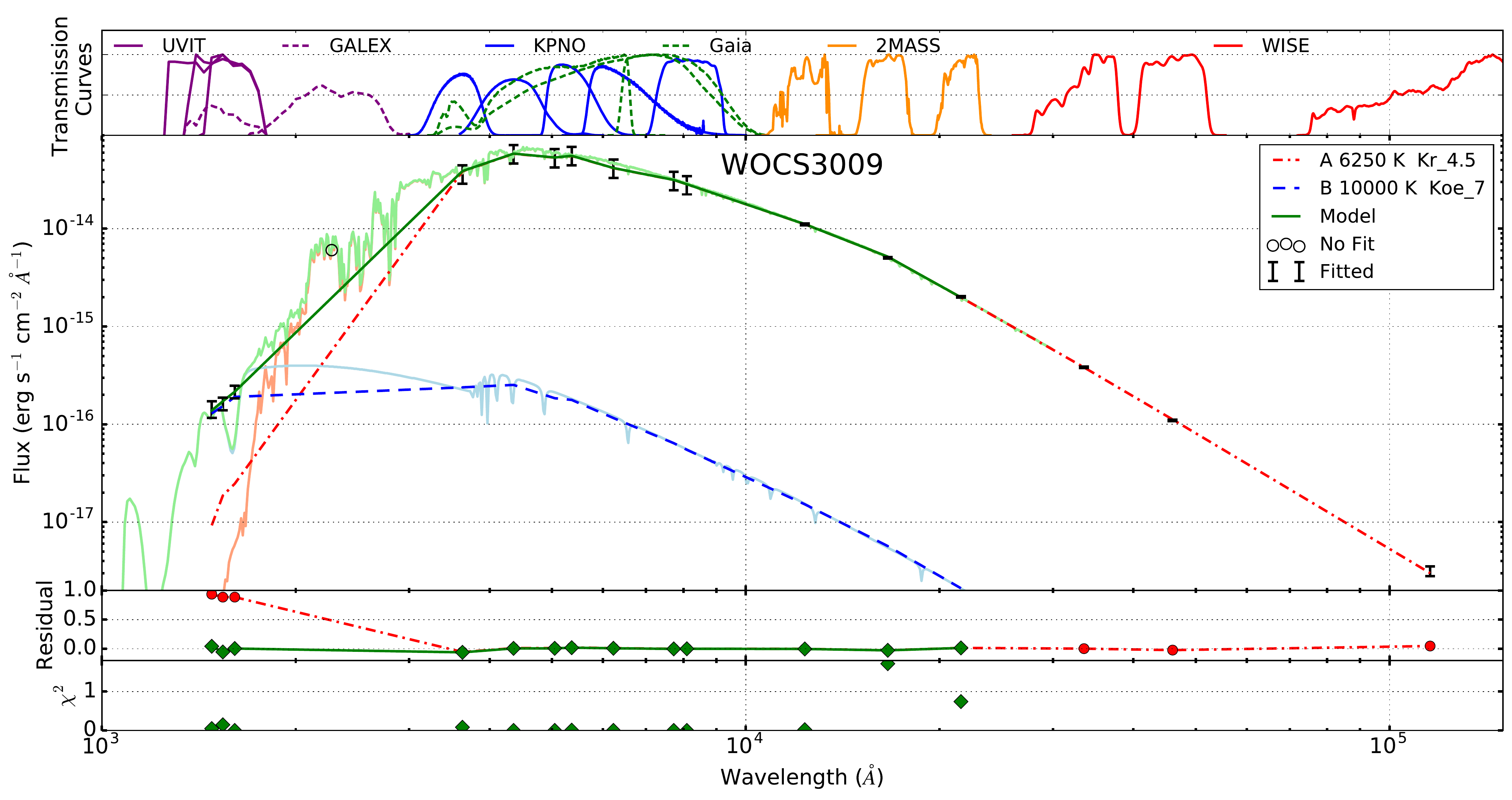}} \\
    \multicolumn{2}{c}{(c)} \\
  \end{tabular}
  \caption{An example of the method used to fit SEDs using WOCS3009. (a) Top panel shows the least $\chi^2$ fit for a single SED over UV-IR data points. The legend notes the $T_{eff}$ of the fit, the model used (Kr: Kurucz, Koe: Koester WD models) and the log $g$. The fitted points are shown with error-bars, while not fitted points are shown as circles. The middle panel shows the residual for the fit. The bottom panel shows the individual $\chi^2$ values
  calculated at each point.
  (b) Same as `a' but fit is done over optical-IR region.
  (c) The top bar shows the transmission curves (not to scale) for the filters of respective telescopes. 
  The second panel shows the composite double fit SED. 
  The figure shows `A component' (Kurucz model; red; dot-dashed), `B component' (Koester WD model; blue; dashed), `Model' i.e. total flux of 2 components (green; solid) and Observed flux (black, as only error-bars to simplify the graph) and unfitted points (hollow circle). The light coloured solid lines in blue, red and green show the high-res spectra corresponding to Kurucz, Koester (WD) and composite models respectively. Third and fourth panels are similar to the residual and $\chi^2$ panels in `a'.}
  \label{fig:SED_3009}
\end{figure*}

In order to characterise the excess UV flux as suggested by the optical and UV CMDs, we performed a detailed study using their SEDs. We compiled the fluxes of 30 sources (23 stars and 7 WDs) from UV to IR. The multi-wavelength SEDs were created and compared with model SEDs to determine their characteristics. The analysis presented here is similar to that presented in \citet{Subramaniam2016} and \citet{Sindhu2019}. 

We used the virtual observatory tool, {\sc VOSA} (VO SED Analyzer, \citealt{Bayo2008}) for SED analysis. {\sc VOSA} calculates synthetic photometry for a selected theoretical model using filter transmission curves. It performs a $\chi^{2}$ minimisation test by comparing the synthetic photometry with observed data to get the best-fit parameters of the SED. We estimated the $\chi^{2}_{red}$ value using the expression given by
\begin{equation}
\small
     \chi^{2}_{red} =\frac{1}{N-N_{f}} \sum_{i=1}^{N}\Big\{\frac{(F_{o,i}-M_{d}F_{m,i})^{2}}{\sigma_{o,i}^{2}}\Big\}
\label{chi2}
\end{equation}

where N is the number of photometric data points, N$_{f}$ is the number of free parameters in the model, $F_{o,i}$ is the observed flux, $M_{d}F_{m,i}$ is the model flux of the star, $\displaystyle{M_{d}=\bigg(\frac{R}{D}\bigg)^{2}}$ is the Scaling Factor (SF) corresponding to the star (where R is the radius of the star and D is the distance to the star) and $\sigma_{o,i}$ is the error in the observed flux. The P-value for each fit was estimated from $\chi^{2}_{red}$ and $N-N_{f}$. The value of N for stellar sources changes from 10 to 16 depending on the number of detections in all available filters. Similarly, N$_{f}$ changes between 2 or 4 on whether the fit is single ($T_{eff}$ and SF) or double fit (two $T_{eff}$s and SFs for two components).

We use the theoretical stellar models which span the UV to IR wavelength coverage as our SEDs cover from 130 nm to 16000 nm. We use updated Kurucz stellar atmospheric models \citep{Castelli1997} for the stellar (non-WD) sources which cover the same wavelength range. The theoretical spectra for WDs of spectral type DA with pure hydrogen atmospheres were obtained from \citet{Koester2010}, which are mentioned as WD models in the rest of the paper. The spectra were converted to synthetic photometry for the required filters using {\sc VOSA} \citep{Bayo2008}, according to the individual filter profiles \citep{Rodrigo2012}. Extinction of $A_V = 0.1736$ and distance 831$\pm$11 pc were used to normalise the SEDs. We used \citet{Fitzpatrick1999} extinction curves to calculate extinction coefficients in all other bands.

\begin{figure*}
  \centering
   \includegraphics[width=0.95\textwidth]{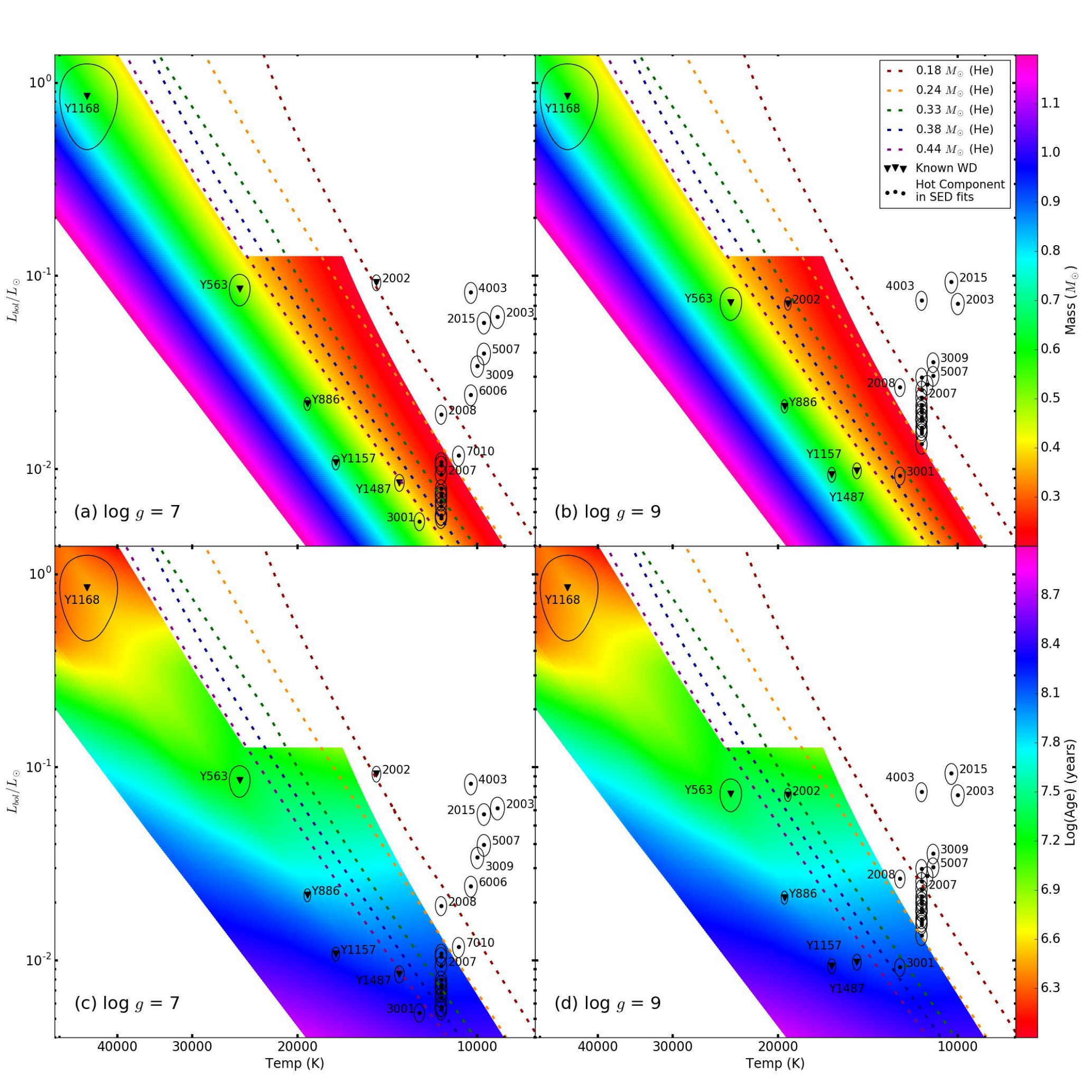}\\
   \caption{HR diagram of WDs and hotter components plotted over interpolated DA type \citep{Bergeron2009} (as solid band) and He-core \citep{Panei2007} (as dashed lines) WD cooling curves. The errors are plotted as ellipses. Legends for all figures are same as in (b).
   (a) The gradient corresponds to the mass of the DA model WDs. The single WDs and the hotter components' parameters from 'log $g$ = 7' fits are over-plotted to estimate their mass.
   (b) Same as 'a' for the SED fits with 'log $g$ = 9'.
   (c) The gradient corresponds to cooling age of the DA model WDs. The single WDs and the hotter components' parameters from 'log $g$ = 7' fits are over-plotted to estimate their age.
   (d) Same as 'c' for the SED fits with 'log $g$ = 9'.
  }
  \label{fig:WD_mass_radius}
\end{figure*}

Out of the detected stars, the study on 9 bright bona fide BSSs (WOCS numbers: 1006, 1007, 2011, 2013, 3005, 3010, 3013, 4006, 5005) is presented in \citet{Sindhu2019}, \citet{sindhu2019b} and Sindhu et al. (in prep.).
We were not able to successfully cross-match WOCS8010 with archival data to create a useful SED due to its closeness to WOCS3010. WOCS9028 lies near the edge of our field of view and was only detected in F169M. We could not analyse WOCS3012 due to it being a triple system with no known individual parameters.
After removing these stars from consideration, we fitted the observed flux distribution of 30 stars with SED models of which the 7 WDs were fitted with WD model SEDs with range of $T_{eff}$ = 5000 to 80000 K. \citet{kepler2015} showed that most WDs have surface gravity near log $g \sim$ 8. We have thus used two surface gravity values (log $g$ = 7 and 9) for the WD models. The results of SED fittings of WDs are discussed in subsection~\ref{sec:WD}. 
The stellar SEDs were fitted with Kurucz model SEDs with solar metallicity and limited the fits to log $g$ = 3 to 5 and T = 3500 to 50000 K. Each fit provides us with the $T_{eff}$ and radius corresponding to the star. The {\it GALEX} DR6 magnitudes are not available for all stars. We also observe variations in {\it GALEX} and {\it UVIT} magnitudes in FUV region for some stars; the reasons may be the non-simultaneous nature of observations or \textit{UVIT}'s superior resolution of 1.5$\arcsec$ in FUV when compared to 4.5$\arcsec$ of \textit{GALEX}. Thus, we did not use the {\it GALEX} photometry for fitting SEDs in the case of stellar sources. We show the {\it GALEX} flux in SED only for comparison.

The $\chi^2_{red}$ values for the stellar SEDs fits are listed in Table~\ref{tab:chi} as $\chi^2_{S1}$. Almost all stars show large $\chi^2$ values. As we suspected excess flux in UV as suggested from UV-optical and UV CMDs, we tried to fit the SEDs again by ignoring the flux below 1800 \AA. The $\chi^2_{red}$ values of modified fits are given in the bracket as $\chi^2_{S2}$. These values are relatively less when compared to $\chi^2_{S1}$ values. 
As the flux in the UV region is ignored, the residual flux which consists mainly of the excess UV flux was then fitted with a WD model, such that the flux due to two models are added up to fit the full range of observed flux.
Details of the double fits are also shown in the table, where the $\chi^2_{red}$ for the double model fit is denoted as $\chi^2_{Dob}$.
Note that these S2 fits also have lesser number of data points when compared to S1 fits and do not cover the full range of wavelength and hence it is better to compare $\chi^2_{S1}$ with $\chi^2_{Dob}$ directly. In most of the SED fits, it can be seen that $\chi^2_{Dob}$ is significantly less than $\chi^2_{S1}$. It is important to note that we are able to fit most of the SEDs satisfactorily using a double fit, as suggested by the low $\chi^2_{Dob}$ values. The interpretation of the changes in $\chi^2$ for each source can be found in section~\ref{sec:results}.

Fig.~\ref{fig:SED_3009} demonstrates the above procedure using WOCS3009 as an example. Panel (a) shows the SED when we fit all available data points with a single Kurucz spectrum ($\chi^2_{red}$ = 2.6), (b) shows the SED when we fit only optical and IR data points ($\chi^2_{red}$ = 1.3). A good SED fit in the optical-IR region and a lower value of $\chi^2_{red}$ supports the presence of a UV excess in WOCS3009. Panel (c) shows the result of fitting two component SED (Kurucz SED of 6250 K and log $g$ = 4.5 and a WD SED of 11000 K and log $g$ = 9) with $\chi^2_{Dob}$ = 0.2. The residuals in (c) and in all SED figures are calculated as
\begin{equation}
\small
     Residual=(Flux_{Obs}-Flux_{Model})/Flux_{Obs}
\label{residual}
\end{equation}
where $Flux_{Model}$ is flux for model (single/composite double) SED. 

\subsection{Mass and Age Estimation}
\label{sec:mass}
The SED fits provided the temperature and radius of all the components. The bolometric Luminosity of the components is calculated using luminosity relation
\begin{equation}
\small
     L=4 \pi R^2 \sigma T_{eff}^4
\label{Luminosity}
\end{equation}

We use DA (pure hydrogen) WD models \citep{Bergeron2009} to estimate the age and mass of the WDs. The model cooling curves were available for 0.2 to 1.2 $M_\odot$ in increments of 0.2 $M_\odot$. We assumed log $T_{eff} \propto$ log $M$,  log $L \propto$ log $M$, log $Age \propto$ log $M$ \citep{Myakutin1995} and linearly interpolated log $L$, log $M$, log $Age$ and log $T_{eff}$. The interpolation was done producing 100 steps in mass range (0.2 to 1.2 $M_\odot$) and 500 steps in luminosity range ($5\times10^{-6}$ to 1.4 $L_\odot$), with each point having corresponding age and temperature. 

Fig.~\ref{fig:WD_mass_radius} (a) and (b) shows the HR diagram of the known WDs and hotter components in double fits with log $g$ = 7 and 9 respectively plotted over interpolated DA WD model with mass as the auxiliary colour. Similarly (c) and (d) are the same data points plotted over DA WD model with age as the auxiliary colour. We also include cooling curves from He-core low mass WD model \citep{Panei2007} for completeness' sake, as many of the hotter components lie near lower mass range.
We used the intrinsic errors in the SED fits to estimate the errors in the mass and age. These errors are plotted as ellipses in the figure.
For the points outside the interpolated DA model, the mass is stated in Table~\ref{tab:All_para} as $<$0.2 $M_\odot$ and upper limit of age is calculated from the vertical intercept to DA model cooling curve at 0.2 $M_\odot$.

\section{Results}
\label{sec:results}

\begin{table*}
\begin{scriptsize}
\centering
\caption{The best-fit parameters of all sources estimated using $\chi^2$ fits. First column has identification from
\newline$^a$ WOCS: \citet{Geller2015}, S: \citet{Sanders1977}, Y: \citet{Yadav2008}, WD: \citet{Williams2018}. }
\label{tab:All_para}
\begin{tabular}{lcccc ccccr} 
\toprule
\textbf{Name$^a$}	&	\textbf{Comp}	&	\textbf{$T_{eff}$}	&	\textbf{log $g$}	&	\textbf{R}	&	\textbf{L}	&	\textbf{$M_{WD}$}	&	\textbf{$Age_{WD}$}	&	\textbf{Comments}	&	\textbf{Remark}	\\	
&	&	(K)	&	(cm s$^{-2}$)	&	(\(R_\odot\))	&	(\(L_\odot\))	&	(\(M_\odot\))	&	(Myr)	&	&	\\	\toprule
WOCS1001	&	A	&	6250	$\pm$	125	&	5	&	1.99$\pm$0.03	&	5.1	&	&	&	BM, SB2,PV,	&	WD/Ch	\\	
(S1024)	&	B	&	11500	$\pm$	250	&	9	&	0.0344$\pm$0.0005	&	0.0190	&	$<$0.2	&	$<$120	&	CX111, X46	&	\\	
&	B	&	11500	$\pm$	250	&	7	&	0.0218$\pm$0.0003	&	0.0075	&	0.27$\pm$0.02	&	178$\pm$13	&	&	\\	\midrule
WOCS11005	&	A	&	6250	$\pm$	125	&	4	&	1.94$\pm$0.03	&	4.8	&	&	&	SM	&	WD?	\\	
(S995)	&	B	&	11500	$\pm$	250	&	9	&	0.0369$\pm$0.0005	&	0.0210	&	$<$0.2	&	$<$120	&	&	\\	
&	B	&	11500	$\pm$	250	&	7	&	0.0224$\pm$0.0003	&	0.0079	&	0.26$\pm$0.02	&	164$\pm$5	&	&	\\	\midrule
WOCS11011	&	A	&	6000	$\pm$	125	&	3.5	&	1.46$\pm$0.02	&	2.3	&	&	&	BLM, SB1, HS Cnc,	&	WD/Ch	\\	
(S757)	&	B	&	11500	$\pm$	250	&	9	&	0.036$\pm$0.0005	&	0.0200	&	$<$0.2	&	$<$120	&	RR, W Uma, PV,	&	\\	
&	B	&	11500	$\pm$	250	&	7	&	0.0218$\pm$0.0003	&	0.0075	&	0.27$\pm$0.02	&	178$\pm$13	&	CX23, NX21	&	\\	\midrule
WOCS2002	&	A	&	5250	$\pm$	125	&	5	&	5.34$\pm$0.07	&	18	&	&	&	BM, SB1, PV,	&	WD+Ch	\\	
(S1040)	&	B	&	19250	$\pm$	250	&	9	&	0.0242$\pm$0.0003	&	0.0720	&	0.31$\pm$0.01	&	25$\pm$1	&	YG, WD,	&	\\	
&	B	&	14750	$\pm$	250	&	7	&	0.0467$\pm$0.0006	&	0.0930	&	$<$0.2	&	$<$110	&	CX6, X10, NX5	&	\\	\midrule
WOCS2003	&	A	&	6250	$\pm$	125	&	4	&	2.17$\pm$0.03	&	6.4	&	&	&	BM, SB2, PV	&	Ch/Sp	\\	
(S1045)	&	B	&	10000	$\pm$	250	&	9	&	0.0896$\pm$0.001	&	0.0720	&	$<$0.2	&	$<$200	&	CX88, X41	&	\\	
&	B	&	9250	$\pm$	250	&	7	&	0.0969$\pm$0.001	&	0.0620	&	$<$0.2	&	$<$240	&	&	\\	\midrule
WOCS2007	&	A	&	6000	$\pm$	125	&	4	&	2.65$\pm$0.04	&	8.2	&	&	&	SM, BSS	&	WD	\\	
(S984)	&	B	&	11500	$\pm$	250	&	9	&	0.0387$\pm$0.0005	&	0.0240	&	$<$0.2	&	$<$120	&	&	\\	
&	B	&	11500	$\pm$	250	&	7	&	0.0245$\pm$0.0003	&	0.0094	&	0.24$\pm$0.02	&	150$\pm$10	&	&	\\	\midrule
WOCS2008	&	A	&	6500	$\pm$	125	&	4.5	&	4.18$\pm$0.06	&	19	&	&	&	BM, BSS, SB1, YG	&	WD/Ch	\\	
(S1072)	&	B	&	12500	$\pm$	250	&	9	&	0.0348$\pm$0.0005	&	0.0270	&	$<$0.2	&	$<$120	&	CX24, X37, NX16	&	\\	
&	B	&	11500	$\pm$	250	&	7	&	0.035$\pm$0.0005	&	0.0190	&	$<$0.2	&	$<$120	&	&	\\	\midrule
WOCS2009	&	Aa	&	7250	$\pm$	250	&	4	&	2.04$\pm$0.1	&	10	&	&	&	BM, BSS, SB2, PV	&	Ch/Sp	\\	
(S1082)	&	Ab	&	6000	$\pm$	250	&	4	&	2.15$\pm$0.1	&	5.4	&	&	&	ES Cnc, Triple, &	\\	
&	B	&	6000	$\pm$	250	&	4.5	&	2.02$\pm$0.03	&	4.8	&	&	&	CX3, X4, NX37	&	\\	\midrule
WOCS2012	&	A	&	6000	$\pm$	125	&	3	&	2.2$\pm$0.03	&	5.3	&	&	&	SM	&	WD?	\\	
(S756)	&	B	&	11500	$\pm$	250	&	9	&	0.0313$\pm$0.0004	&	0.0150	&	$<$0.2	&	$<$120	&	&	\\	
&	B	&	11500	$\pm$	250	&	7	&	0.0188$\pm$0.0002	&	0.0056	&	0.34$\pm$0.03	&	219$\pm$14	&	&	\\	\midrule
WOCS2015	&	A	&	6250	$\pm$	125	&	3.5	&	2.81$\pm$0.04	&	9.7	&	&	&	SM, BSS	&	Ch	\\	
(S792)	&	B	&	10250	$\pm$	250	&	9	&	0.0969$\pm$0.001	&	0.0930	&	$<$0.2	&	$<$200	&	&	\\	
&	B	&	9750	$\pm$	250	&	7	&	0.0842$\pm$0.001	&	0.0570	&	$<$0.2	&	$<$240	&	&	\\	\midrule
WOCS3001	&	A	&	6750	$\pm$	125	&	4.5	&	1.32$\pm$0.02	&	3.1	&	&	&	BM, SB1	&	WD	\\	
(S1031)	&	B	&	12500	$\pm$	250	&	9	&	0.0205$\pm$0.0003	&	0.0093	&	0.30$\pm$0.02	&	149$\pm$9	&	&	\\	
&	B	&	12500	$\pm$	250	&	7	&	0.0157$\pm$0.0002	&	0.0054	&	0.45$\pm$0.04	&	235$\pm$22	&	&	\\	\midrule
WOCS3009	&	A	&	6250	$\pm$	125	&	4.5	&	2.46$\pm$0.03	&	7.8	&	&	&	SM, BSS	&	WD?	\\	
(S1273)	&	B	&	11000	$\pm$	250	&	9	&	0.0522$\pm$0.0007	&	0.0360	&	$<$0.2	&	$<$150	&	&	\\	
&	B	&	10000	$\pm$	250	&	7	&	0.0618$\pm$0.0008	&	0.0340	&	$<$0.2	&	$<$200	&	&	\\	\midrule
WOCS4003	&	A	&	6500	$\pm$	125	&	4.5	&	1.78$\pm$0.02	&	4.6	&	&	&	BM, BSS, SB1, RR,	&	Ch/Sp	\\	
(S1036)	&	B	&	11500	$\pm$	250	&	9	&	0.069$\pm$0.0009	&	0.0750	&	$<$0.2	&	$<$120	&	EV Cnc, PV	&	\\	
&	B	&	10250	$\pm$	250	&	7	&	0.0912$\pm$0.001	&	0.0820	&	$<$0.2	&	$<$200	&	CX19, X45, NX20	&	\\	\midrule
WOCS4015	&	A	&	6250	$\pm$	125	&	4.5	&	1.99$\pm$0.03	&	5.1	&	&	&	SM	&	WD?	\\	
(S1456)	&	B	&	11500	$\pm$	250	&	9	&	0.0338$\pm$0.0004	&	0.0180	&	$<$0.2	&	$<$120	&	&	\\	
&	B	&	11500	$\pm$	250	&	7	&	0.0209$\pm$0.0003	&	0.0068	&	0.29$\pm$0.02	&	191$\pm$7	&	&	\\	\midrule
WOCS5007	&	A	&	6250	$\pm$	125	&	5	&	1.92$\pm$0.03	&	4.5	&	&	&	SM	&	WD?	\\	
(S1071)	&	B	&	11000	$\pm$	250	&	9	&	0.0481$\pm$0.0006	&	0.0300	&	$<$0.2	&	$<$150	&	&	\\	
&	B	&	9750	$\pm$	250	&	7	&	0.07$\pm$0.0009	&	0.0400	&	$<$0.2	&	$<$240	&	&	\\	\midrule
\end{tabular}
\end{scriptsize}
\end{table*}
\addtocounter{table}{-1}
\begin{table*}
\begin{scriptsize}
\centering
\caption{\textit{Continued...}}
\begin{tabular}{lcccc ccccr} 
\toprule
\textbf{Name$^a$}	&	\textbf{Comp}	&	\textbf{$T_{eff}$}	&	\textbf{log $g$}	&	\textbf{R}	&	\textbf{L}	&	\textbf{$M_{WD}$}	&	\textbf{$Age_{WD}$}	&	\textbf{Comments}	&	\textbf{Remark}	\\	
&	&	(K)	&	(cm s$^{-2}$)	&	(\(R_\odot\))	&	(\(L_\odot\))	&	(\(M_\odot\))	&	(Myr)	&	&	\\	\toprule
WOCS5013	&	A	&	6250	$\pm$	125	&	5	&	1.68$\pm$0.02	&	3.6	&	&	&	SM	&	WD?	\\	
(S1230)	&	B	&	11500	$\pm$	250	&	9	&	0.0354$\pm$0.0005	&	0.0200	&	$<$0.2	&	$<$120	&	&	\\	
&	B	&	11500	$\pm$	250	&	7	&	0.0215$\pm$0.0003	&	0.0073	&	0.28$\pm$0.02	&	181$\pm$13	&	&	\\	\midrule
WOCS6006	&	A	&	6250	$\pm$	125	&	4.5	&	1.82$\pm$0.02	&	4.3	&	&	&	SM	&	WD	\\	
(S1271)	&	B	&	11250	$\pm$	250	&	9	&	0.0437$\pm$0.0006	&	0.0270	&	$<$0.2	&	$<$120	&	&	\\	
&	B	&	10250	$\pm$	250	&	7	&	0.0495$\pm$0.0007	&	0.0240	&	$<$0.2	&	$<$200	&	&	\\	\midrule
WOCS7005	&	A	&	6000	$\pm$	125	&	4	&	2.19$\pm$0.03	&	5.3	&	&	&	SM	&	WD?	\\	
(S1274)	&	B	&	11500	$\pm$	250	&	9	&	0.0293$\pm$0.0004	&	0.0130	&	$<$0.2	&	$<$120	&	&	\\	
&	B	&	11500	$\pm$	250	&	7	&	0.0192$\pm$0.0003	&	0.0058	&	0.33$\pm$0.03	&	209$\pm$11	&	&	\\	\midrule
WOCS7009	&	Aa	&	6250	$\pm$	250	&	4.5	&	1.3$\pm$0.1	&	2.8	&	&	&	BLM, SB1, RR,	&	Ch+WD?	\\	
(S1282)	&	Ab	&	6250	$\pm$	250	&	4.5	&	0.68$\pm$0.05	&	0.64	&	&	&	AH Cnc, W Uma,	&	\\	
&	B	&	11500	$\pm$	250	&	9	&	0.0324$\pm$0.0004	&	0.0160	&	$<$0.2	&	$<$120	&	CX16, X40, NX10	&	\\	
&	B	&	11500	$\pm$	250	&	7	&	0.0202$\pm$0.0003	&	0.0064	&	0.306$\pm$0.03	&	198$\pm$14	&	&	\\	\midrule
WOCS7010	&	A	&	6750	$\pm$	125	&	3.5	&	1.97$\pm$0.03	&	4.8	&	&	&	SM	&	WD?	\\	
(S1083)	&	B	&	11500	$\pm$	250	&	9	&	0.0338$\pm$0.0004	&	0.0180	&	$<$0.2	&	$<$120	&	&	\\	
&	B	&	10750	$\pm$	250	&	7	&	0.0313$\pm$0.0004	&	0.0120	&	$<$0.2	&	$<$170	&	&	\\	\midrule
WOCS8005	&	A	&	6000	$\pm$	125	&	4.5	&	2.18$\pm$0.03	&	5.1	&	&	&	SM	&	WD?	\\	
(M5951)	&	B	&	11500	$\pm$	250	&	9	&	0.0317$\pm$0.0004	&	0.0160	&	$<$0.2	&	$<$120	&	&	\\	
&	B	&	11500	$\pm$	250	&	7	&	0.0192$\pm$0.0003	&	0.0058	&	0.33$\pm$0.03	&	209$\pm$11	&	&	\\	\midrule
WOCS8006	&	A	&	6750	$\pm$	125	&	4	&	1.56$\pm$0.02	&	4.1	&	&	&	SM, BSS	&	WD?	\\	
(S2204)	&	B	&	11500	$\pm$	250	&	9	&	0.0404$\pm$0.0005	&	0.0260	&	$<$0.2	&	$<$120	&	&	\\	
&	B	&	11500	$\pm$	250	&	7	&	0.0258$\pm$0.0003	&	0.0100	&	0.22$\pm$0.02	&	139$\pm$10	&	&	\\	\midrule
WOCS9005	&	A	&	6500	$\pm$	125	&	4.5	&	1.86$\pm$0.02	&	5.6	&	&	&	BM, BSS, SB1	&	WD?	\\	
(S1005)	&	B	&	11500	$\pm$	250	&	9	&	0.0437$\pm$0.0006	&	0.0300	&	$<$0.2	&	$<$120	&	&	\\	
&	B	&	11500	$\pm$	250	&	7	&	0.0264$\pm$0.0003	&	0.0110	&	0.22$\pm$0.02	&	136$\pm$10	&	&	\\	\midrule
Y1157	&	&	16250	$\pm$	250	&	9	&	0.0122$\pm$0.0002	&	0.0094	&	0.66$\pm$0.04	&	180$\pm$17	&	&	\\	
(WD30)	&	&	17250	$\pm$	250	&	7	&	0.0117$\pm$0.0002	&	0.011	&	0.70$\pm$0.04	&	163$\pm$15	&	DB	&	\\	\midrule
Y1168,  	&	&	45000	$\pm$	5000	&	9	&	0.0152$\pm$0.0002	&	0.85	&	0.66$\pm$0.20	&	2.6$\pm$0.9	&	&	\\	
(WD15)	&	&	45000	$\pm$	5000	&	7	&	0.0152$\pm$0.0002	&	0.85	&	0.66$\pm$0.20	&	2.6$\pm$0.9	&	DA	&	\\	\midrule
Y1487	&	&	14750	$\pm$	250	&	9	&	0.0152$\pm$0.0002	&	0.0098	&	0.50$\pm$0.03	&	151$\pm$12	&	&	\\	
(WD1)	&	&	13500	$\pm$	250	&	7	&	0.0169$\pm$0.0002	&	0.0085	&	0.41$\pm$0.03	&	157$\pm$12	&	DB	&	\\	\midrule
Y563	&	&	24000	$\pm$	1000	&	9	&	0.0157$\pm$0.0002	&	0.073	&	0.52$\pm$0.08	&	21$\pm$2	&	&	\\	
(WD2)	&	&	25000	$\pm$	1000	&	7	&	0.0157$\pm$0.0002	&	0.086	&	0.52$\pm$0.07	&	18$\pm$2	&	DA	&	\\	\midrule
Y701	&	&	15750	$\pm$	250	&	9	&	0.00941$\pm$0.0001	&	0.0049	&	&	&	&	\\	
(WD9)	&	&	16250	$\pm$	250	&	7	&	0.00934$\pm$0.0001	&	0.0055	&	&	&	DA	&	Unrel	\\	\midrule
Y856	&	&	14500	$\pm$	250	&	9	&	0.0113$\pm$0.0001	&	0.0051	&	&	&	&	\\	
(WD10)	&	&	12750	$\pm$	250	&	7	&	0.0128$\pm$0.0002	&	0.0039	&	&	&	DA, DD &	Unrel	\\	\midrule
Y886	&	&	19500	$\pm$	250	&	9	&	0.0128$\pm$0.0002	&	0.021	&	0.65$\pm$0.02	&	85$\pm$4	&	&	\\	
(WD25)	&	&	19250	$\pm$	250	&	7	&	0.0133$\pm$0.0002	&	0.022	&	0.61$\pm$0.03	&	77$\pm$7	&	DA	&	\\	\bottomrule
\multicolumn{10}{c}{\textit{WD:Candidate WD, WD?: possible WD, ch: chromospheric, Sp: hot-spots, Unrel:Unreliable}} \\
\multicolumn{10}{c}{\textit{PV:Pulsating Variable, RR:Rapid Rotator, DD: Double Degenerate WD, DA/DB: WD spectral types}} \\
\multicolumn{10}{c}{\textit{BM:Binary Member, BLM:Binary Likely Member, SM:Single Member}} 

\end{tabular}

\end{scriptsize}

\end{table*}

The parameters estimated from the best SED fits are listed in Table~\ref{tab:All_para}. All 'A' components (including Aa and Ab) are fitted with Kurucz Model SEDs while all 'B' components (except WOCS2009, a triple system, where B component is also fitted with Kurucz SED) are fitted with WD model SEDs suitable for hotter companions. The temperature $T_{eff}$, log $g$ and radii are obtained from SED fit parameters.
The `Comments' column in Table~\ref{tab:All_para} includes the comments by \citet{Geller2015} and X-ray detections identifiers from \textit{ROSAT} (X, \citealt{Belloni1998}), \textit{Chandra} (CX, \citealt{Van2004}) and \textit{XMM-Newton} observations (NX, \citealt{Mooley2015}).
In the case of WDs, the SEDs with fewer data points are noted as Unreliable (Unrel).

\begin{figure*}
  \centering
  \begin{tabular}{cc}
    \includegraphics[width=0.45\textwidth]{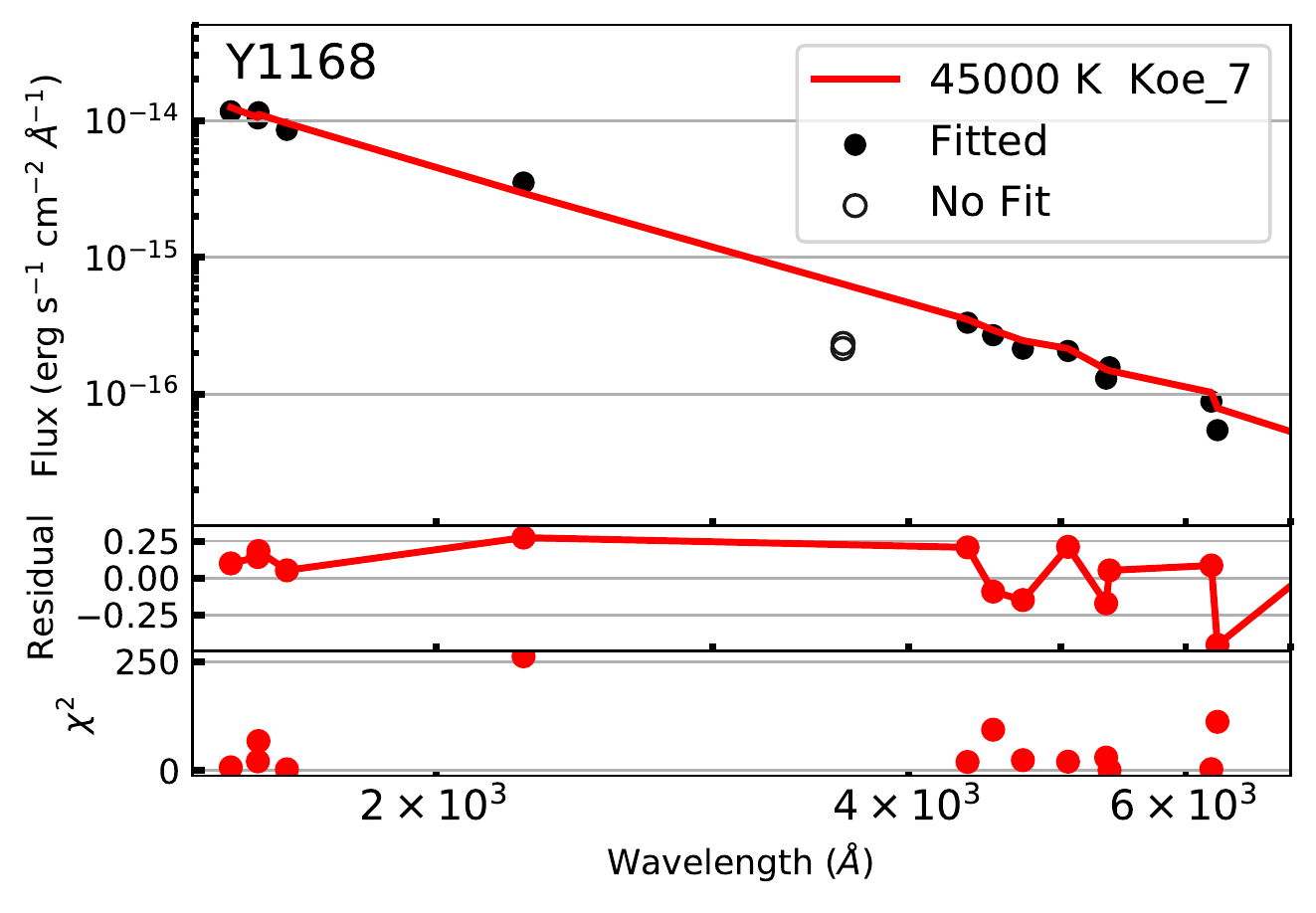}&
    \includegraphics[width=0.45\textwidth]{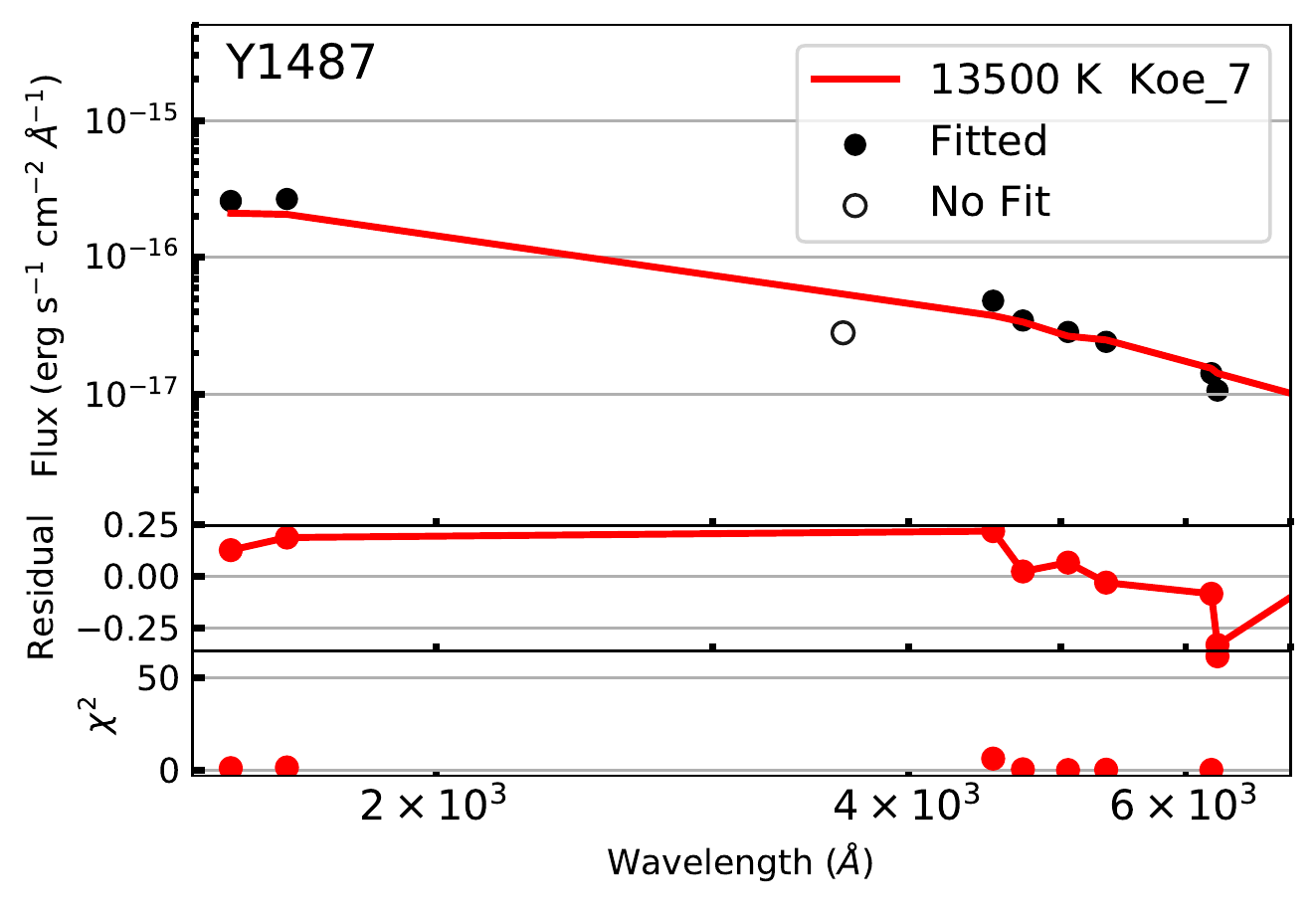} \\
    \includegraphics[width=0.45\textwidth]{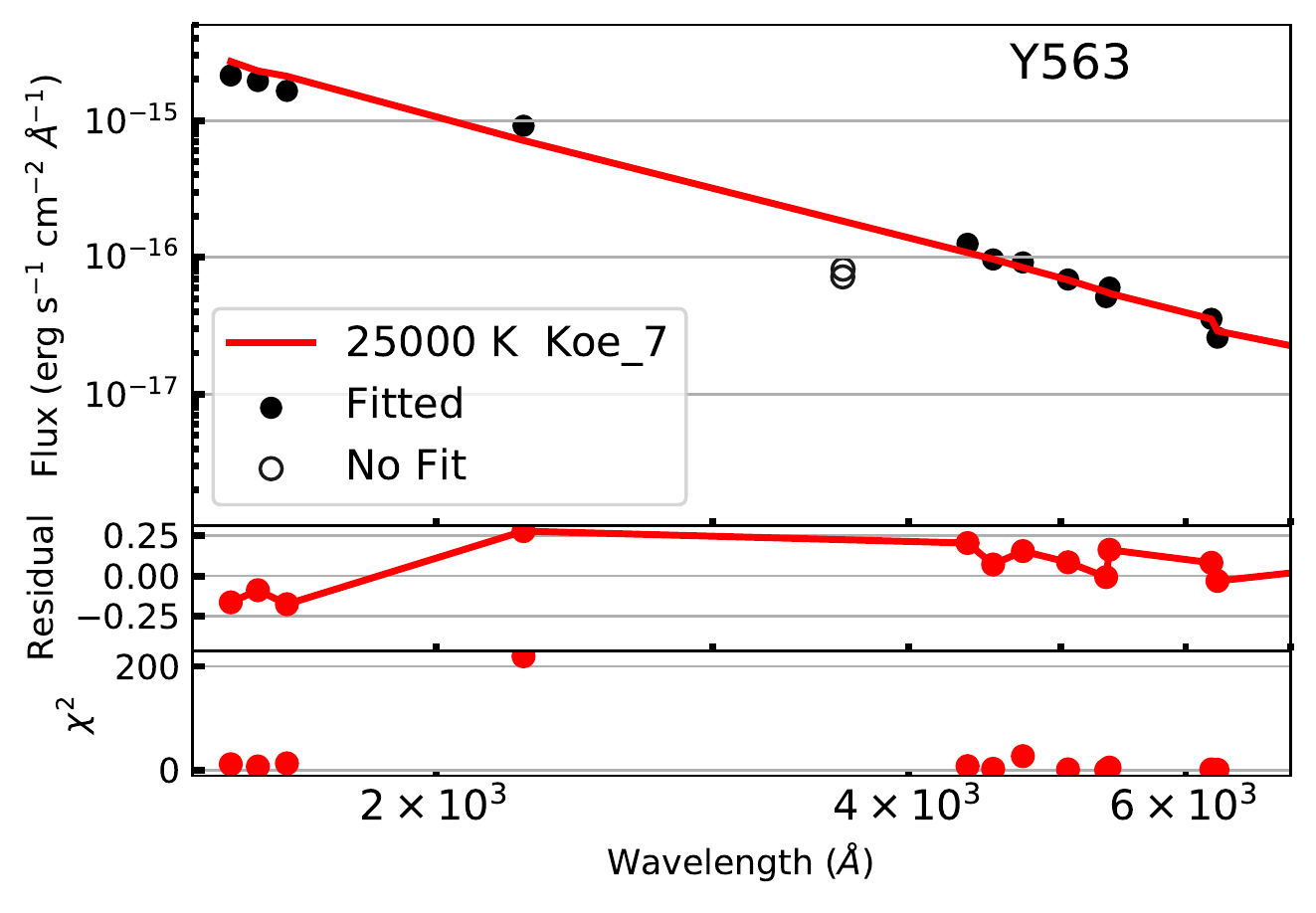} &
    \includegraphics[width=0.45\textwidth]{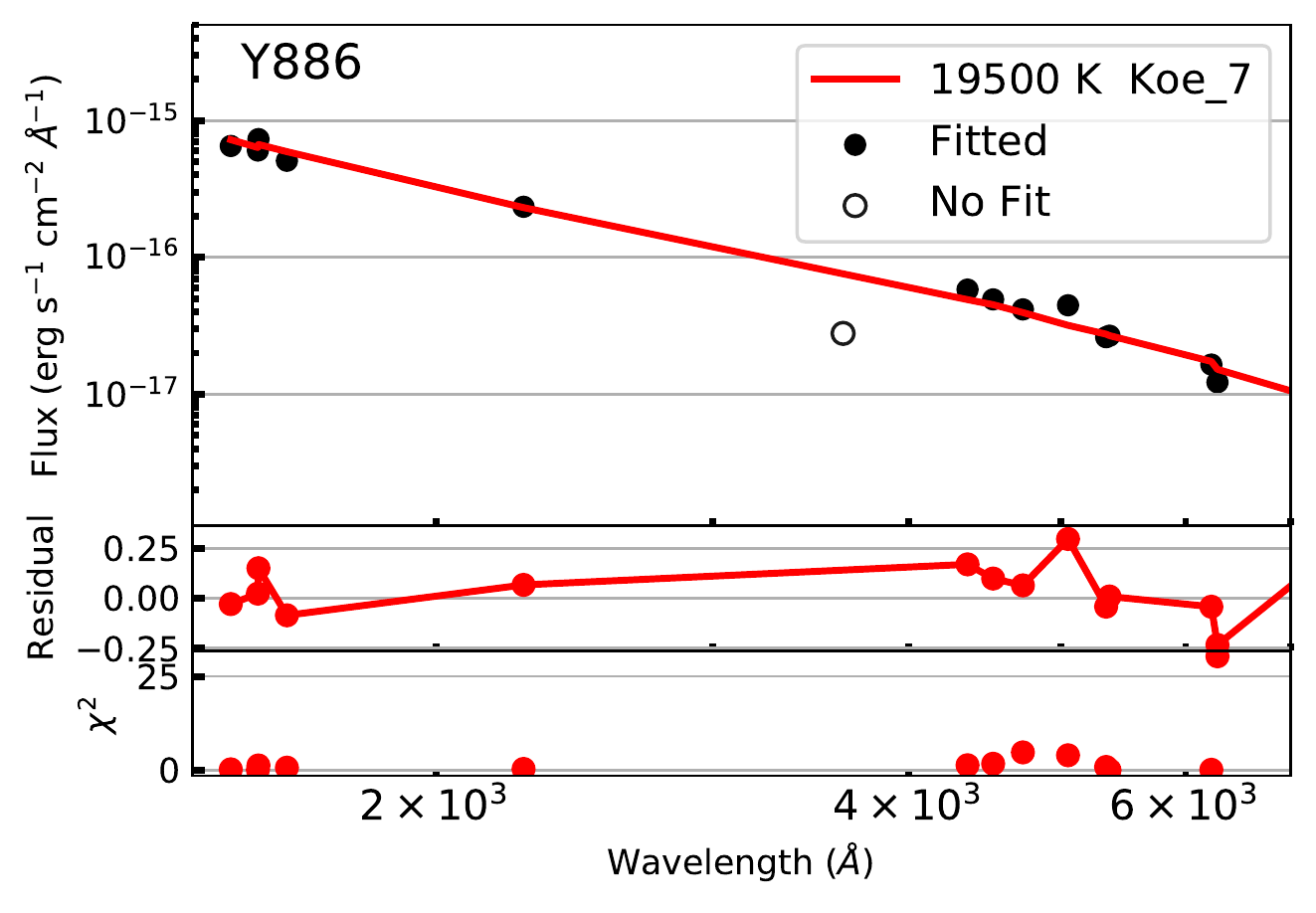} \\
  \end{tabular}
  \begin{tabular}{ccc}
    \includegraphics[width=0.3\textwidth]{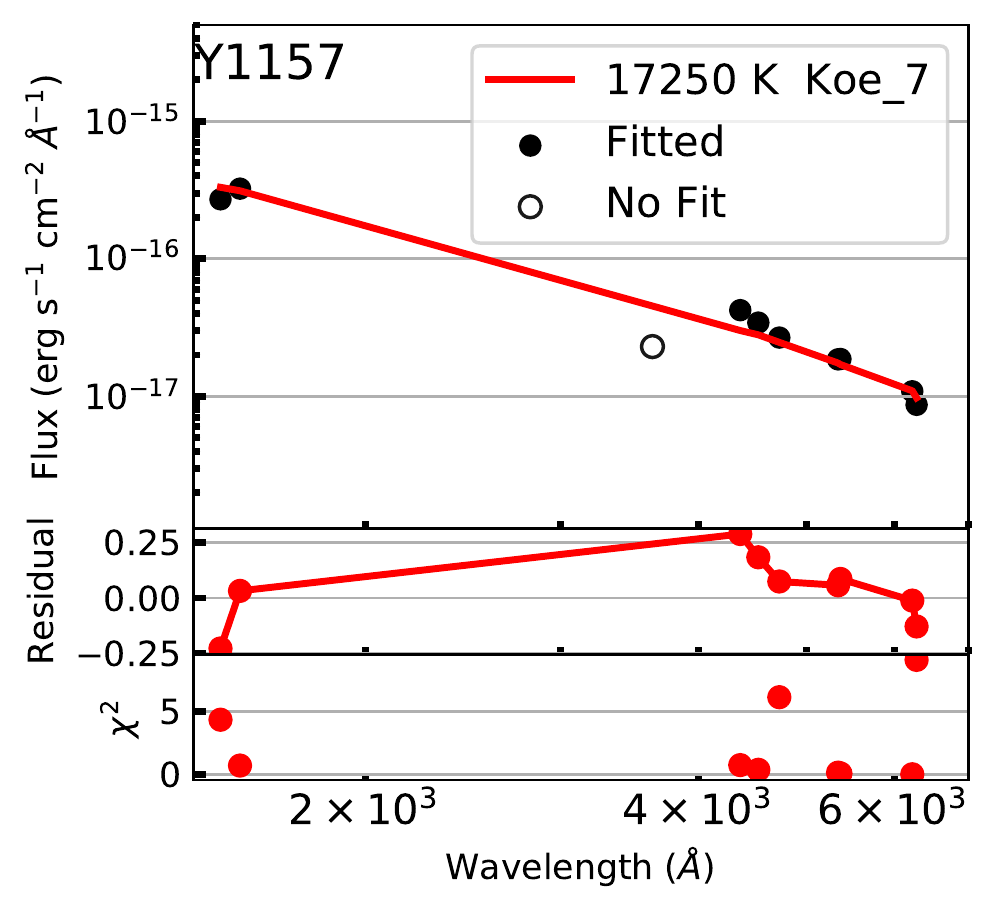} &
    \includegraphics[width=0.3\textwidth]{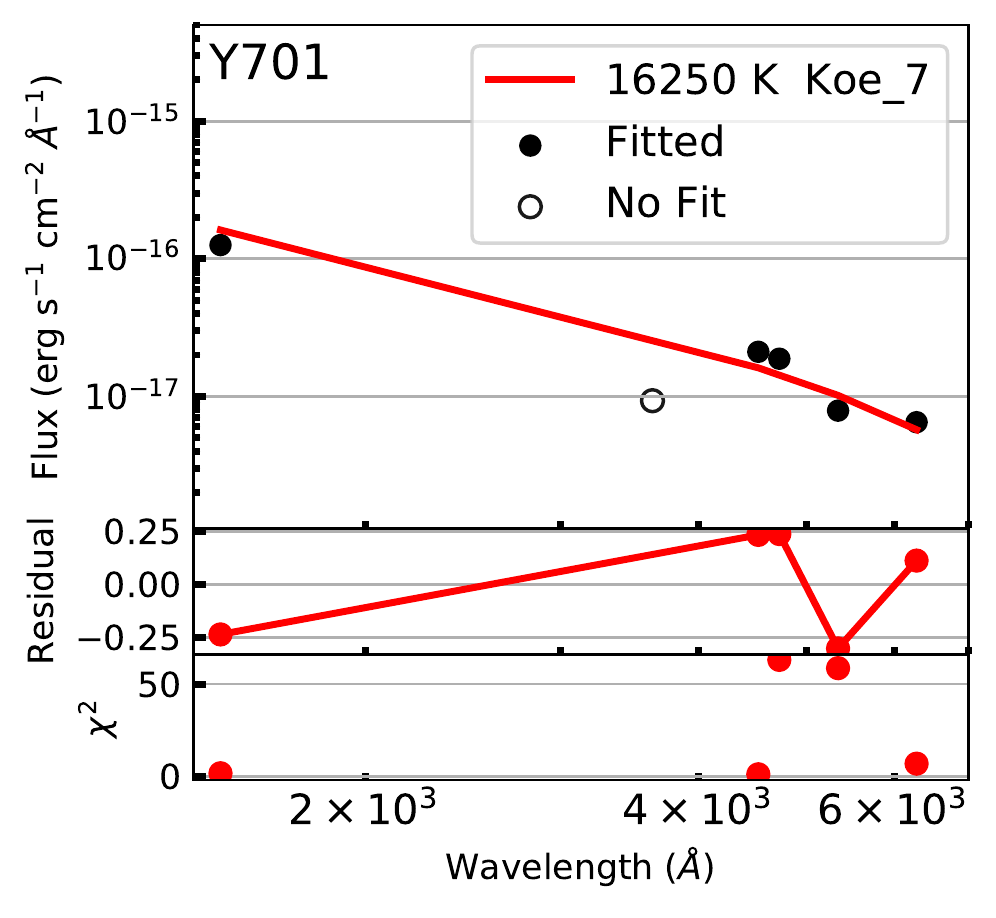} &
    \includegraphics[width=0.3\textwidth]{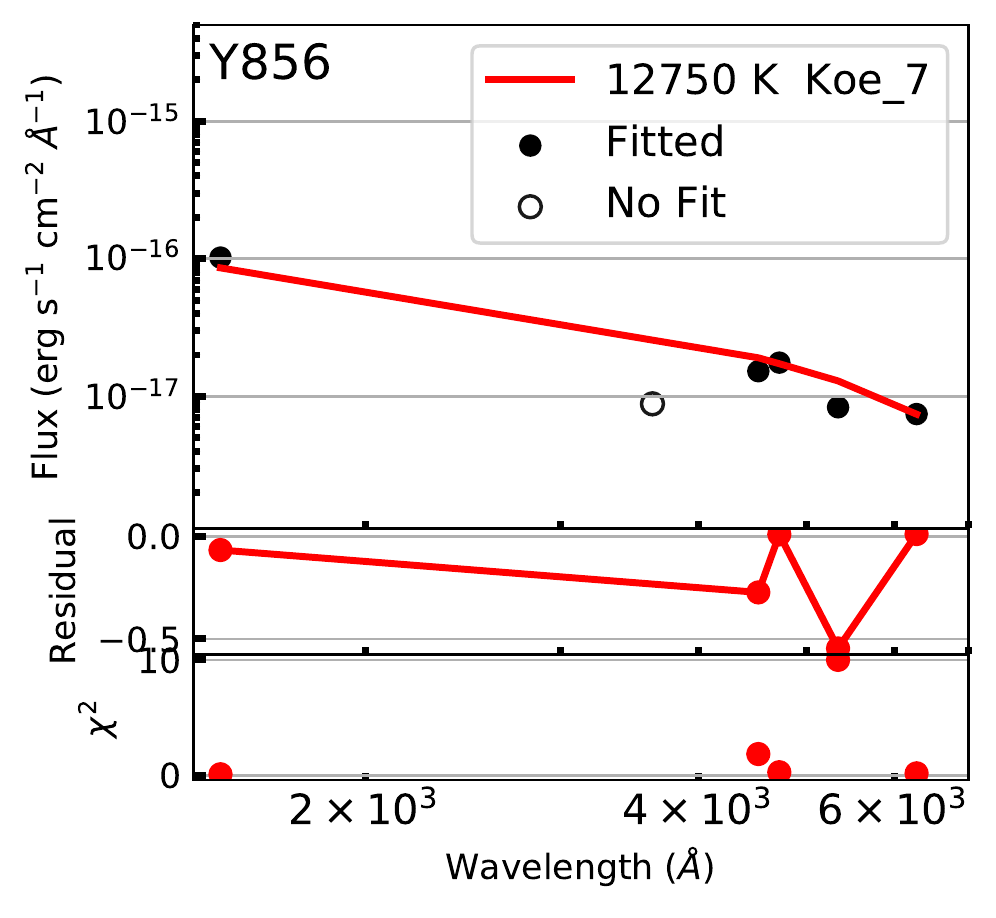} \\
  \end{tabular}
   \caption{SED fits of all isolated WDs with Koester WD model SEDs of log $g$ = 7.}
  \label{fig:SED_WD}
\end{figure*}

The single WD fits are shown in Fig.~\ref{fig:SED_WD}. The stellar SEDs comprising of double fits are shown in Fig.~\ref{fig:SED_double_2002} and  Fig.~\ref{fig:SED_double_1001} to Fig.~\ref{fig:SED_double_7010} and Triple systems in Fig.~\ref{fig:SED_triples} in respective sections.

\subsection{Method to interpret the UV properties}
\label{sec:What can we say from the results}

We present the analysis of individual sources in the following manner. We compiled all relevant information from the literature regarding X-ray detections, periods, eccentricities, temperatures, radii etc. and discussed the UV flux in the light of above properties.
Then we analysed the changes in positions of the sources in various CMDs and their implications. The results of the SED fits are summarised thereafter.

The nature of the UV flux is discussed in accordance with the classification of the source as given below:

\textbf{Comparison with WD models:} We compare the parameters of hotter companions to DA and He-core models in  Fig.~\ref{fig:WD_mass_radius}. The sources which deviate from the WD models are mostly active binaries/triples: WOCS2003 (SB2, RS CVn), 2015 (likely an evolving YG), 4003 (W Uma), 5007 (SM). Among the stars which are not already classified, if the companion parameters deviate significantly from the models, we propose the source of the UV flux not to be a WD. 

\textbf{X-ray detection:} X-rays indicate the presence of some surface activity on the stars (hot-spots, chromospheric or coronal activity). This can contaminate the UV flux to some degree. Thus, the residual UV flux can be the result of these activities or a hotter companion. Even if there is a hotter companion, the $T_{eff}$ and radius obtained via SED fitting may not be accurate. Thus, cannot comment on the presence of a WD.

\textbf{Number of detections in the {\it UVIT}:} The number of data points is an important variable in the SED fitting. Less UV data points lead to multiple SED fits with a relatively similar $\chi^2$. Thus, the fit parameters of stars detected in only 1 filter may not be entirely accurate and hence can only suggest the possibility of hotter companion depending on the previously known information.

\textbf{SB2/Triples:} In this case, the single fits over the optical and IR points are not entirely trivial. Thus, fitting multiple components SEDs to optical-IR part and its implications are explained in appendix section~\ref{sec:triples}.

\subsection{White Dwarfs}
\label{sec:WD}

We detected 7 WDs in the {\it UVIT} images. All 7 WDs have photometry available from \citet{Yadav2008} and \citet{Williams2018}. Y1168, Y563, Y886 and Y1157 (star numbers from \citealt{Yadav2008}) are also detected by \citet{Montgomery1993}. 
We created SEDs using this photometry along with {\it Gaia} DR2 measurements.
The results of single Koester model SED fits are shown in Fig.~\ref{fig:SED_WD}. The U band fluxes in KPNO and MMT filters were found to be consistently lower than models, hence were not considered for the SED fitting. Including these points would create higher $\chi^{2}$ values with slightly lower $T_{eff}$.

A single star at the MSTO, with mass $\sim$ 1.3 \(M_\odot\), should evolve in to a WD of $\sim$ 0.4 \(M_\odot\) \citep{Cummings2018}. Thus, a young WD with mass $>$ 0.4 \(M_\odot\) would require a heavier progenitor i.e. a BSS.
As all the WDs we detect have cooling age $<$200 Myr, the mass alone could be an indicator of a possible BSS progenitor.
\citet{Williams2018} argued that a WD (WD29) of 0.7 \(M_\odot\) should be a product of 3 \(M_\odot\) BSS, assuming the WD mass from a BSS is similar to a WD produced by a single star. 
\citet{Landsman1998} estimated the number of WDs expected with a cooling age $<$ 60 Myr and $<$ 200 Myr as 8 and 25 respectively. We detect 2 and 7 WDs in the respective age ranges with \textit{UVIT}. 
Assuming core radius of 5.2' and tidal radius of 75', the \textit{UVIT} field of view (radius of 14') should contain $\sim$80\% of the WDs. We are detecting $\sim$30\% of the expected WDs \citep{King1966}. The off-centre pointing of UVIT and unknown membership of WDs in literature are the reasons for the lower than expected WDs.
The luminosity of WDs increases with mass for a particular age (Fig.~\ref{fig:WD_mass_radius} (c) and (d)), which supports the higher fraction of high mass WDs detected in this study.
We discuss each of the WD detected by \textit{UVIT} below:

\textbf{Y1157 (WD30, MMJ6126):} \citet{Williams2018} identified it as DB spectral type WD. From Fig.~\ref{fig:WD_mass_radius} we found this to be a WD with 0.66 - 0.7 \(M_\odot\), formed in last 200 Myr, demanding a BSS progenitor.

\textbf{Y1168 (WD15, MMJ5670):} 
Our estimate of $T_{eff}=$ 45000 K is considerably lower than the estimate of 68230 K by \citet{Fleming1997}. 
According to Wein's law, the peak radiation of 60000 K star would be at 48 nm, which is not covered in our observations. Thus, the SED results which depend on the spectral slope may not be as accurate as the results from spectroscopy.
Our mass estimate of 0.66 \(M_\odot\) is comparable to 0.55 \(M_\odot\) as estimated by \citet{Williams2018}, which also demands a BSS progenitor.

\textbf{Y1487 (WD1):} \citet{Williams2018} categorised it as a DB type WD. We calculated the $T_{eff}$ as 13500 - 14750 K with a mass of 0.4 - 0.5 \(M_\odot\). The progenitor could be a BSS as the mass is slightly greater than 0.4 \(M_\odot\).

\textbf{Y563 (WD2, MMJ5973):} \citet{Fleming1997} calculated T$_{eff}=$ 17150 K whereas our estimation is much higher, T$_{eff}=$ 24000 - 25000 K. In this temperature range, our observations in FUV filters is capable of producing a better estimate as the peak of 17150 K lies as 170 nm. As we detect a rising flux in the \textit{UVIT} filters, a hotter temperature estimate is to be preferred.
Mass calculated by \citet{Williams2018} at 0.69 \(M_\odot\) is higher than our estimation at $\sim$ 0.52 \(M_\odot\). The difference in masses is likely the result of a difference in log $g$ between \citet{Williams2018} and our study. The mass again demands a BSS progenitor. 

\begin{figure*}
  \centering
  \begin{tabular}{c c}
    \includegraphics[width=.43\textwidth]{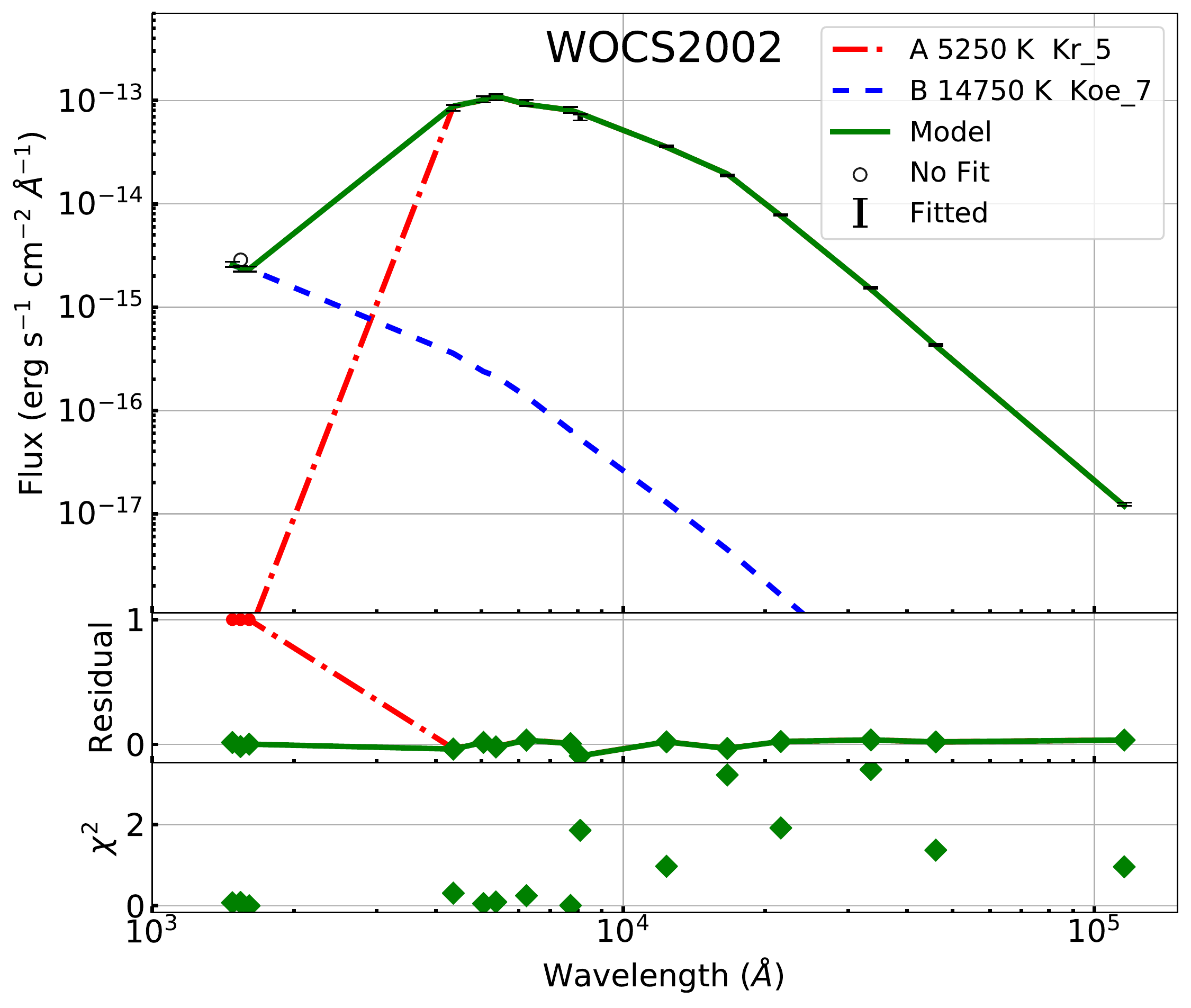} &
    \includegraphics[width=.43\textwidth]{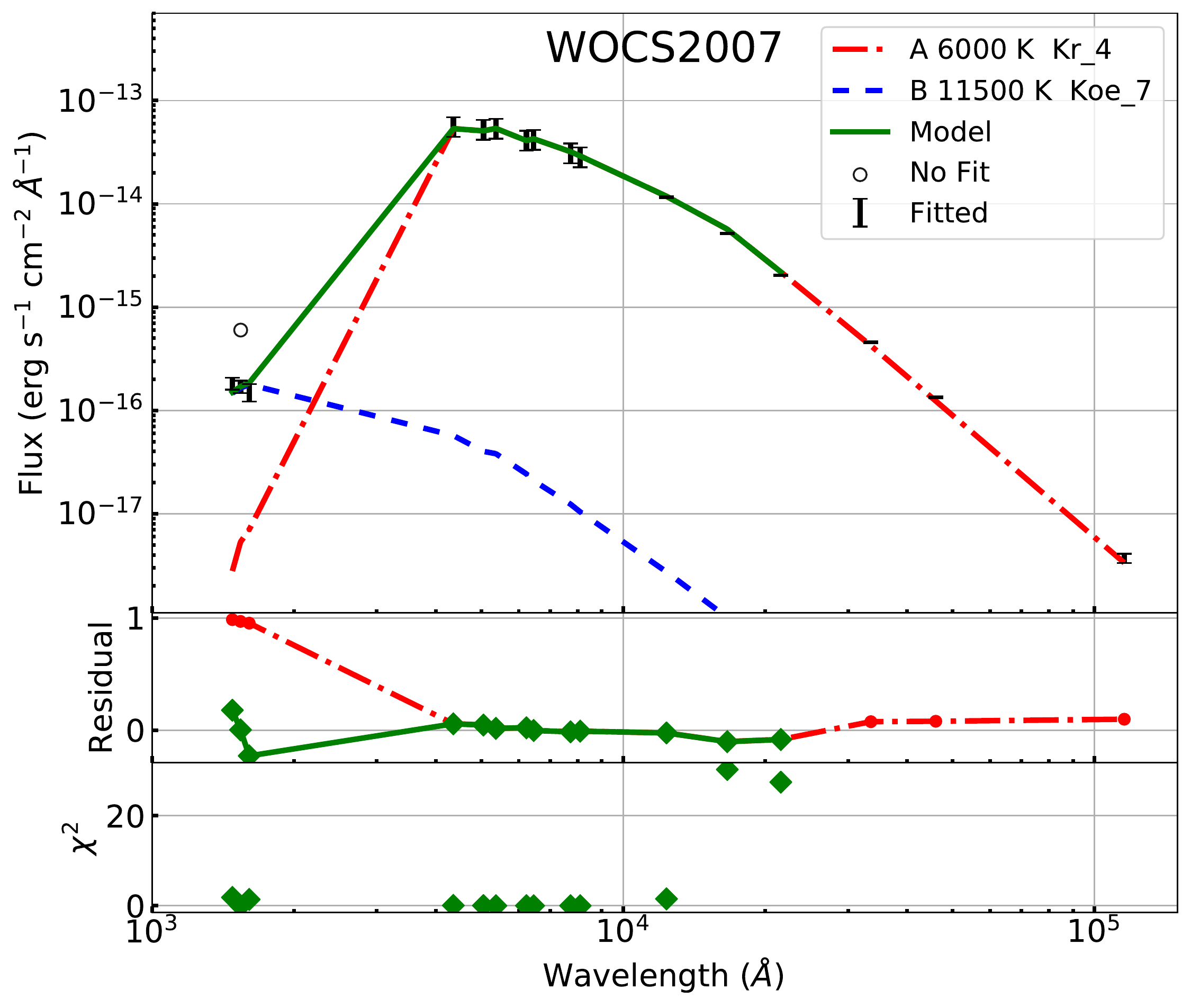} \\
    \includegraphics[width=.43\textwidth]{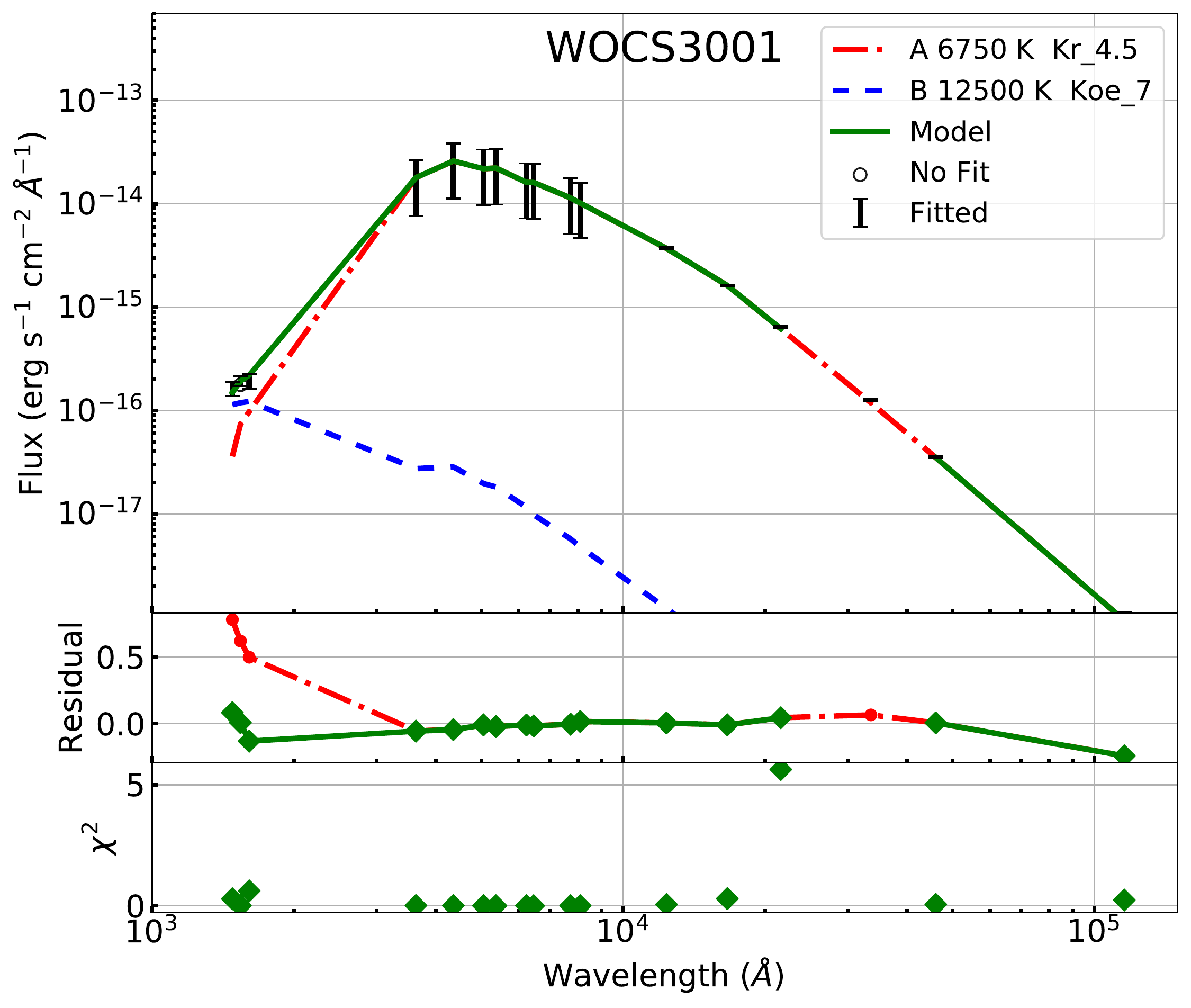}&
    \includegraphics[width=.43\textwidth]{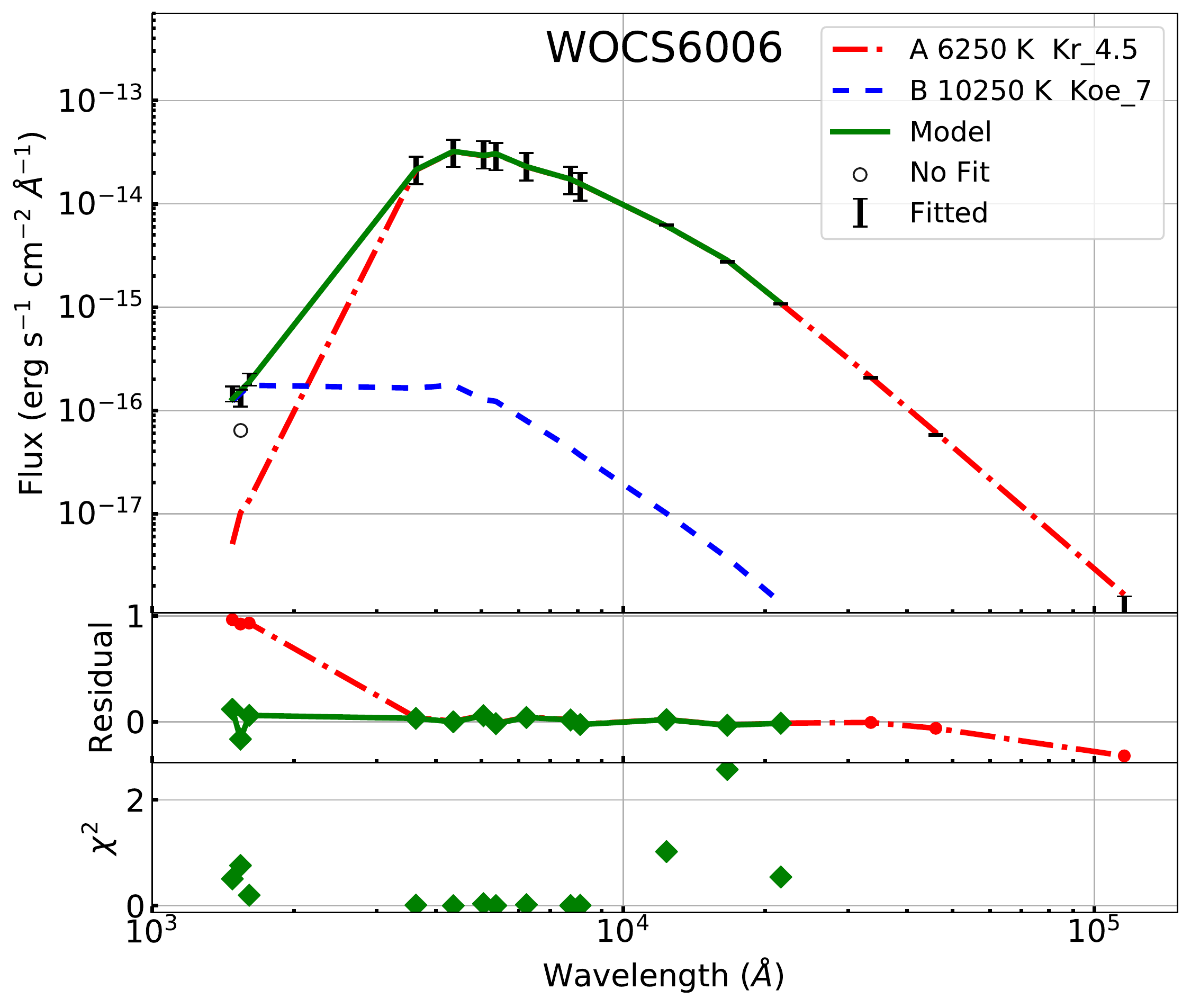} \\
  \end{tabular}
    \caption{Double SED fits of WOCS2002, WOCS2007, WOCS3001 and WOCS6006, same as \ref{fig:SED_3009}c} 
  \label{fig:SED_double_2002}
\end{figure*} 

\textbf{Y701 (WD9) and Y856 (WD10):}
Both WDs were detected in one {\it UVIT} filter. The optical data points along with a single UV data point were not sufficient to fit the SED satisfactorily, thus the $T_{eff}$ and radii estimations are unreliable therefore we do not estimate the mass and cooling age.
\citet{Williams2018} calculated the mass of Y701 $\sim$ 0.56 \(M_\odot\). They also noted Y856 is a possible double degenerate (DD) WD system, thus making the single fit SED unreliable.

\textbf{Y886 (WD25, MMJ6061):} \citet{Williams2018} identified this as a DA type WD with 0.61 \(M_\odot\) which is same as our results at 0.61 - 0.65 \(M_\odot\). The mass and cooling age of 85 Myr suggests that this WD also evolved from a BSS.

\subsection{WOCS2002/S1040}
\label{sec:WOCS2002}
WOCS2002 contains a YG and a WD with a circularised orbit of $P=$ 42.83 days and $e=0.027 \pm 0.028$ \citep{Latham1992}. \citet{Belloni1998} detected the star in X-rays and suggested the X-ray emissions are due to chromospheric activity, as the WD is too cold to produce X-rays. \citet{Mooley2015} found this to be a variable in X-rays. \citet{Landsman1997} studied the source in detail using Goddard High-Resolution Spectrograph (GHRS) and the Faint Object Spectrograph on the \textit{HST} and estimated $T_{eff}=16160 K$, log $g$ = 6.7 and M $\sim$ 0.22 \(M_\odot\). They proposed a mass transfer scenario where the donor in a short period ($\sim$ 2 days) binary began mass transfer while on the lower giant branch leading to the formation of a longer period BSS + helium WD binary. \citet{Van2004} stated that their spectral fits indicate the source of X-ray flux is coronal.

We reproduced the characteristics ($T_{eff}$, mass and luminosity) of the WD companion to WOCS2002 using SED fits for two log $g$ values (Fig.~\ref{fig:SED_double_2002}). Our calculations suggest the WD component of WOCS2002 has $T_{eff}=14750$ to 19250 K. which is in agreement with the estimate of \citet{Landsman1997}.
We estimate the cooling age as 25 - 110 Myr, which suggests the WD have formed recently. The estimated mass of 0.2 - 0.3 \(M_\odot\) indicates that it is an ELM WD and is formed due to interaction with its close companion. As the system demands MT from the WD progenitor to the YG progenitor, the present YG is likely to be an evolving BSS.

\subsection{WOCS2007/S984}
\label{sec:WOCS2007}
This is classified as a BSS and an SM \citep{Geller2015}.
\citet{Shetrone2000} observed the radial velocity variations consistent with a circular orbit with $P=$ 1.5 day, and suggested the possibility of it being a non-interacting close binary. 
As they detected Li abundance similar to a single turn off star, hence concluded it is most likely a result of a collisional merger and has a subluminous companion.
\citet{Sandquist2003b} found no variability in the light curve. \citet{Bertelli2018} calculated the temperature of the star as $T_{eff}=6118 K$, rotational velocity as $v\ sin i=8 km s^{-1}$ and suggested that it is a long period binary.

We detect large and consistent UV excess in all three {\it UVIT} filters while the {\it GALEX} FUV flux was found to be higher than {\it UVIT} estimates.  
The SED fits suggest the presence of $<$ 150 Myr old 11500 K hotter component along with a 6000 K cooler component with luminosities of 0.01 and 8 \(L_\odot\) respectively (Fig.~\ref{fig:SED_double_2002}).

The parameters of the hot component are consistent with the WD models and it does not have X-ray emissions. Therefore, this is likely to be a BSS + WD system. As the WD mass is $<$ 0.24 \(M_\odot\), it should be an ELM WD which has undergone MT.

The surface Li is reduced by both merger \citep{Lombardi2002} and mass transfer \citep{Hobbs1991}. \citet{Shetrone2000} suggested the high Li observed in BSS supports its  formation via merger instead of MT, but the mass of companion WD strongly indicates MT. The high Li abundance found in WOCS2007 is indeed puzzling for either merger or MT formation. The study of Li abundance variation in stellar interactions is beyond the scope of this paper, but WOCS2007 could be an interesting case study for tracing the chemical signatures of BSS formation.

\subsection{WOCS3001/S1031}
\label{sec:WOCS3001}
\citet{Geller2015} listed this system as an SB1 binary member. It is bluer than the MS in optical CMD. \citet{Leiner2019} calculated $v\ sin i=14.7 km s^{-1}$, $P=128.14$ days, $e=0.04$ and a binary mass function = 0.0143. \citet{Leiner2019} suggested that the system is formed through MT as inferred from its rapid rotation and circularised orbit.

We detect a small amount of excess UV flux in all the 3 filters of \textit{UVIT}. We could fit a cooler companion (6750 K) and a hotter companion (12500 K) of 0.3 - 0.45 \(M_\odot\) (Fig.~\ref{fig:SED_double_2002}). 

The hotter component parameters are consistent with WD models. The mass of the possible WD is also compatible with the binary mass function estimated by \citet{Leiner2019}. The absence of X-ray suggests a chromospherically inactive cooler component. Thus, we propose the UV flux is indeed the result of a WD companion to WOCS3001. Although the 0.3 - 0.45 \(M_\odot\) WD may or may not require MT, the circularised orbit and rapid rotation are tracers of MT in close binaries.

\subsection{WOCS6006/S1271}
\label{sec:WOCS6006}
This single member \citep{Geller2015} was observed in all three {\it UVIT} filters. It shifts from MSTO in the optical CMD to the beginning of the BSS sequence in the UV-optical CMD. \citet{Bertelli2018} found T$_{eff}=$ 6360 K.

The large UV flux in all three filters results in a good double fit with a cooler (6500 K) and a hotter (10250 to 11250 K) component (Fig.~\ref{fig:SED_double_2002}). There is a small IR deficiency according to the fitted models in WISE.W2 and WISE.W3 filters.

The hotter companion's parameters match well with the model WD parameters in Fig.~\ref{fig:WD_mass_radius}. Without any X-ray detection, the large UV flux leads us to conclude that there is a WD companion of $<$0.2 \(M_\odot\) to WOCS6006. The presence of an ELM WD signals an MT between the two components, where one component has evolved into an ELM WD, while the other has remained on the MS instead of jumping to the BSS sequence. Therefore, this is a post-MT system and a blue lurker candidate \citep{Leiner2019}. 

\subsection{Other members}
\label{sec:other_members}
We also analysed 7 other \textit{UVIT} detected members (excluding WOCS2002) which are also detected in X-rays. The X-ray emitting phenomenon may or may not contaminate in the UV flux, but as a precaution, we do not confirm any WD companion to these sources. 
Among these, WOCS2009 is a triple system and WOCS7009 is a suspected triple system. We characterised the third components based on the known flux of two components.

Similarly, there are 12 sources with no X-ray detection, and 1 or 2 detections in \textit{UVIT} filters. As fewer data points reduce the significance of SED fits, the parameters derived from the SED fits may not be definitive. Thus, we only suggest a possibility of WD companions to these stars.

The SEDs and detailed discussions of these 19 sources can be found in the  Appendix. The parameters of these fits are listed in Table~\ref{tab:All_para}.

\section{Discussion}
\label{sec:Discussion}

The SED analysis of 30 stars enabled us to characterise these UV bright M67 members. We assessed the nature of UV flux in each system using the SED fit parameters, known binarity and X-ray detections.

The SED fit of stars with multiple {\it UVIT} detections are vital to characterise any hotter component. WOCS2007, WOCS3001 and WOCS6006 have 3 {\it UVIT} detections with large UV excess and are shifted blueward in the UV-optical CMD. The parameters of their possible hotter components more or less match with the WD models. 

WOCS2007 is a short period BSS with a possible ELM WD companion. This makes it the second BSS to be identified to have formed via MT in M67, and also to have an ELM WD as a companion. The first one being WOCS1007, which is recently found to have an ELM WD companion \citep{Sindhu2019}. These systems are post-MT systems where the MT should have happened while the primary is in the sub-giant phase or earlier (Case A/B MT, when the core mass is still $<$ 0.2 \(M_\odot\)).
 
We estimated the mass and age of WDs by comparing them to WD models in Fig.~\ref{fig:WD_mass_radius}. Among 5, 4 WDs have an estimated mass of $>$ 0.5 \(M_\odot\) indicating a BSS progenitor. Y1487 has a mass of 0.4 - 0.5 \(M_\odot\), which may or may not require a BSS progenitor. These massive WDs could be the product of single massive BSSs, or mergers in close binaries or triples (via Kozai-cycle-induced merger). Including the detection from \citet{Williams2018}, the number of WDs which demand a BSS progenitor is on the rise (5 to 6 WDs).

\begin{figure}
  \centering
    \includegraphics[width=0.45\textwidth]{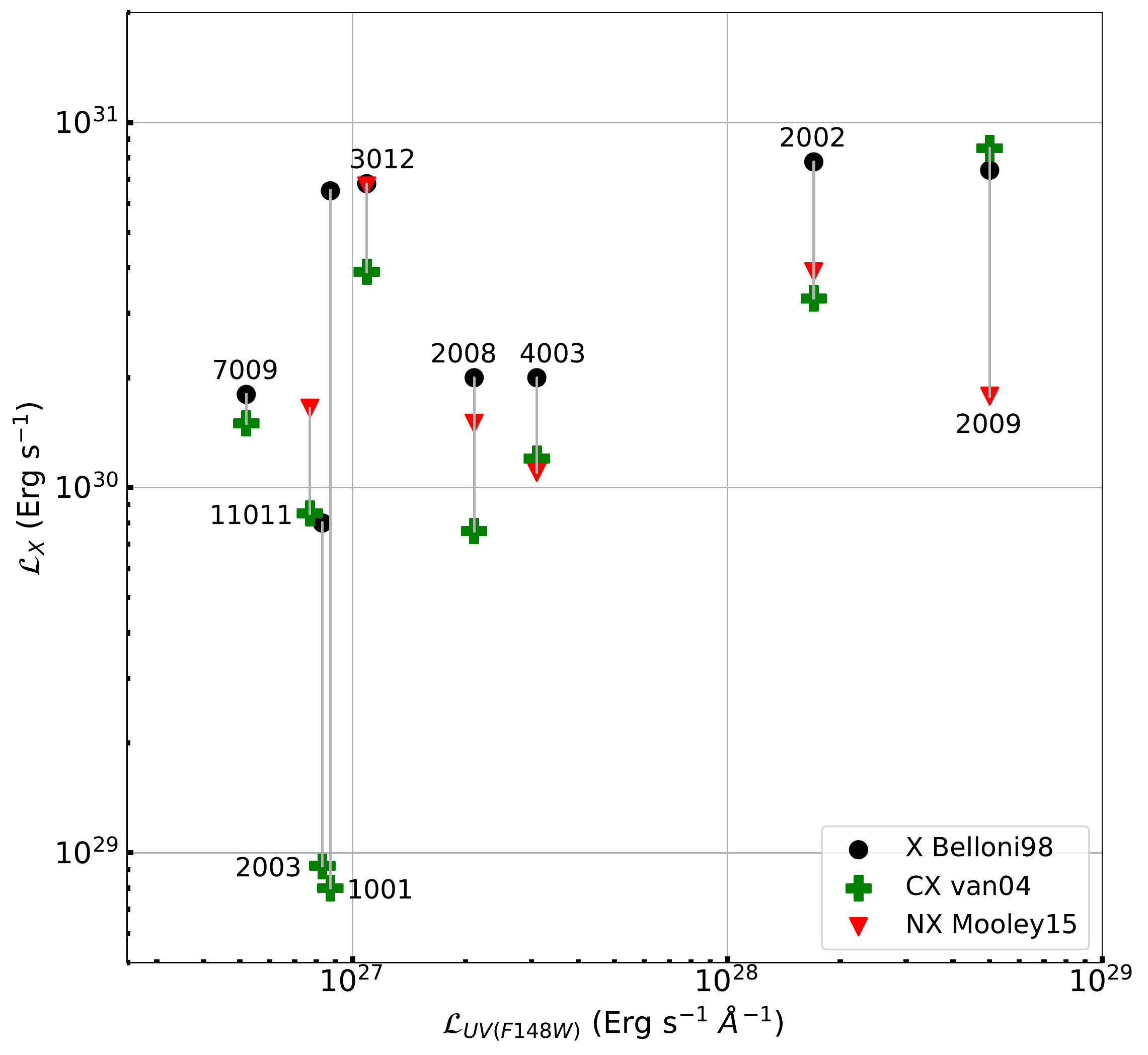} \\
  \caption{Comparison of UV and X-ray luminosity for all X-ray detected stars with X-ray luminosity taken from \citet{Belloni1998}, \citet{Van2004} and \citet{Mooley2015}}
  \label{fig:UV-X}
\end{figure}

9 {\it UVIT} detected members were seen in X-ray by at least one of the missions, \citet{Belloni1998}, \citet{Van2004} or \citet{Mooley2015}. All the sources are spectroscopic binary systems. Close binaries which are spun up by tidal interactions are known sources of intrinsic X-ray emissions in old open clusters \citep{Belloni1993, Van2004, van2013}. The X-ray emissions in stellar flares are also known to cause UV emissions \citep{Mitra2005}. Although we did not find any explicit correlation in UV and X-ray flux as seen in Fig.~\ref{fig:UV-X} (which was not expected due to non-simultaneous observations), the order of UV flux is similar to that of flares.

Acknowledging the possible UV excess flux due to X-ray activity, we fitted a hotter SED to compensate for the residual UV flux, resulting in possible companions with $T_{eff}$ ranging from 9000 to 12000 K.
Comparing the SED fit parameters to WD models, we found potential companions of 2 stars (WOCS2003, WOCS4003) to vary significantly from the WD models. Both systems are contact binaries with known periods and eccentricities, indicating the UV flux is the result of binary interactions or surface activity, and not due to any hotter companions. 

The SED fit of WOCS2009 shows the third component is not a compact hot object, thus the UV flux is due to stellar interactions/activity similar to the contact binaries. We reproduced the $T_{eff}$ of the third component of WOCS2009 consistent with the results by \citet{Sandquist2003}.
The suspected triple system, WOCS7009, showed minimal optical excess. The X-ray emissions and similarities of SED fit parameters with WD models do not conclusively comment on the possibility of a third hotter component or stellar activity.

Our estimates of ELM WD parameters in WOCS2002 are similar to \citet{Landsman1997}, even though it has X-ray emission. This suggests that the X- ray emitting phenomenon does not necessarily contaminate the UV flux. If we extend this argument to other X-ray emitting sources with UV excess, some of them could also have a WD companion with parameters mentioned in Table~\ref{tab:All_para}, such as WOCS11011, WOCS2008 and WOCS7009.

This study has increased the number of post-MT systems in M67 to at least 5 (from the previously known 2; WOCS1007 and WOCS3001). Among these, two are BSSs, one is probably an evolved BSS and two are on the MS. It will be interesting to explore the presence of progenitors and predecessors of these types of systems. The close binary in the triple system WOCS7009 has a low mass companion and therefore could be a potential progenitor of the post-MT system with a low mass WD. Similarly, the contact binaries (WOCS1001, WOCS11011, WOCS2003, WOCS7009) could also evolve through an MT process to a BSS/MS + WD system. Y856 is suspected to be a  double-degenerate WD, which is a potential predecessor. All these points to an increasing amount of evidence suggesting that there is a significant number of systems in M67, which follow the binary evolution pathway through MT, potentially of case A/B type, in close binaries. Since our study is limited to stars near and brighter than the MS turn-off, there could be many more such fainter sources on the MS. Therefore, M67 is likely to have a relatively large number of post-MT systems. 

Some of the short period post-MT systems may also merge with the evolution of the secondary, resulting in single stars with a relatively large mass. Such formation scenario may explain the existence of massive BSSs (e. g. WOCS1010) in M67. Our finding of massive WDs with a range of cooling age (3 - 200 Myr) requiring BSS progenitors indicates that the BSSs have been forming and evolving off, in M67 in the recent past as well. This is in agreement with the suggestion by \citet{Sindhu2018} that this cluster has been producing BSSs more or less continuously. 

ELM WDs degenerate objects are thought to be the products of common envelope binary evolution \citep{Marsh1995} and are the signposts of gravitational wave and/or possible supernova progenitors. 
Most of the low mass WDs are in double degenerate systems or compact binaries \citep{Brown2013, Istrate2014}.
M67 contains at least 2 ELM WD + BSS systems 
The detection of a large number of ELM WDs as companions to MS and BSSs will open up a new window to understand the formation pathways of these WDs. The ELM WDs in practice may be formed either through a Roche lobe overflow or common-envelope ejection event.
Thus, the detection of ELM WDs in binaries is also an important tracer for MT. Binaries in open clusters provide further constraints, which help in determining the evolution of close binaries.

\section{Summary and Conclusions}
\label{sec:Summary}
With the aim of understanding products of binary evolution in open clusters, the results from our study of M67 using \textit{UVIT} FUV images is presented in this paper. Below, we summarise the highlights of this study.

\subsection{Method}
Our observations of M67 using {\it UVIT} detected 41 members which include MS stars, WDs, BSSs and YGs. The UV-Optical and UV CMDs overlaid with isochrones indicate a large number of members having UV excess. We used the SEDs to characterise 30 members (including 7 WDs) by fitting double component SEDs to 21 members and a single component SED to the WDs.

\subsection{Results}
We detect ELM WD companions to WOCS2007 and WOCS6006, hence these are post-MT systems. WOCS3001 also has a WD companion and most likely has undergone MT. We also estimate the mass of ELM WD companion to WOCS2002 suggesting it to also be a post-MT binary.
WOCS11005, WOCS2012, WOCS3009, WOCS4015, WOCS5013, WOCS7005, WOCS7010, WOCS8005, WOCS8006 and WOCS9005 require further observations to confirm the presence of hotter companions.
M67 is therefore likely to have a relatively large number of post-MT systems.
9 sources show X-ray flux and excess flux in \textit{UVIT} filters and are therefore classified as sources with activity (chromospheric/ hot-spots/ coronal/ ongoing-MT). 
5 out of 7 are WDs characterised by SED fitting and 4 of them have mass $>$ 0.5 \(M_\odot\) and cooling age of less than 200 Myr, thus demanding BSS progenitors. The massive WDs detected in M67 require BS progenitors. These massive WDs could be the product of single massive BSSs, or mergers in close binaries or triples (via Kozai-cycle-induced merger).
The SED confirms the presence of the third component in WOCS2009. It is comparable to the cooler star in the inner binary. 

\subsection{Conclusions}
\begin{itemize}
\item The UV-optical and UV-UV CMD of M67 are not as elementary as the optical CMD. The position of stars is heavily impacted by intrinsic and extrinsic factors such as surface activities and binarity. X-ray detections play an important role in identifying stellar activity.

\item This study brings out the importance of deep imaging in the UV to detect and characterise WDs, and WD/ELM WD companions in non-degenerate systems.

\item As many as 12 sources need deeper UV imaging to confirm the presence of a WD companion. Spectroscopic analysis in the FUV region is generally necessary to confirm the existence of all optically sub-luminous low mass WD companions and determine log $g$, mass and age with more certainty. We plan to make deeper observations in FUV filters in {\it UVIT} to identify more sources with potential WD companions as well as to confirm the candidates identified in this study.

\item The detection of ELM WD companions to BSSs (WOCS1007 and WOCS2007) and YG (WOCS2002) shines a light on the formation pathways of these systems. The low mass of WDs signifies that the MT happened before the donor reaching core mass of 0.3 \(M_\odot\), indicating a case A/B MT. Such systems require close binaries as progenitors. Contact binaries (WOCS11011, WOCS4003) and close binaries (WOCS1001, WOCS2003) are likely progenitors of such BSS + ELM WD systems. 

\item Similarly, ELM WD detection along with an MS star (WOCS6006) indicates that other MT systems can be present in the MS of M67 and probably other similar clusters.  They will masquerade as an MS star whose MT will be only decipherable via the presence of an ELM WD or unusually high rotation. Both MS + ELM WD and BSS + ELM WD systems can evolve to form double degenerate WD systems which could remain strong emitters of mHz gravitational waves for Gyr \citet{Brown2016}.

\item This study demonstrates that UV observations are key to detect and characterise the ELM WDs in non-degenerate systems. The presence of systems such as  ELM WD + BSS, ELM WD + MS, WD + MS and hot massive WDs, and evidence of MT on MS show that there are constant stellar interactions going on in M67, which is likely the case for more similar OCs.

\end{itemize}

\acknowledgments
\textit{UVIT} project is a result of the collaboration between IIA, Bengaluru, IUCAA, Pune, TIFR, Mumbai, several centres of ISRO, and CSA. This publication makes use of data products from the Two Micron All Sky Survey, which is a joint project of the University of Massachusetts and the Infrared Processing and Analysis Center/California Institute of Technology, funded by the National Aeronautics and Space Administration and the National Science Foundation. This publication makes use of data products from the Wide-field Infrared Survey Explorer, which is a joint project of the University of California, Los Angeles, and the Jet Propulsion Laboratory/California Institute of Technology, funded by the National Aeronautics and Space Administration. This publication makes use of {\sc VOSA}, developed under the Spanish Virtual Observatory project supported from the Spanish MICINN through grant AyA2008-02156. SN acknowledges support from CSIR for the grant 09/890(0005)/17 EMR-I. We thank the anonymous referee for the invaluable comments.

%





\appendix
\section{Individual Member Discussion}
\label{sec:IMD}

\subsection{WOCS1001/S1024}
\label{sec:WOCS1001}
This is an SB2 source with a period, $P=$ 7.15961 days and eccentricity, $e=$ 0.005$\pm$0.005 \citep{Latham1992}. The source was not detected in X-rays by \citet{Belloni1998}. \citet{Van2004} identified this star as the counterpart of CX111/X46. Hence this star has X-ray emission. \citet{Mathieu1990} find this to be a double-lined spectroscopic binary with nearly identical stars of mass 1.18M$_\odot$. \citet{Yakut2009} detected amplitude variations in its light curve. The orbital period and extremely circular orbit strongly suggest the possibility of an MT event in the system's past.

\begin{figure*}
  \centering
  \begin{tabular}{c c}
    \includegraphics[width=.43\textwidth]{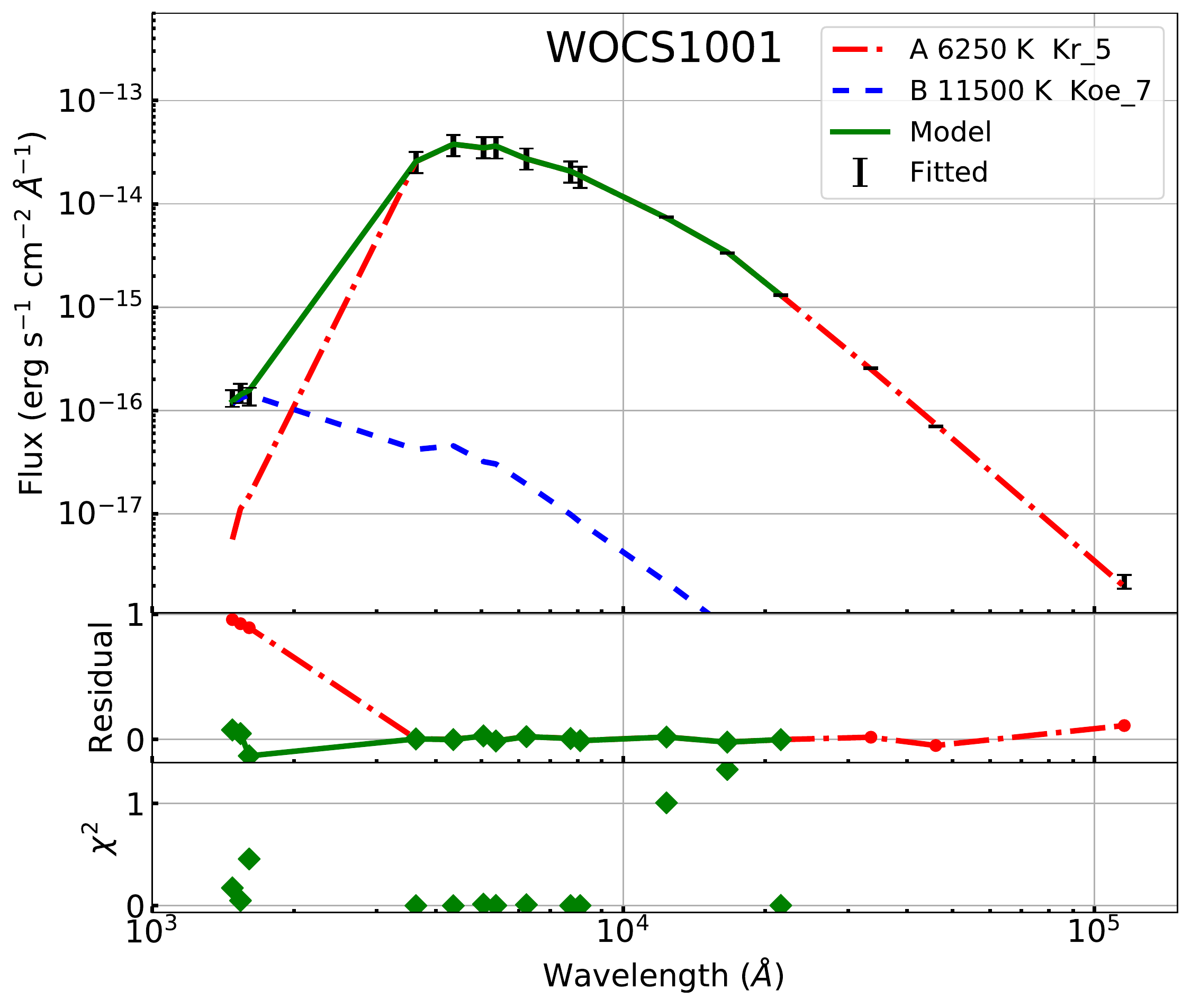} &
    \includegraphics[width=.43\textwidth]{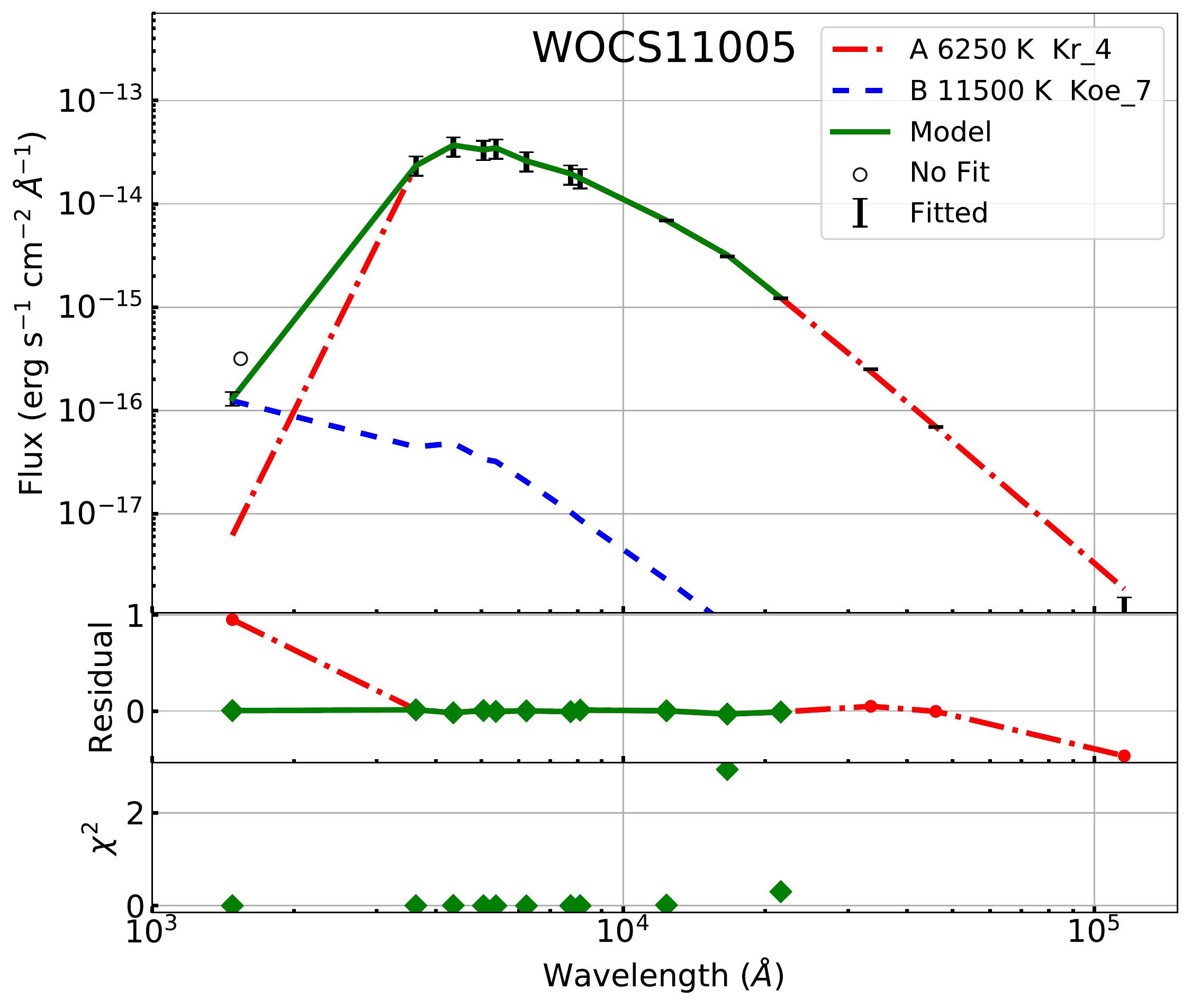} \\
    \includegraphics[width=.43\textwidth]{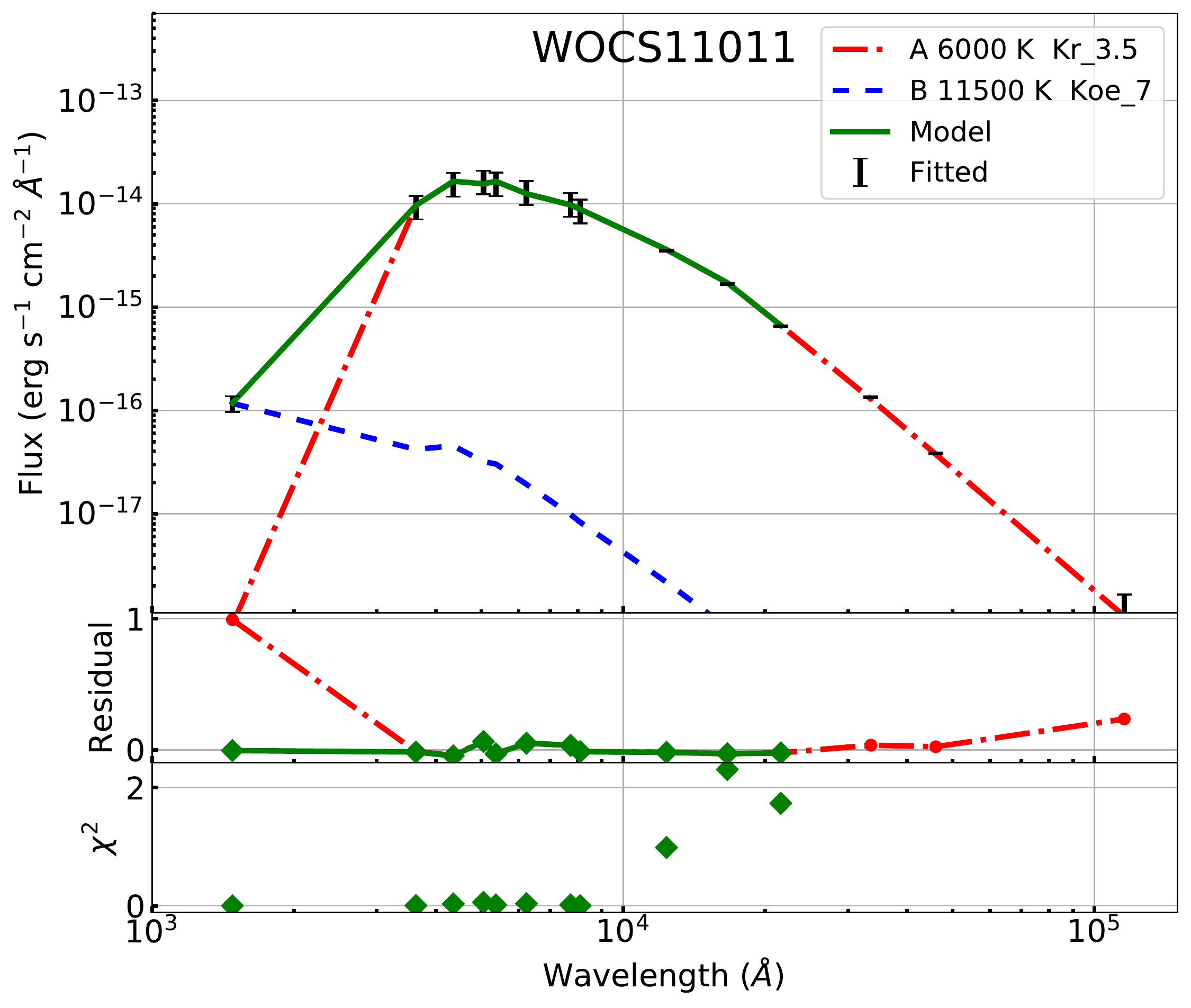} &
    \includegraphics[width=.43\textwidth]{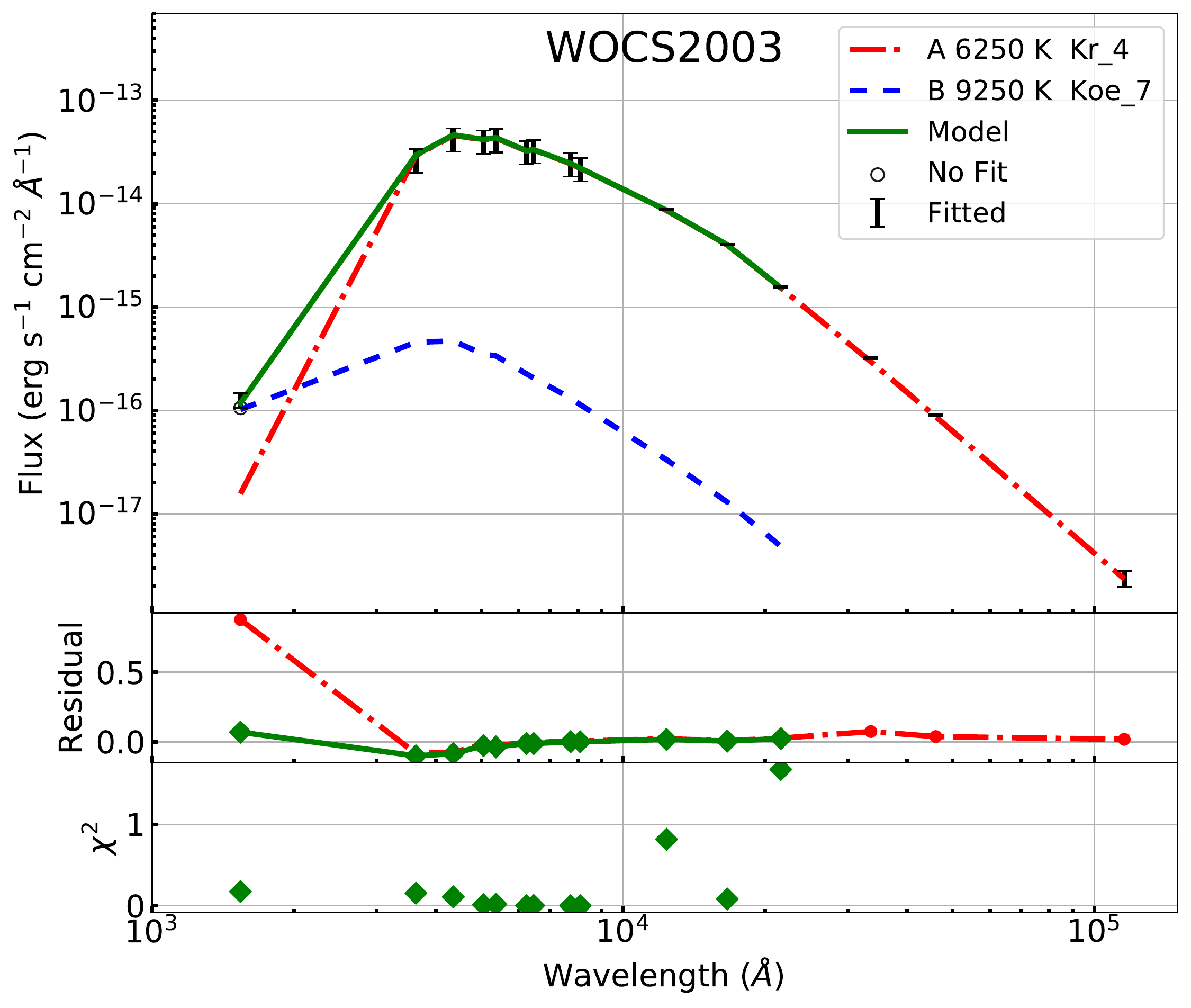} \\
  \end{tabular}
    \caption{Double SED fits of WOCS1001, WOCS11005, WOCS11011 and WOCS2003} 
  \label{fig:SED_double_1001}
\end{figure*}

In the CMDs, the star shifted its location from near MSTO (Fig.~\ref{fig:CMD} (a)) to beginning of the BSS branch (Fig.~\ref{fig:CMD} (b)) of the isochrone, with an FUV excess of about 3 mag. We calculated large $\chi^2$ for the single fit, this along with the consistent UV flux from all three {\it UVIT} filters lead us to perform a double component fit of one hotter (11500 K) and one cooler component (6250 K). Fig.~\ref{fig:SED_double_1001} shows the resultant SED fit for the Koester model hotter component with log $g$ = 7. 
We can establish one hotter and one cooler component from the SED fit, The hotter component of 11500 K is compatible with WD models in Fig.~\ref{fig:WD_mass_radius}. The luminosity are 5 and 0.02 \(L_\odot\) for cooler and hotter components.

Though the results are compatible with the presence of an optically subluminous WD companion to WOCS1001, the presence of X-ray flux, precisely known period suggests the UV source is not a third hotter component. Although the estimated $T_{eff}$ from the SED (11500K) is quite high for a chromospheric activity, it is possible that there could be MT between the two stars creating a hot spot. Then the UV, as well as the X-ray emission, would be from the hot spot on one of the binaries. It is also possible that there is a hot corona for this pair and the detected emission could be due to coronal activity, which is normally seen in contact binaries \citep{Brickhouse1998}.

\begin{figure*}
  \centering
  \begin{tabular}{c c}
    \includegraphics[width=.43\textwidth]{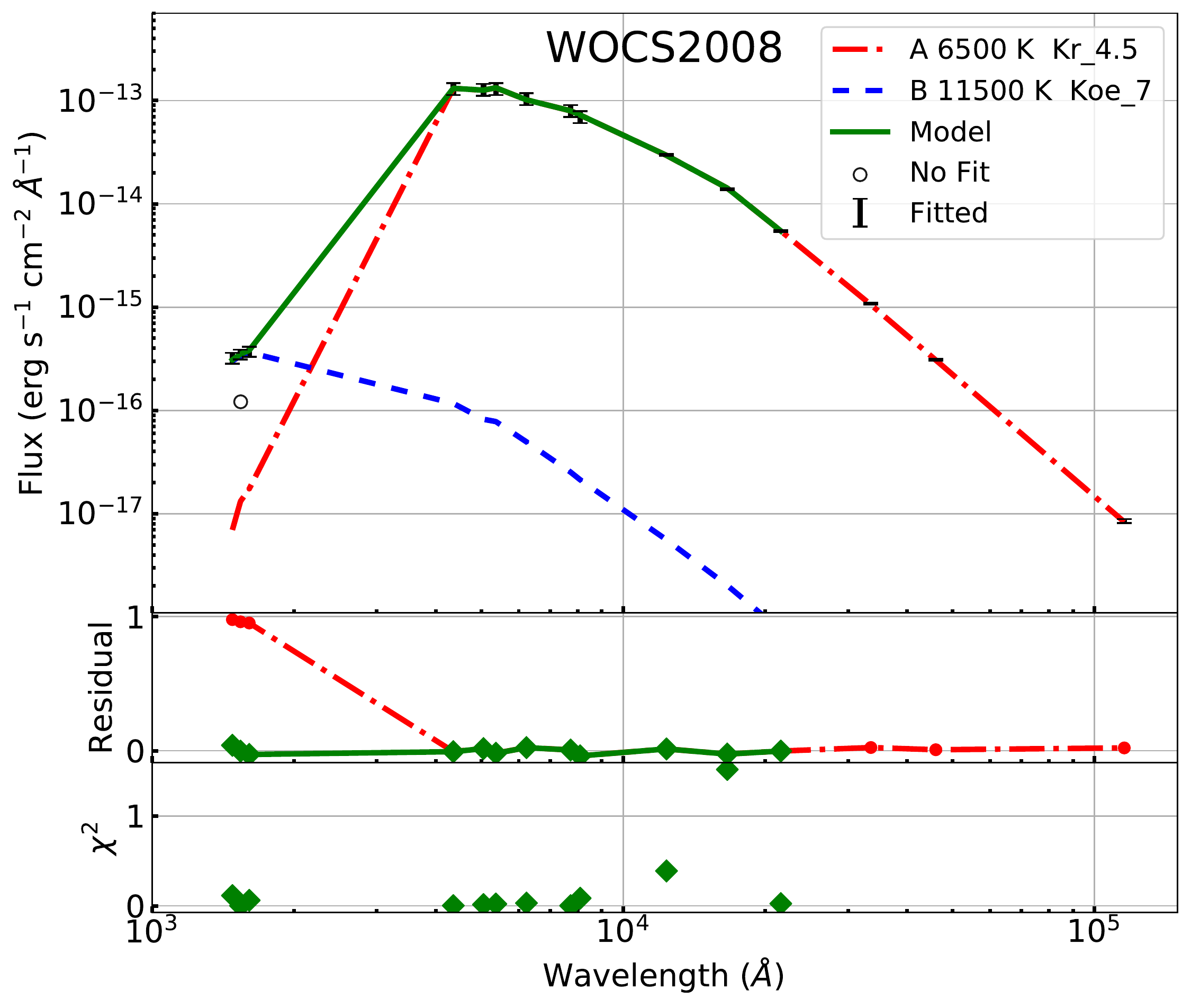} &
    \includegraphics[width=.43\textwidth]{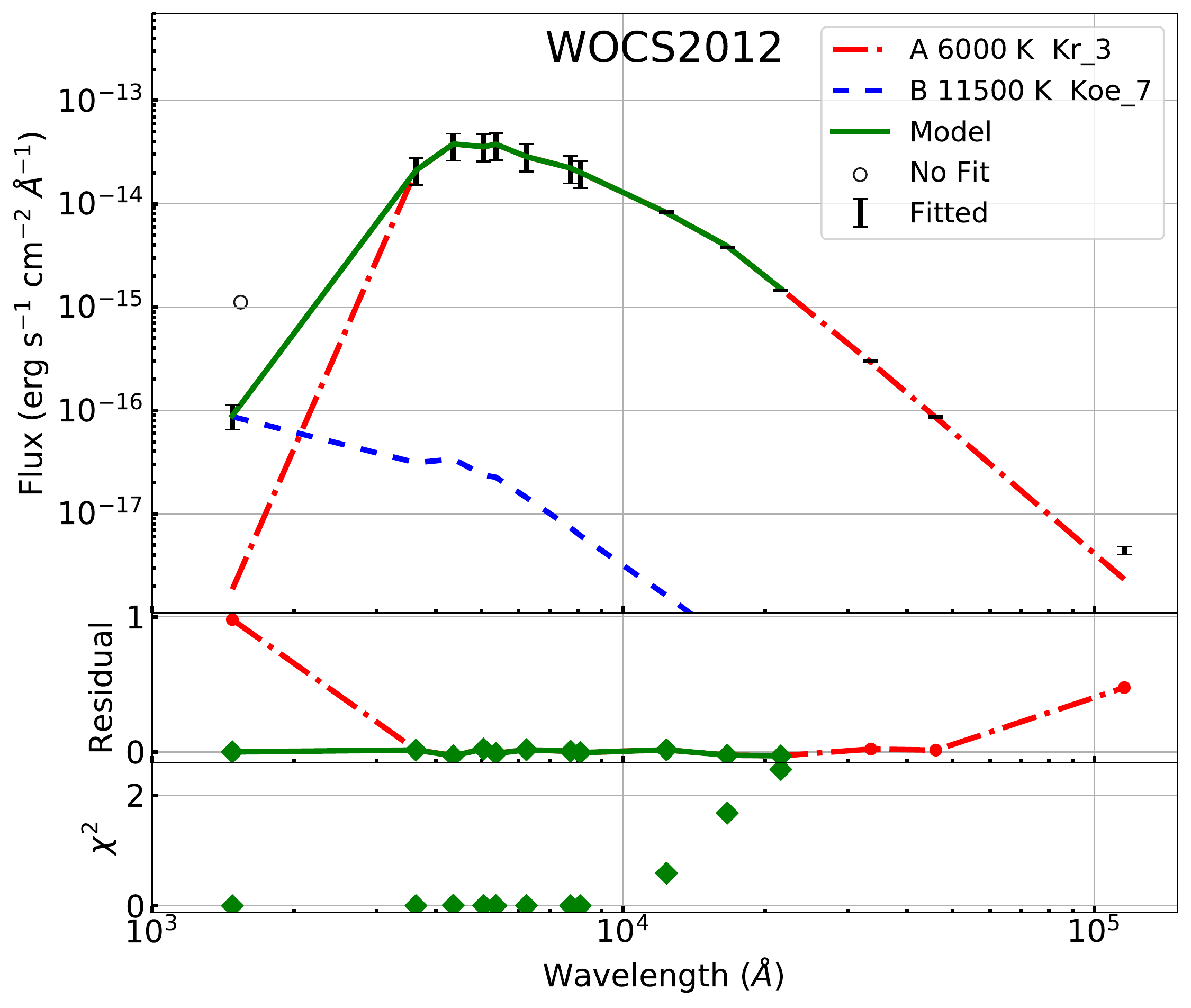} \\
     \includegraphics[width=.43\textwidth]{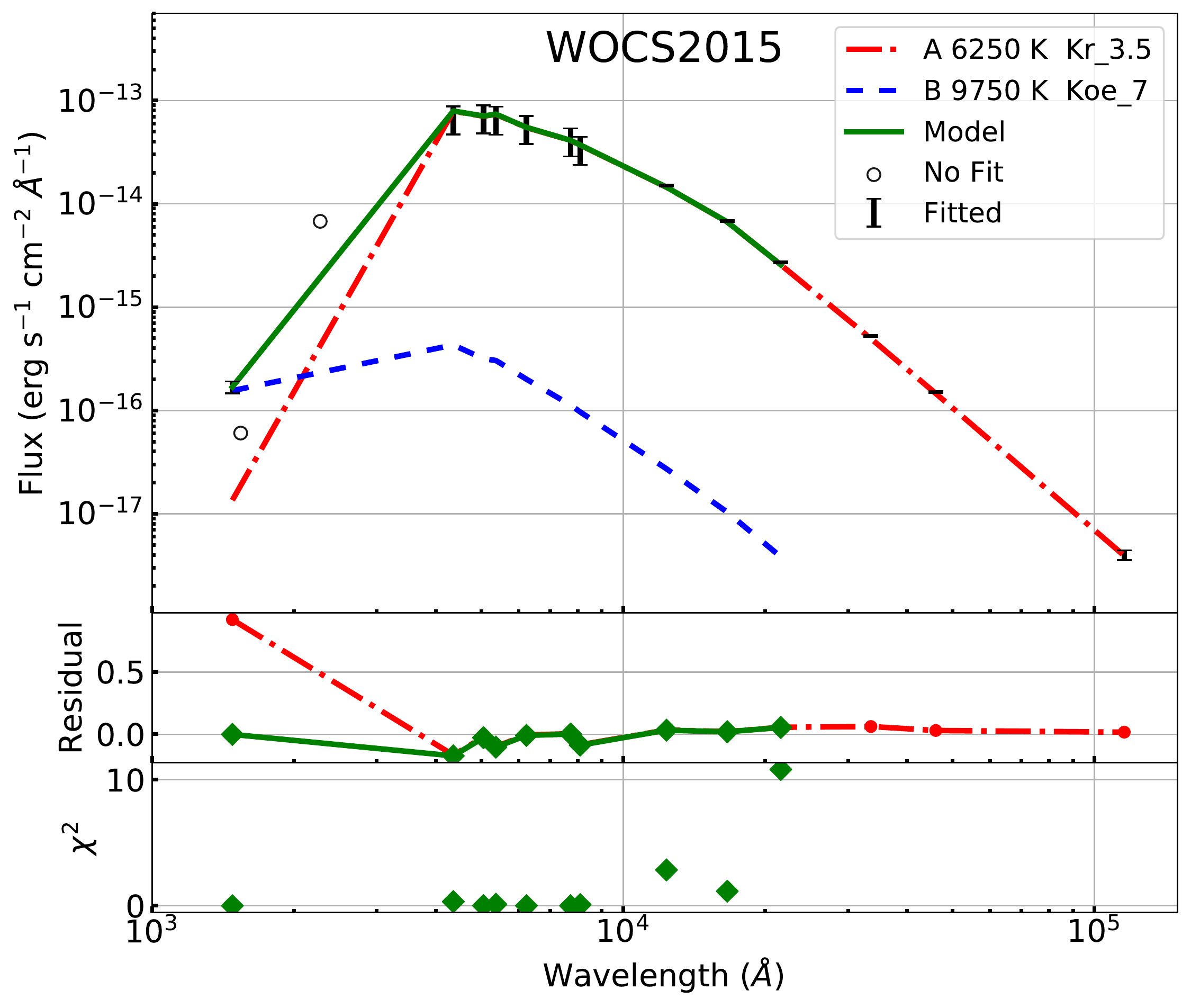} &
    \includegraphics[width=.43\textwidth]{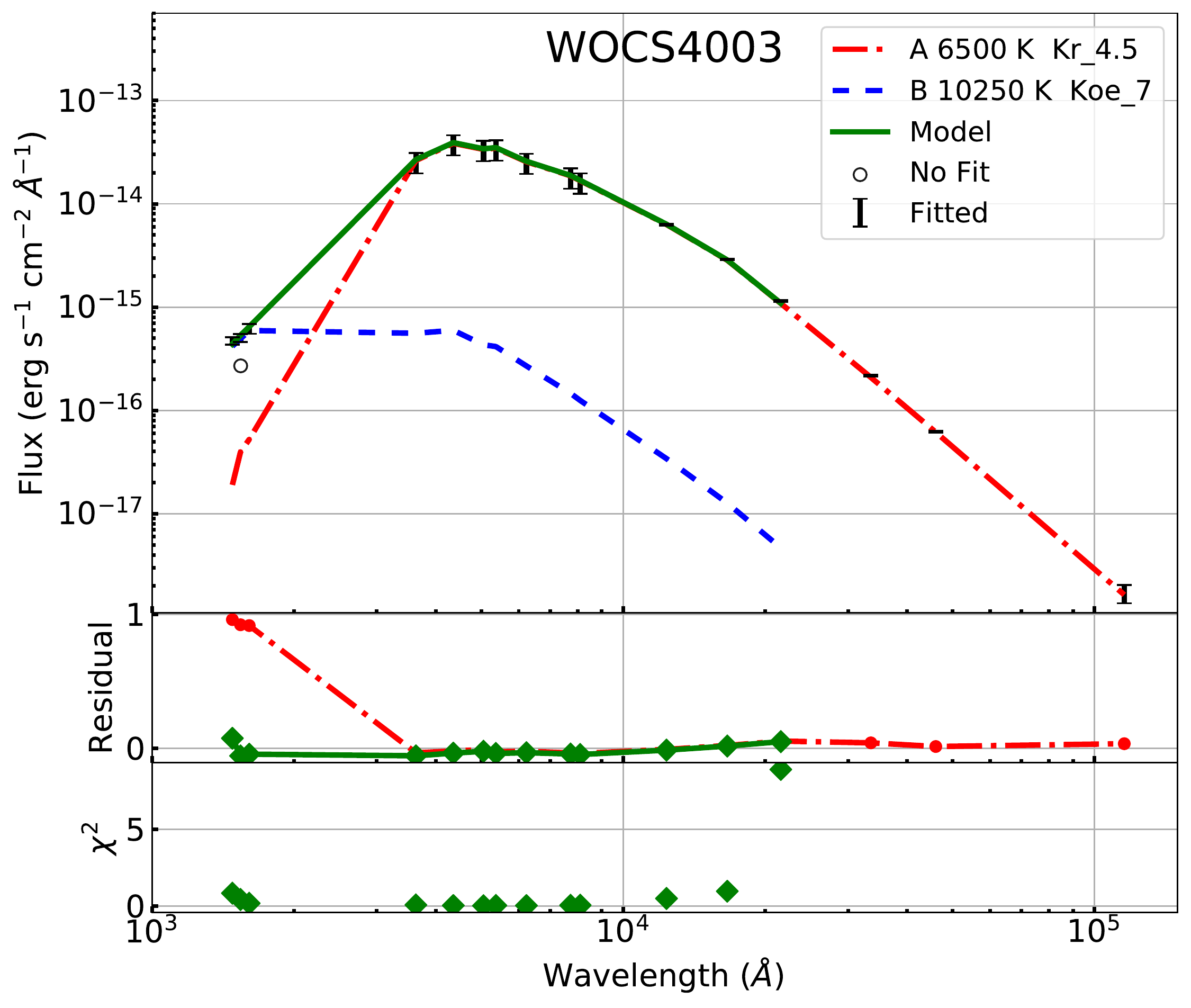} \\
  \end{tabular}
    \caption{Double SED fits of WOCS2008, WOCS2012, WOCS2015 and WOCS4003} 
  \label{fig:SED_double_2008}
\end{figure*} 

\begin{figure*}
  \centering
  \begin{tabular}{c c}
    \includegraphics[width=.43\textwidth]{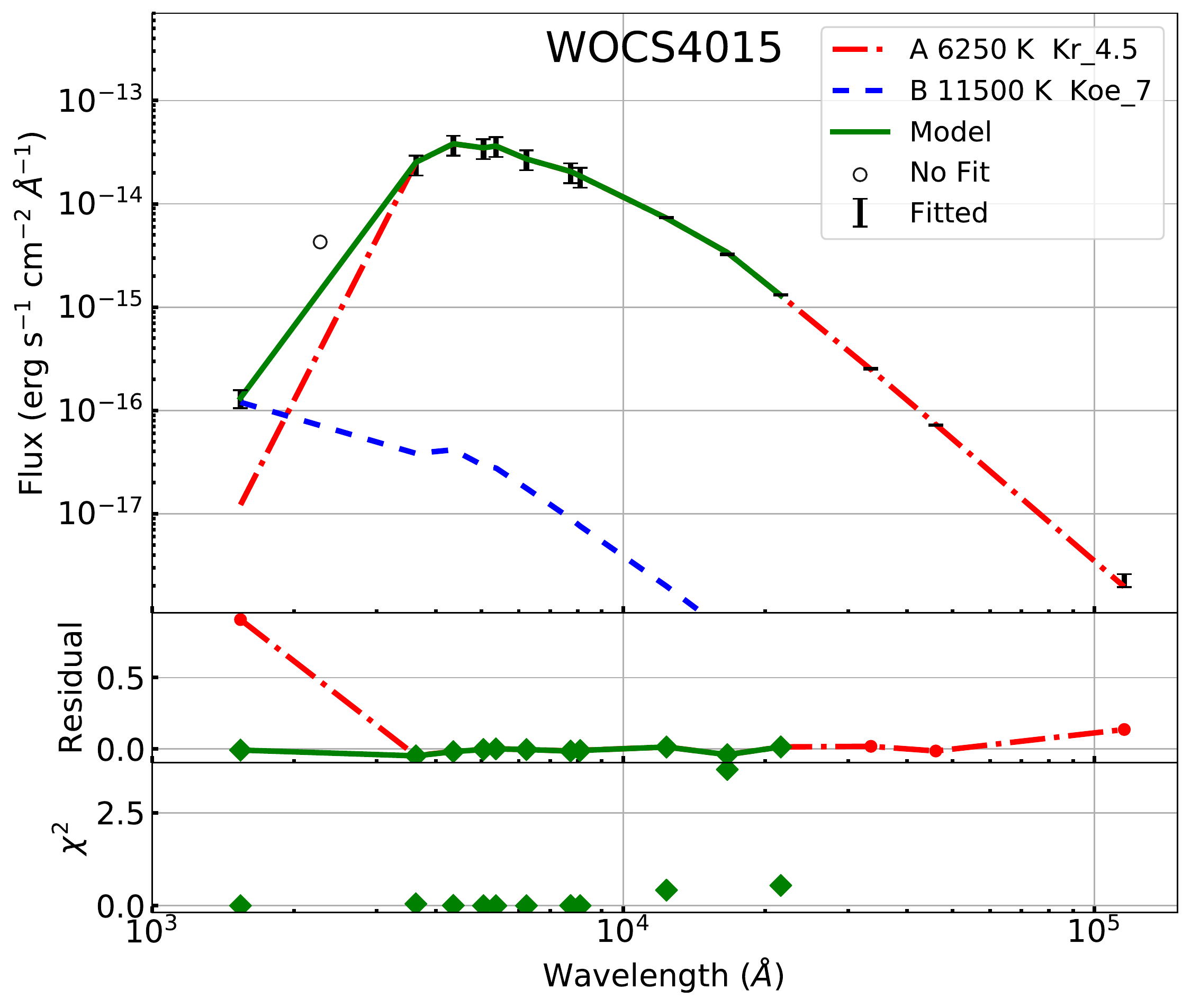} &
    \includegraphics[width=.43\textwidth]{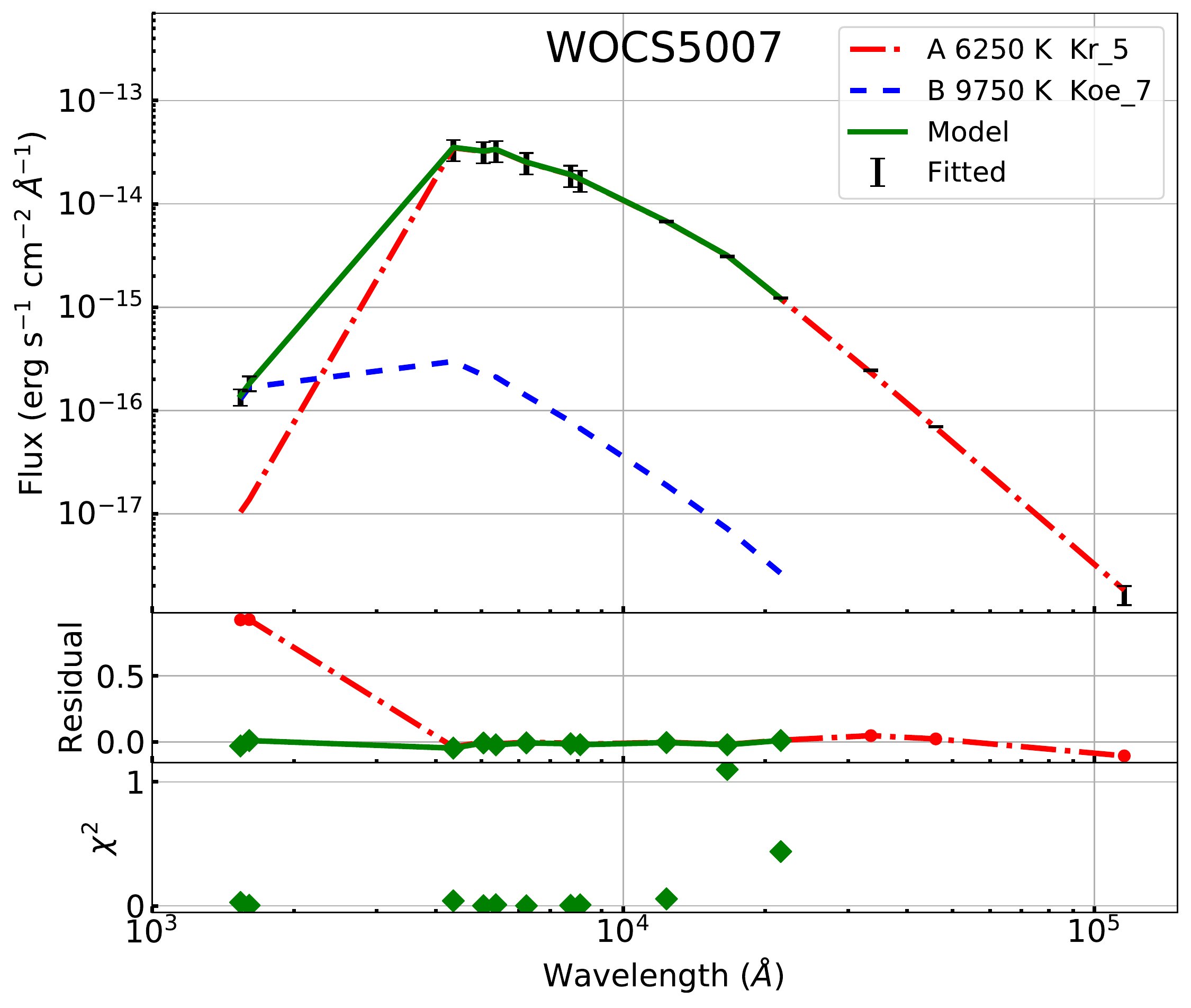} \\
    \includegraphics[width=.43\textwidth]{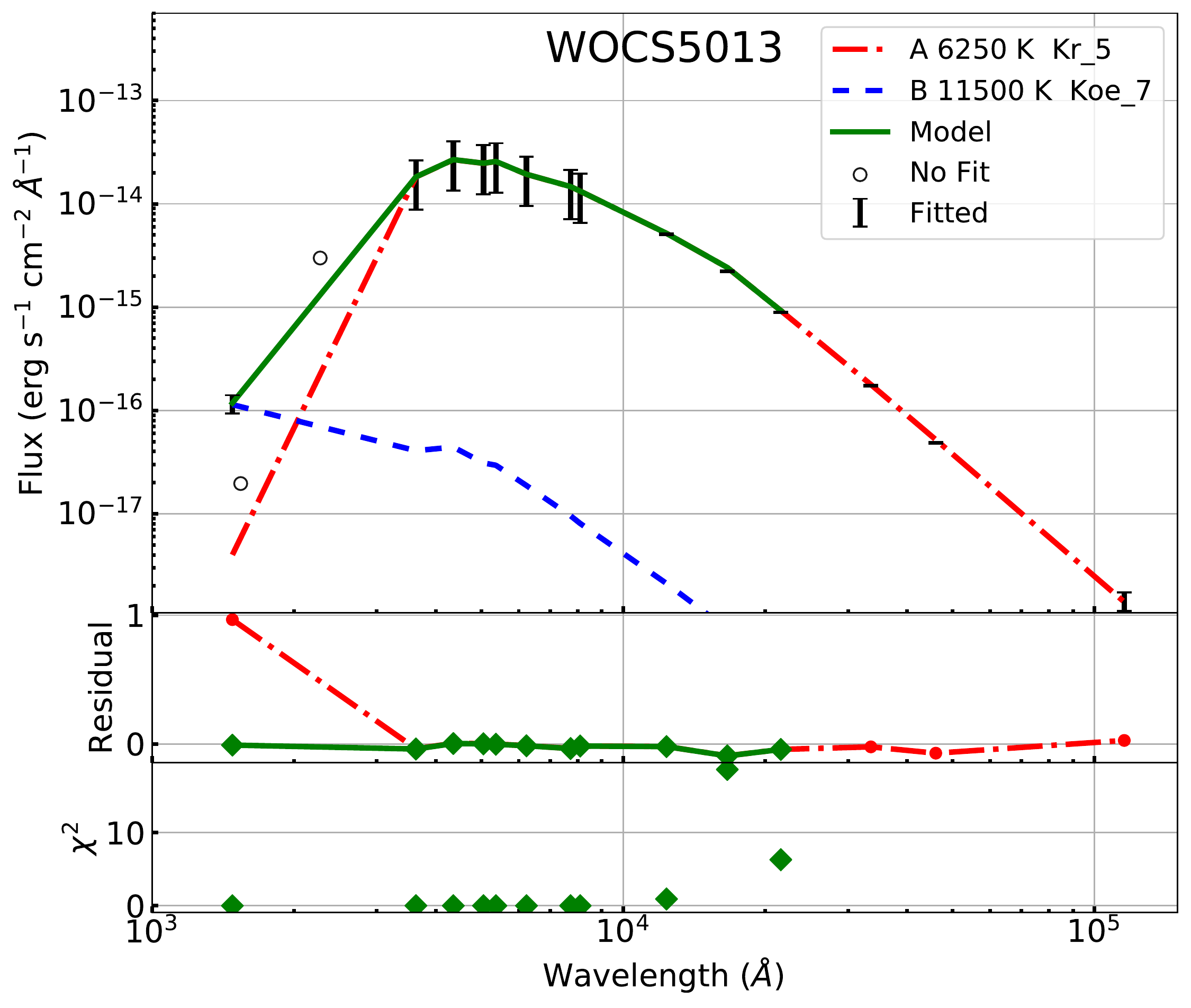} &
    \includegraphics[width=.43\textwidth]{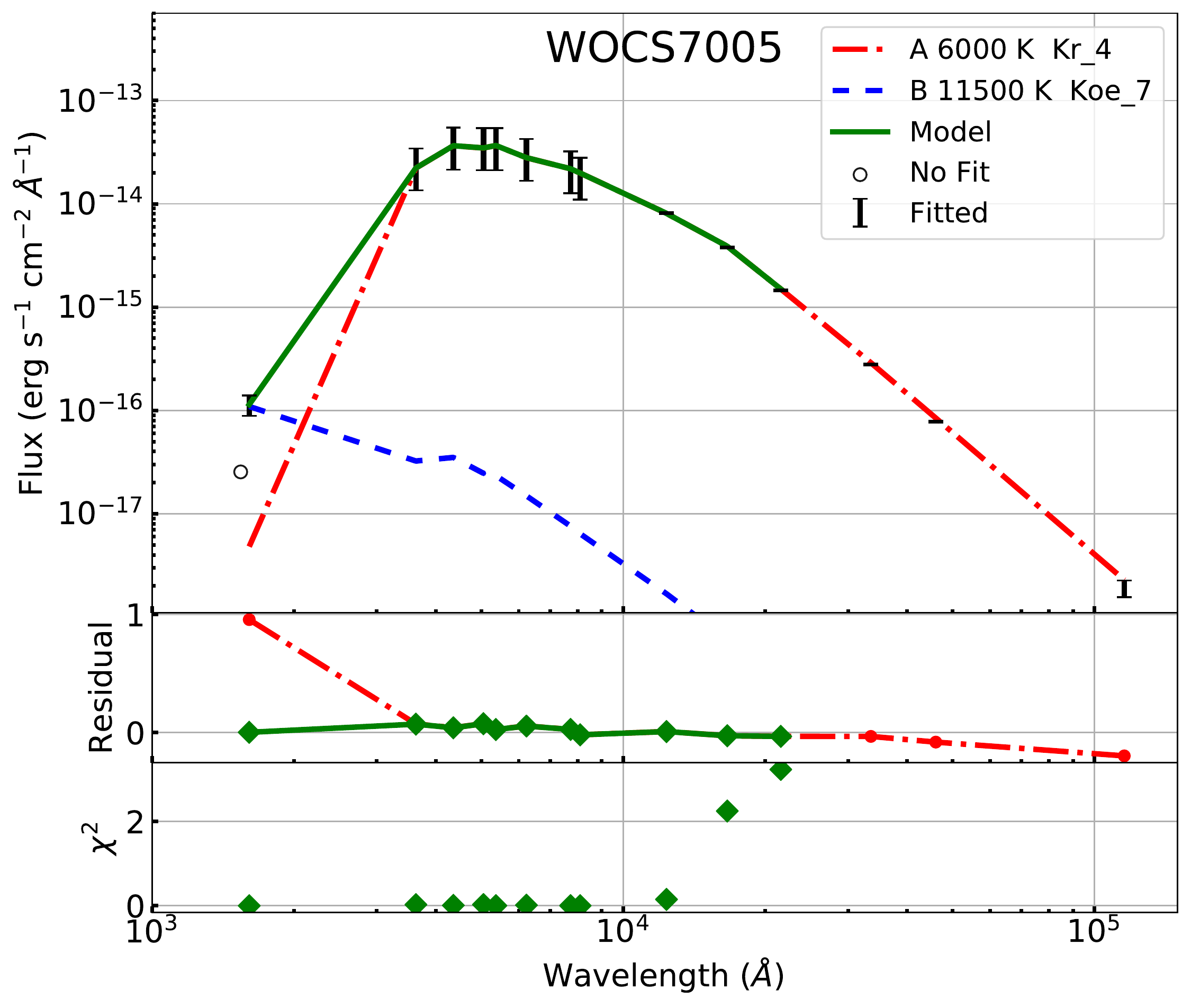} \\
  \end{tabular}
    \caption{Double SED fits of WOCS4015, WOCS5007, WOCS5013 and WOCS7005} 
  \label{fig:SED_double_4015}
\end{figure*}

\begin{figure*}
  \centering
  \begin{tabular}{c c}
    \includegraphics[width=.43\textwidth]{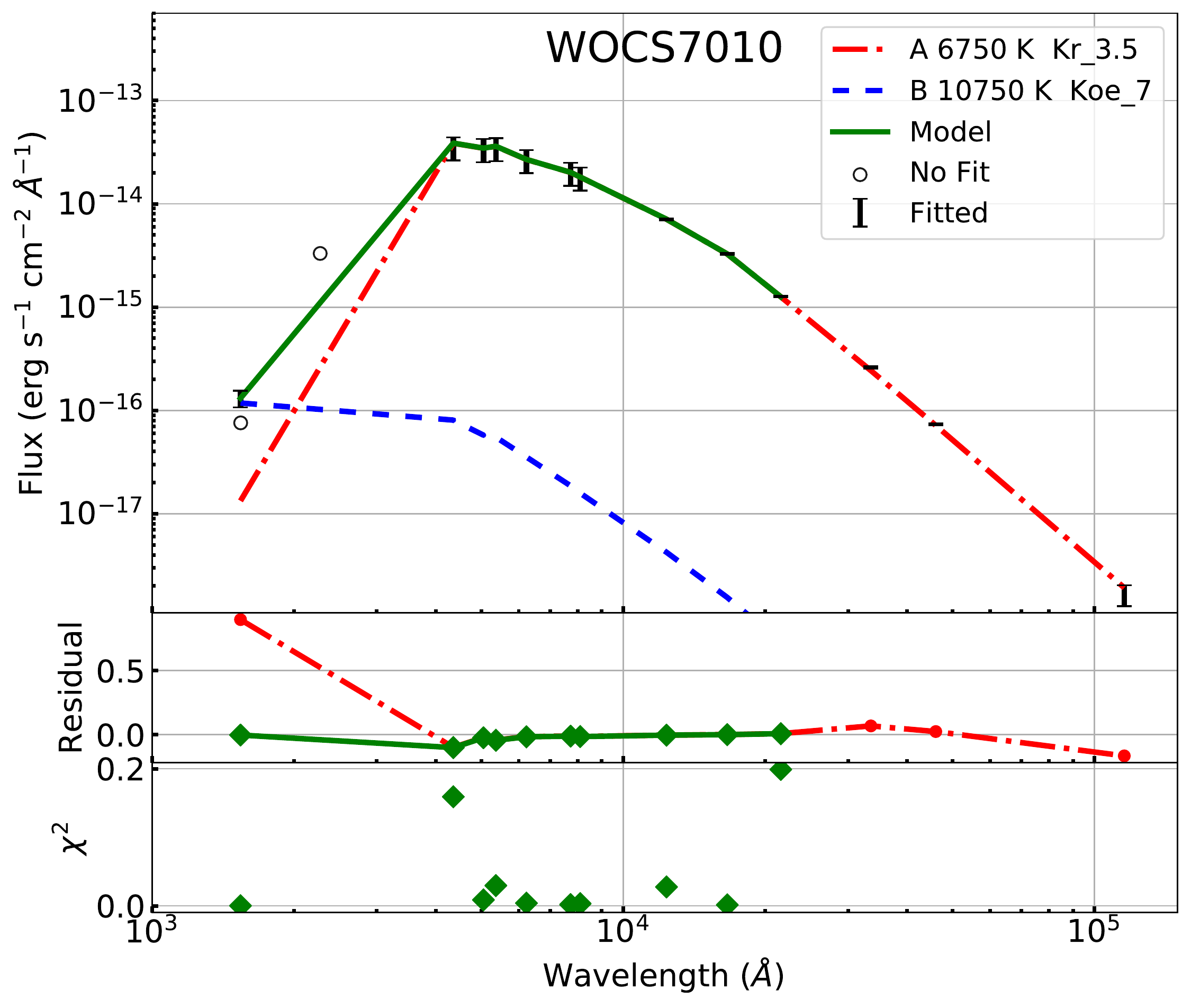} &
    \includegraphics[width=.43\textwidth]{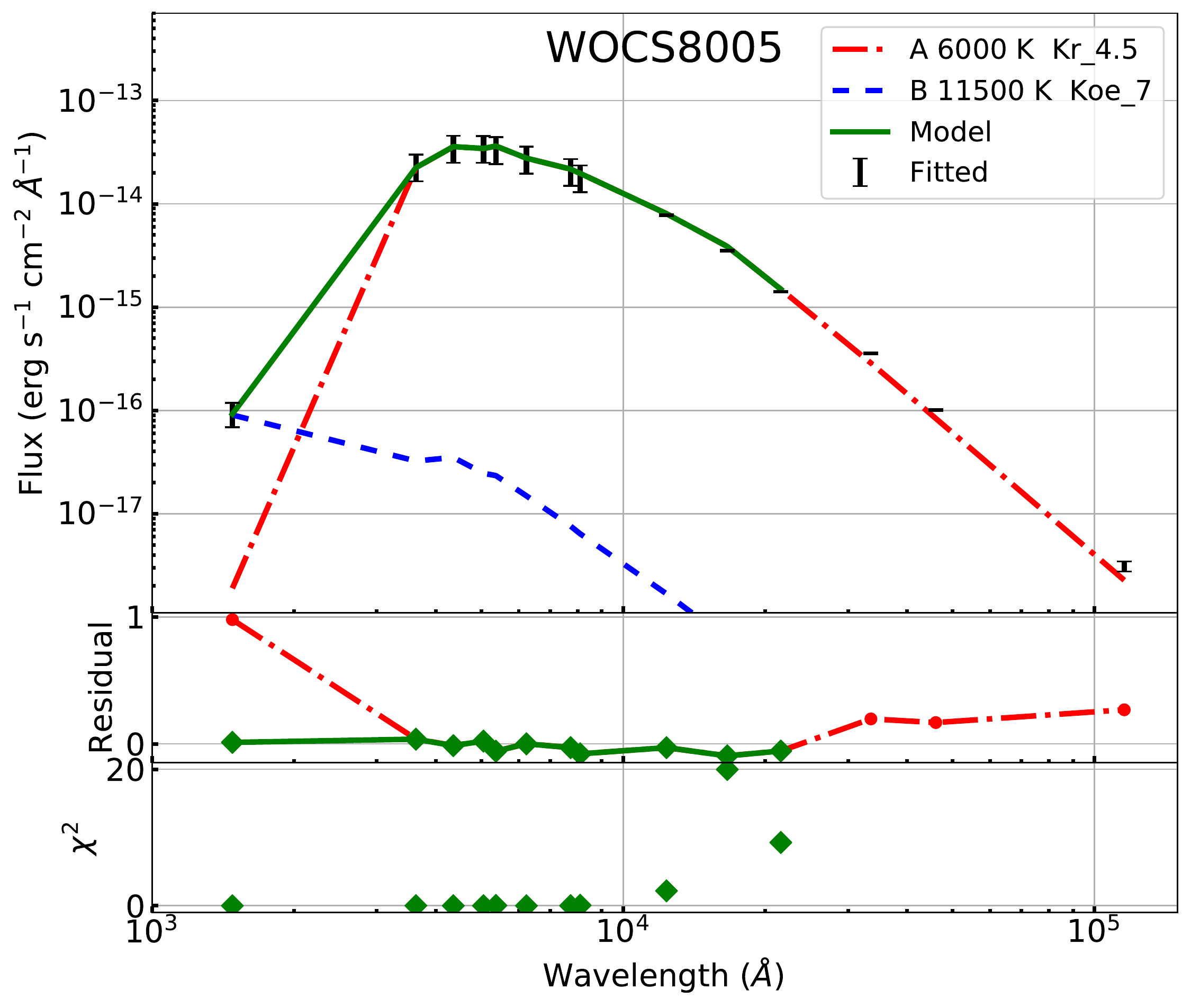} \\
    \includegraphics[width=.43\textwidth]{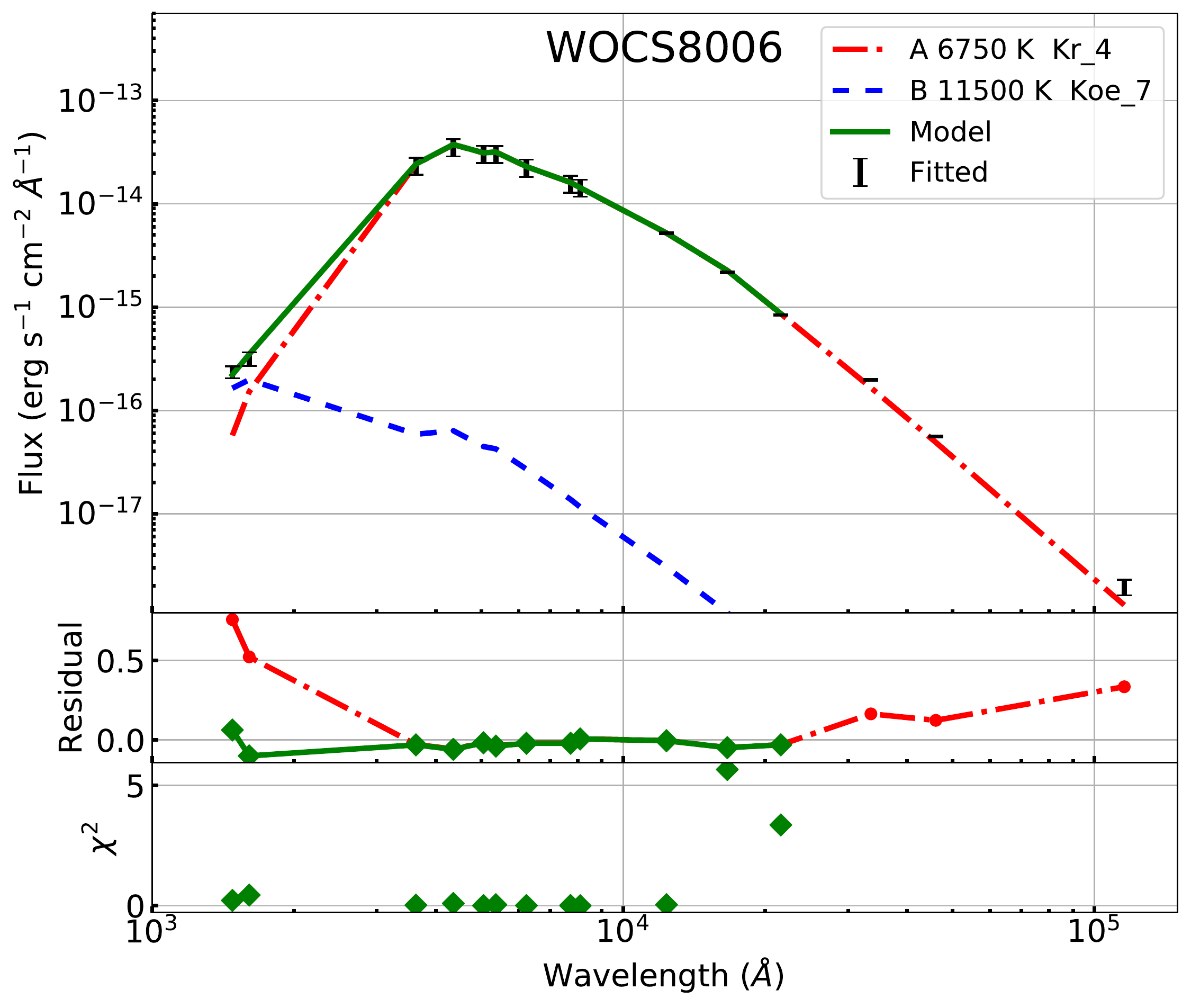} &
    \includegraphics[width=.43\textwidth]{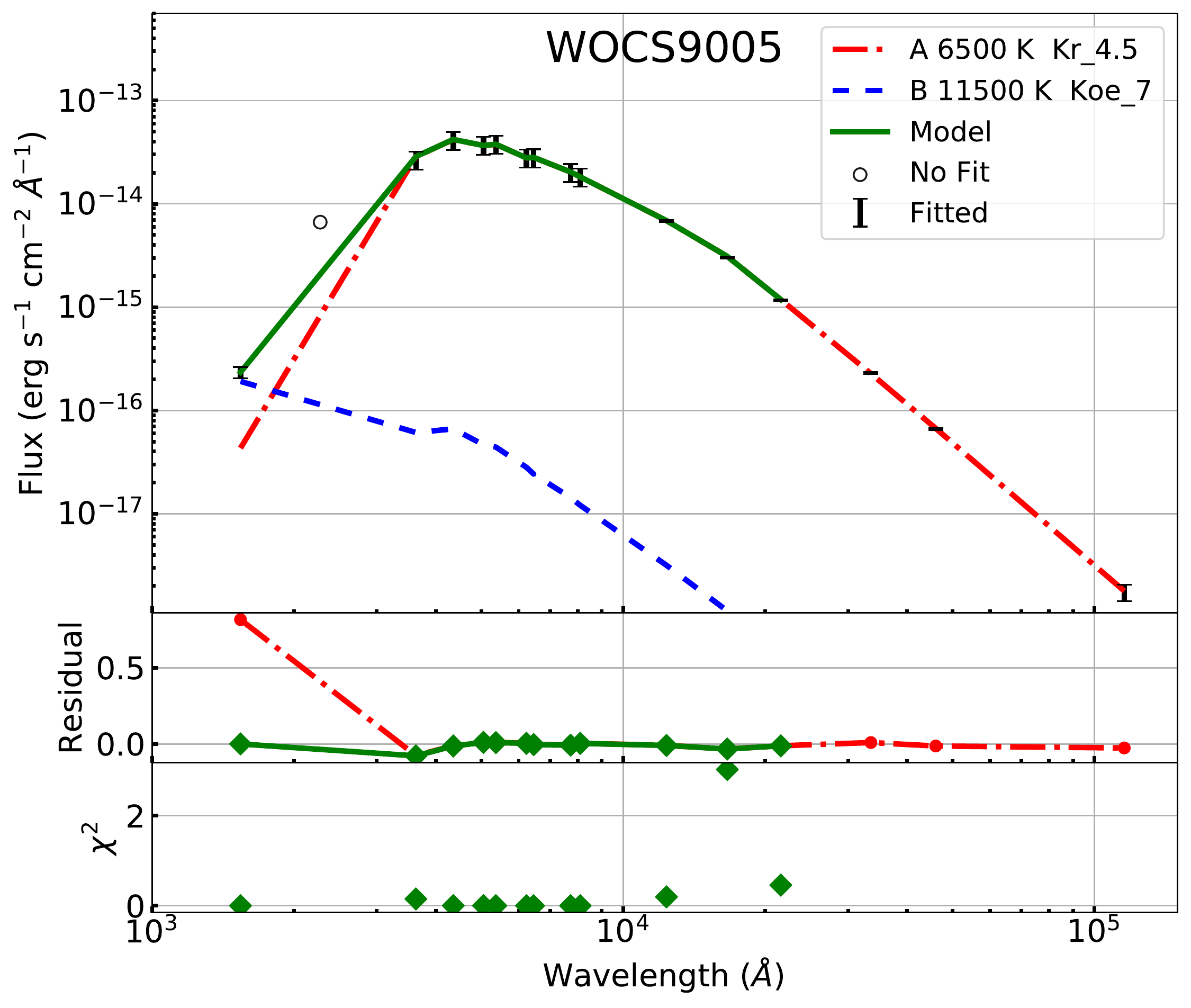} \\
  \end{tabular}
    \caption{Double SED fits of WOCS7010, WOCS8005, WOCS8006 and WOCS9005} 
  \label{fig:SED_double_7010}
\end{figure*}

\subsection{WOCS11005/S995}
\label{sec:WOCS11005}
This is described as a single member of M67 \citep{Geller2015}. It lies near the MSTO in optical the CMD and near the beginning of the BSS sequence in the UV-optical CMD. \citet{Melo2001} estimated a slow rotation of $v\ sin i=7.9km s^{-1}$.

The source was detected in one {\it UVIT} filter (F148W). The {\it GALEX}  FUV flux is more or less consistent with the F148W flux from the {\it UVIT} and provides support to the {\it UVIT} detection. The parameters of hotter companion lie within the predicted values of WD models in Fig.~\ref{fig:WD_mass_radius}. 

The absence of detection in X-ray suggests minimal chromospheric activity and therefore the source of the high UV flux could be a possible WD.
We cannot confirm the presence of WD due to only one detection in the \textit{UVIT}, thus it is noted as `WD?' in the Table~\ref{tab:All_para}.

\subsection{WOCS11011/S757}
\label{sec:WOCS11011}
\citet{Van2004} commented that the X-ray luminosity of the star is the result of coronal activity and is comparable with other known contact binaries of similar colour. The source is also observed by \citet{Mooley2015} in X-ray commenting it to be a W Uma type source. \citet{Geller2015} listed this source as SB1, HS Cnc, RR, W Uma and PV. The source lies well below the MSTO in the optical CMD and is the faintest optical source detected by \textit{UVIT}. 

We detected the source in only F148W filter in {\it UVIT}. 
Single SED fit over the optical and IR region results in a star with $T_{eff}=6000$ K and shows large UV excess flux (Fig.~\ref{fig:SED_double_1001}).
The double component fit suggested a hotter component of 11500 K to compensate for the UV flux.

The hotter companion's parameters are compatible with low mass WDs (Fig.~\ref{fig:WD_mass_radius}). The active nature of the binary could be the reason for the excess flux in the UV as well as X-ray. We cannot confirm the presence of a hotter WD component.

\subsection{WOCS2003/S1045}    
\label{sec:WOCS2003}
This source is very similar to WOCS1001/S1024 in terms of binary properties and mass of $>$1.18 \(M_\odot\) for each component of the SB2 system \citep{Mathieu1990}. \citet{Belloni1998} detected this RS CVn system in X-rays with $P=$ 7.65 days and $e = 0.007 \pm 0.005$ \citep{Latham1992}. They expected the system to be chromospherically active. \citet{Geller2015} described it as SB2 and PV.

We detected the source in only F154W filter and fitted the SED (Fig.~\ref{fig:SED_double_1001}) with one cooler and one hotter companion. We estimated the $T_{eff}$ = 6250 K cooler component and 9250 - 10000 K for hotter component with very low $\chi^2_{red}$. 
We suggest that this source is a binary with similar temperature stars and the excess flux in UV can be the result of chromospheric/coronal activity or hot-spot on the RS CVn system.

\subsection{WOCS2008/S1072}
\label{sec:WOCS2008}
\citet{Belloni1998} detected the source in X-rays but the X-ray emissions are not credited to any WD. The system has $P=$ 1495 days and $e=$ 0.32 \citep{Mathieu1990}. \citet{Geller2015} described the source as a YG, SB1 and a BSS candidate. \citet{Bertelli2018} calculated $T_{eff}=5915 K$ and found the chemical abundances consistent with a turn off stars and no signatures of recent MT in C abundance. They suggested it is a product of 3 stars formed by Kozai-cycle-induced merger in a hierarchical triple system \citep{Perets2009}. 

We observe a large UV flux consistently in three \textit{UVIT} filters.
Double component SED fitting resulted in a hotter companion of 11500 to 12500 K and cooler companion of 6500 K (Fig.~\ref{fig:SED_double_2008}). The hotter companion's cooling age is $<$120 Myr and mass $<$ 0.2 \(M_\odot\).

The WD parameters fit well within 0.18 and 0.20 cooling curves of \citet{Panei2007} model (Fig.~\ref{fig:WD_mass_radius}). The {\it GALEX} FUV flux was lower than {\it UVIT}. The variation in FUV flux and X-ray detection could be the result of flares. 
The X-ray detection also means that there may be contamination in UV flux and thus making the parameters of hotter companion unreliable. 
On the other hand, the matching WD temperatures estimated by us and \citet{Landsman1997}, in case of WOCS2002 (YG + WD system), points to the possibility that the X-ray flux not contaminating the UV flux significantly. Hence, there may still be a WD companion with the estimated parameters present in the system.

\subsection{WOCS2012/S756}
\label{sec:WOCS2012}
\citet{Geller2015} listed this star as a single member. This is also one of the faintest source observed (22.3 mag in F148W). This has not been detected in X-rays.

We detected the source in only F148W filter. After fitting a cooler component, we observed excess flux in the UV region consistent with a WD companion of 11500 K. 
The {\it GALEX} FUV observation also showed UV flux larger than \textit{UVIT} detection (Fig.~\ref{fig:SED_double_2008}).
In the absence of any contradicting information and large UV flux, we suggest the source is comprised one MS star and one WD, although the parameters of WD will not be entirely accurate due to a single \textit{UVIT} data point. 

\subsection{WOCS2015/S792}
\label{sec:WOCS2015}
\citet{Geller2015} considered this as a single member and a possible BSS but suggested the possibility of a very long period binary or a BSS formed by collision. The absence of X-ray detection decreases the possibility of an interacting close binary or chromospheric activity. There was no variability detected in the light curves by \citet{Sandquist2003b}. \citet{Bertelli2018} found the APOGEE rotational velocity to be $v\ sin i=3.63 km s^{-1}$ and $T_{eff}=5943 K$. This star lies between the MSTO and YGs (WOCS2002, WOCS2008) in the optical CMD.

We detected the star in only F148W filter. We could fit the observed SED with one hotter (9750-10250 K) and one cooler (6250 K) components (Fig.~\ref{fig:SED_double_2008}). The large $\chi^2_{red}$ value for the fit is mostly due to a very small error in 2MASS.Ks filter magnitude. The {\it GALEX} detection in NUV is consistent with a single star while {\it GALEX} FUV has smaller excess than {\it UVIT}. The hotter companion parameters differ from the models of \citet{Bergeron2009} and \citet{Panei2007}, thus the excess UV flux may not be due to a WD. 

\subsection{WOCS3009/S1273}
\label{sec:WOCS3009}
\citet{Geller2015} listed it as a single member of M67 and a possible BSS. We did not find any X-ray detections or photometric variability in the literature.  

We observed significant excess UV flux in all three filters. The resultant SED fit, in Fig.~\ref{fig:SED_3009} (c), shows the existence of a 10000 - 11000 K hotter companion. The mass and age were estimated to be $<$0.2 \(M_\odot\) and $<$ 200 Myr respectively. The estimated WD parameters are not fully compatible with the WD models. Therefore, we are not confirming the presence of a WD in the system.

\subsection{WOCS4003/S1036}
\label{sec:WOCS4003}
This source is an EV Cnc of W Uma type with $P=$ 0.44 days and $e=0.00$. \citet{Belloni1998} detected it in X-rays and related the detection to chromospheric activity and rapid rotation. \citet{Yakut2009} found the light curve to be unusual for a contact binary and estimated the temperature of two components as $T_{hotter}=6900$ K and $T_{cooler}=5200-5830$ K. The system lies slightly blueward of MSTO in the optical CMD.

We found the source to have a very large UV flux in 3 \textit{UVIT} filters, not explained by both these components' continuum flux.
Due to the unavailability of individual parameters, we performed the SED fit assuming a single cooler component (6500 K) and a hotter component (10250 K, Fig.~\ref{fig:SED_double_2008}).

The Fig.~\ref{fig:WD_mass_radius} shows that the parameters of hotter component strongly deviate from the WD models. Thus, we deduce the UV flux is the result of chromospheric activity or spots. 

\subsection{WOCS4015/S1456}
\label{sec:WOCS4015}
This star was only categorised as a single member by \citet{Geller2015}. This source lies just below the MSTO of optical CMD but shifts bluer in UV-optical CMD. 

We detected the star in only F148W filter. The double fit has high $\chi^2_{red}$ (Fig.~\ref{fig:SED_double_4015}) but most of it can be attributed to small errors in 2MASS filter magnitudes. The WD parameters lie within the expected region in Fig.~\ref{fig:WD_mass_radius}, but the single detection in {\it UVIT} bars us from confirming the presence and parameters of the hotter companion.

\subsection{WOCS5007/S1071}
\label{sec:WOCS5007}
\citet{Geller2015} listed this as a single member of M67. There is a small blueward shift from the optical to the UV-optical CMD. We detected the source in F154W and F169M filters, both near the limiting magnitude. The double SED fit gives us the parameters of a hotter companion of 9750 to 11000 K (Fig.~\ref{fig:SED_double_4015}).

The absence of any X-ray detections suggests that there is insignificant chromospheric activity. The estimated parameters of the hotter companion lie not too far from the WD the models (figure). We do not confirm the presence of a WD companion due to non-detection in F148W filter.

\subsection{WOCS5013/S1230}
\label{sec:WOCS5013}
This is a single member as described by \citet{Geller2015}. Similar to WOCS5007, there is a small shift in the position of the star from optical to UV-optical CMD.

We detected the source in only F148W filter. The SED fit shows high UV flux compared to a star with T$_{eff}=$ 6250 K (Fig.~\ref{fig:SED_double_4015}). The double fit shows the presence of a hotter companion ($T_{eff}=$ 11500 K) with temperature and mass consistent with the WD models. On the other hand, we need deeper observations in multiple filters to confirm the presence and parameters of the hotter component.

\subsection{WOCS7005/S1274}
\label{sec:WOCS7005}
\citet{Geller2015} described the source as a single member. It lies near the MSTO in the optical CMD. In the V, (F169M-V) CMD it shifts near the beginning of the BSS branch of the isochrone (This CMD is not shown in Fig.~\ref{fig:CMD}).

We detected the source in only F169M filter. Our double fit (Fig.~\ref{fig:SED_double_4015}) shows a large UV flux resulting in a hotter companion of 0.2 to 0.33 \(M_\odot\). According to Fig.~\ref{fig:WD_mass_radius}, the companion parameters are similar to WD parameters. But the single filter detection and lower {\it GALEX} FUV flux bars us from decisively claiming the presence of a WD companion. 

\subsection{WOCS7010/S1083}
\label{sec:WOCS7010}
\citet{Geller2015} listed this as a single member source. It lies just below MSTO in the optical CMD but it is much bluer in the V, (F154W-V) CMD (not shown in Fig.~\ref{fig:CMD}). 

We detected the star in F154W filter alone. The resultant single SED suggests excess UV flux (Fig.~\ref{fig:SED_double_7010}). {\it GALEX} observations are consistent with double fit with a cooler component (6750 K) and a hotter component (10750 - 11500 K).
The mass and radius of the hotter companion are compatible with the WD models but due to the single filter detection, we only hint at the possibility of the presence of a WD. 

\subsection{WOCS8005/MMJ5951}
\label{sec:WOCS8005}
This is a single member \citep{Geller2015} located near the MSTO in the optical CMD, but bluer in the UV-optical CMD.

We detected a large UV flux residual and a small IR residual after fitting a cooler companion SED (Fig.~\ref{fig:SED_double_7010}). The double fit gives the hotter companion parameters as 120 to 210 Myr old WD with a mass of 0.2 to 0.33 \(M_\odot\). Fig.~\ref{fig:WD_mass_radius} shows the obtained parameters are compatible with a WD. The only caveat is that the detection is only in F148W {\it UVIT} filter near the limiting magnitude of the observations. Deeper observations in UV are required for further characterising the source.

\subsection{WOCS8006/S2204}
\label{sec:WOCS8006}
According to \citet{Geller2015}, this is a single member BSS candidate. We notice a smaller blue shift from the optical to the UV-optical CMD when compared to other detected members. \citet{Bertelli2018} calculated a T$_{eff}=$ 6650 K from APOGEE spectra.

We created the SED using detections in 2 {\it UVIT} filters. The final double fit shows a mild IR excess in WISE filters (Fig.~\ref{fig:SED_double_7010}). Similar to WOCS8005, the high $\chi^2$ value of the fit is mostly due to a small error in 2MASS filter magnitudes. Without any known activity on the surface and consistency with the models in Fig.~\ref{fig:WD_mass_radius}, we propose the possibility of a WD companion of $T_{eff}=$ 11500 K with a mass of $\sim$ 0.2 \(M_\odot\) indicating an MT in last 140 Myr. The cooler star's temperature of T$_{eff}=$ 6750 K matches with \citet{Bertelli2018}.

\subsection{WOCS9005/S1005}
\label{sec:WOCS9005}
\citet{Geller2015} listed the source as an SB1 BSS, but also noted that it is not a good candidate for a BSS due to its closeness to MSTO. \citet{Leiner2019} found the orbital properties as $P= 2769$ days, $e=0.15$ and a binary mass function = 0.0368.

We detected the source in only F154W filter. The resulting best fit suggests the source is composed of one hotter and one cooler component (Fig.~\ref{fig:SED_double_7010}). The hotter component of 11500 K and $\sim$ 0.2 \(M_\odot\) is a possible WD candidate due to the similarity to the WD models in Fig.~\ref{fig:WD_mass_radius}.

\subsection{Triple Systems}
\label{sec:triples}
\begin{figure}
  \centering
  \begin{tabular}{c c}
    \includegraphics[width=.45\textwidth]{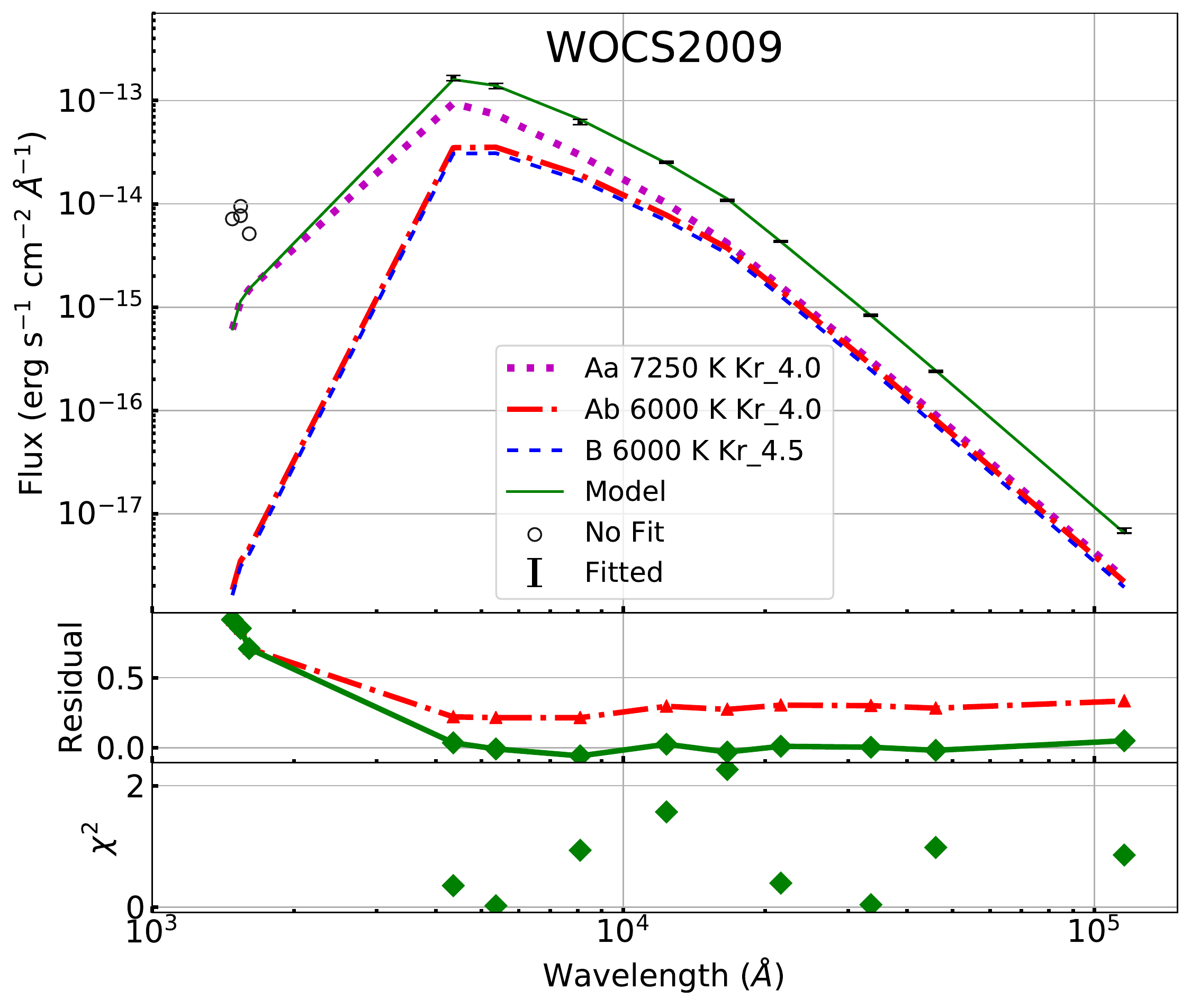} &
     \includegraphics[width=.45\textwidth]{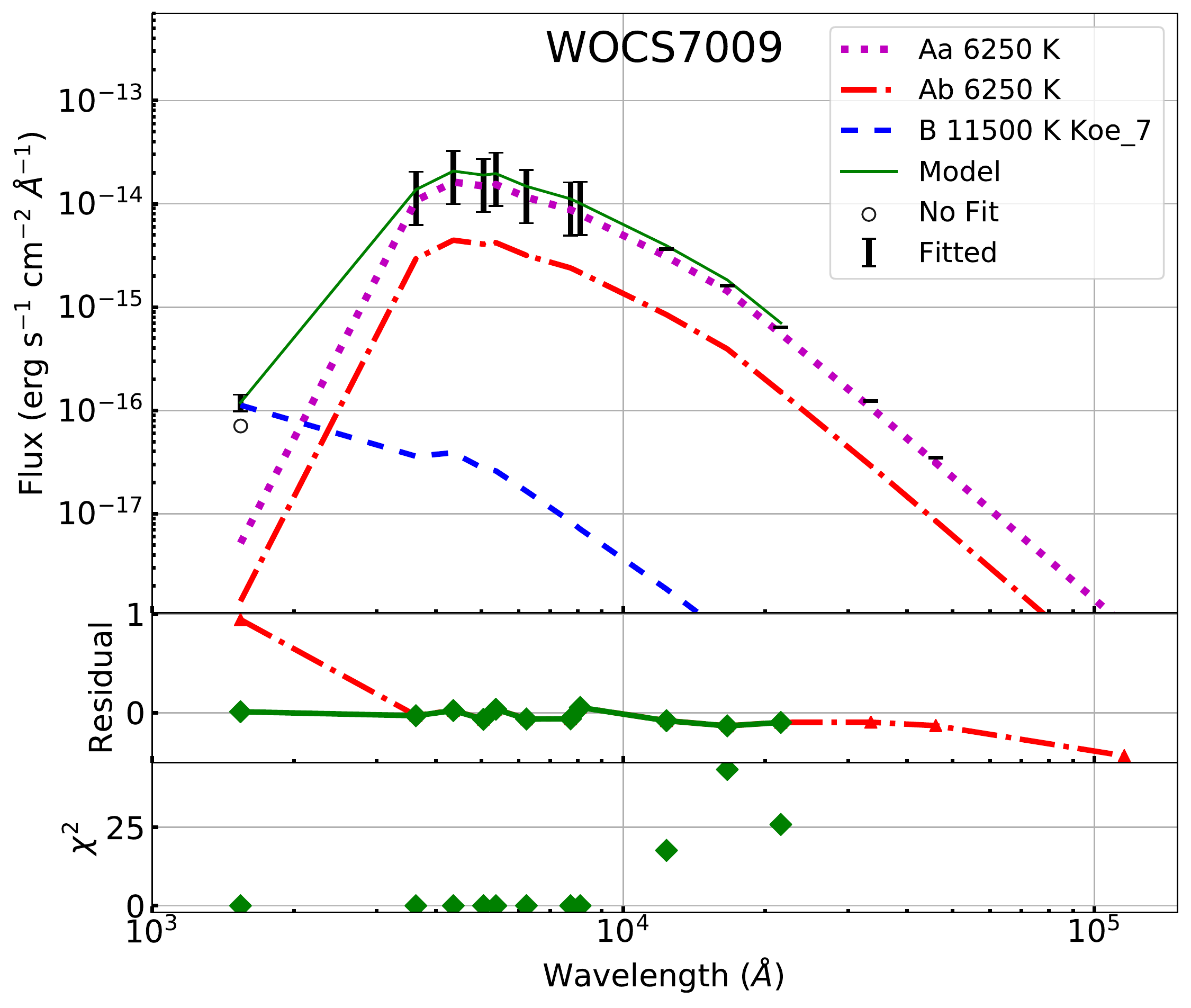} \\
 \end{tabular}  
  \caption{SEDs of triple systems, \textbf{WOCS2009:} Model SED using two known components (Aa and Ab) with third component B fitted only to the optical to IR residual. 
  \textbf{WOCS7009:} System with 2 known components (Aa and Ab) and a possible third component fitted to the UV residual.
  }
  \label{fig:SED_triples}
\end{figure}

\subsubsection{WOCS2009/S1082}
\label{sec:WOCS2009}
\citet{Goranskij1992} determined that WOCS2009 is an eclipsing close binary system with $P=$ 1.0677978$\pm$0.0000050 days. \citet{Belloni1998} detected the RS CVn system in X-ray suggesting active regions on the surface of the close binary. \citet{Shetrone2000} found the radial velocity variations of the eclipsing binary to be not compatible with the period of a close binary.
\citet{Van2001} studied the light curves of the eclipsing binary assuming a circular orbit (for close binary) and comprising of non-spotted stars and proposed the presence of the third component with a longer period. Further study by \citet{Sandquist2003} categorised the system as an ES Cnc, re-estimated the parameters of the close binary and found the orbital parameters of the third component as $P=$ 1188.5$\pm$6.8 days and $e=$ 0.568$\pm$0.076. \citet{Leigh2011} analysed this object using energy conservation in stellar interactions and concluded that the total mass of WOSC2009 is about 5.8 \(M_\odot\) which demands a 3+3 encounter for the formation of the present triple system.

We detected this star in all the three \textit{UVIT} filters and the \textit{GALEX} FUV flux is similar to the {\it UVIT} fluxes.
Hereafter the close binary components will be referred to as Aa and Ab while the third component will be referred to as B. \citet{Sandquist2003} calculated the radius and the temperature of Aa and Ab, while they only estimated the temperature of B. We calculated the flux from Aa and Ab components and found the combined SED flux to have a $\sim$ 0.25 residual in optical and IR region.

We fitted a third MS star (Kurucz model) SED to the residual in the optical and IR region. The resultant best fit gave the temperature of B component as 6000 K, which is consistent with the estimate of 6850 K by \citet{Sandquist2003}. 
Our estimates of L and $T_{eff}$ for the B component are found to be similar to the Ab component.
We detect a significant excess in the UV flux, even after fitting the three-component SED. The excess UV flux is likely to be the result of spot activities in the close binary which is an RS CVn system.

\subsubsection{WOCS7009/S1282}
\label{sec:WOCS7009}
\citet{Belloni1998} detected the AH Cnc contact binary in X-rays. \citet{Qian2006} and \citet{Pribulla2006} suggested the existence of a third component. \citet{Yakut2009} calculated the MT rate of $9.4\times 10^{-8} M_{\odot} yr^{-1}$ for the binary with a mass ratio of 0.17, P = 0.3604360 $\pm$ 0.0000001 days. \citet{Peng2016} estimated the masses and radii of the components of the contact binary as 1.188$\pm$0.061 \(M_\odot\), 1.332$\pm$0.063 \(R_\odot\) and 0.185$\pm$0.032 \(M_\odot\), 0.592$\pm$0.051 \(R_\odot\) respectively through period analysis. 
 
We used parameters from \citet{Yakut2009} to calculate the contribution from Aa ($T_{eff}$ = 6300 K, log $g = 4.31$, R = 1.40 \(R_\odot\)) and Ab ($T_{eff}$ = 6275 K, log $g = 4.17$, R = 0.68 \(R_\odot\)) components. Due to model constraints we fitted the Aa and Ab components with $T_{eff}=$ 6250 K and log $g = 4.0$. The combined model flux of these 2 components matches with observations in optical region suggesting these two components are sufficient to account for the observed flux in the optical-IR SED. We detect an excess flux in F148W, which we fit with a WD model in the SED (Fig.~\ref{fig:SED_triples}).

The parameters of hotter component lie within the model predictions in Fig.~\ref{fig:WD_mass_radius}, but the contact binary itself can be responsible for the UV flux as seen in WOCS11011, WOCS2003 and WOCS2009. The results from SED fit does not find the presence of any cooler third component, whereas a hotter component may or may not be present.

\end{document}